# Efficient Implementation of ADER Schemes for Euler and Magnetohydrodynamical Flows on Structured Meshes – Comparison with Runge-Kutta Methods

By

Dinshaw S. Balsara[1], Chad Meyer[1], Michael Dumbser[3], Huijing Du[2]

and Zhiliang Xu[2]

(dbalsara@nd.edu, cmeyer8@nd.edu, michael.dumbser@iag.uni-stuttgart.de,

hdu@nd.edu, zxu2@nd.edu)

[1] Physics Department, University of Notre Dame, 225 Nieuwland Science Hall,

Notre Dame, IN, 46556, USA

[2] Applied and Computational Mathematics and Statistics Department, University of Notre

Dame, Hayes-Healey Hall,

Notre Dame, IN, 46556, USA

[3] Laboratory of Applied Mathematics, University of Trento, Via Mesiano 77, I-38100

Trento, Italy

## Abstract

ADER (Arbitrary DERivative in space and time) methods for the time-evolution of hyperbolic conservation laws have recently generated a fair bit of interest. The ADER time update can be carried out in a single step, which is desirable in many applications. However, prior papers have focused on the theory while downplaying implementation details. The purpose of the present paper is to make ADER schemes accessible by providing two useful formulations of the method as well as their implementation details on three-dimensional structured meshes. We therefore provide a detailed formulation of ADER schemes for conservation laws with non-stiff source terms in nodal as well as modal space along with useful implementation-related details. A good implementation of ADER requires a fast method for transcribing from nodal to modal space and vice versa and we provide innovative transcription strategies that are computationally efficient. We also provide details for the efficient use of ADER schemes in obtaining the numerical flux for conservation laws as well as electric fields for divergence-free magnetohydrodynamics (MHD). An efficient WENO-based strategy for obtaining zone-averaged magnetic fields from face-centered magnetic fields in MHD is also presented. Several explicit formulae have been provided in all instances for ADER schemes spanning second to fourth orders.

The schemes catalogued here have been implemented in the first author's RIEMANN code. The speed of ADER schemes is shown to be almost twice as fast as that of strong stability preserving Runge-Kutta time stepping schemes for all the orders of accuracy that we tested. The modal and nodal ADER schemes have speeds that are within ten percent of each other. When a linearized Riemann solver is used, the third order ADER schemes are half as fast as the second order ADER schemes and the fourth order ADER schemes are a third as fast as the third order ADER schemes. The third order ADER-WENO scheme, either with an HLL or linearized Riemann solver, represents an excellent upgrade path for scientists and engineers who are working with a second order Runge-Kutta based total variation diminishing (TVD) scheme. We demonstrate the ADER scheme's order of accuracy for hydrodynamical and MHD problems. Several stringent test problems have also been catalogued.

## I) Introduction

The accurate simulation of Euler and magnetohydrodynamical (MHD) flows has received considerable recent attention. Simulation strategies for hyperbolic systems that go beyond second order in space and time have started reaching maturity. Robust methods for reconstructing the solution with better than second order accuracy in space were achieved rather early (Harten *et al.* [41], Shu & Osher [54], [55], Barth & Frederickson [13], Liu, Osher & Chan [45], Jiang & Shu [44], Balsara & Shu [5], Suresh & Huynh [58], Dumbser & Käser [28], Balsara et al. [11], Colella & Sekora [21]). Methods that evolve some or all spatial moments of the solution have also been developed (Cockburn & Shu [19], Qiu & Shu [48], Schwartzkopff, Dumbser & Munz [53], Balsara et al. [9], Dumbser et al. [30]) but even they have to resort to limiters and reconstruction at discontinuities (Xu, Liu & Shu [66], [67]). When the solution becomes discontinuous, a stable spatial reconstruction again becomes important even for schemes that evolve moments of the solution. Early attempts to achieve temporal accuracy that goes beyond second order relied almost exclusively on a method of lines approach using strong stability preserving Runge-Kutta (RK) time stepping (Shu & Osher [54], [55], Spiteri & Ruuth [56], [57]). In such approaches, the solution has to be reconstructed at each fractional sub-step of the multi-stage RK method. Likewise, the Riemann problem has to be solved at each of the stages of the RK method. This adds to the cost of the method. RK time discretizations that go beyond third order also require additional steps owing to the Butcher barriers. The RK methods do, however, have the advantage of simplicity; the same, well-known reconstruction strategy and Riemann solver can be invoked at each stage of the method. In recent years, ADER (Arbitrary DERivative in space and time) methods have emerged as competitors to the RK methods (Titarev & Toro [60], [61], Toro & Titarev [62], Dumbser, Enaux & Toro [31]; Dumbser *et al.* [30]; Balsara *et al.* [11]). In the course of the above-mentioned papers, the ADER schemes have undergone a substantial amount of revision. Thus instead of formally solving the generalized Riemann problem (van Leer [64], Ben-Artzi & Falcovitz [14]) at zone boundaries, the newer ADER methods (Dumbser, Enaux & Toro [31]; Dumbser *et al.* [30]; Balsara *et al.* [11]) evolve the spatially accurate solution in time so as to obtain a space-time accurate representation of the solution within a zone. Very efficient methods are then used (Dumbser *et al.* [29]) to solve the Riemann problem only once at each zone boundary while retaining the requisite space-time accuracy. In this paper we focus on a continuous Galerkin variant of ADER schemes, referred to as ADER-CG. These schemes are very efficient in their treatment of hyperbolic conservation laws with non-stiff source terms. Their efficiency is achieved by evaluating a majority of the fluxes and source terms only once at the beginning of each time step.

While the ADER methods have been detailed for unstructured meshes (Dumbser *et al.* [30]) and structured meshes (Balsara *et al.* [11]), they still remain somewhat mysterious to most practitioners. Part of the difficulty stems from the fact that these early papers sought to elaborate the conceptual basis for the method and, therefore, did not pay much attention to the implementation details. The first goal of this paper is to demystify ADER schemes on structured meshes by providing efficient and explicit recipes for implementing ADER schemes. The implementation details provided in the present paper are designed to be reasonably self-contained. Several innovations have been devised here to make the implementation of ADER schemes as efficient as possible. Explicit closed-form formulae are provided here for second, third and fourth orders. At second order, the ADER schemes indeed reduce to a predictor-corrector method. The present paper focuses on implementation details and downplays the conceptual basis of ADER methods. Please look up (Dumbser *et al.* [30]; Balsara *et al.* [11]) for conceptual details. Another part of the difficulty stems from the fact that there is not a single, unique ADER formulation. The methods can be formulated in nodal space if an elaborate nodal basis set is constructed (Dumbser *et al.* [30]). They can also be formulated in modal space using basis sets that are natural for the specific mesh topology (Balsara *et al.* [11]). Specifically, one uses tensor product Hermite basis for three-dimensional structured meshes with hexahedral zones and Dubiner [27] basis for unstructured meshes with tetrahedral zones. Nodal space formulations may have a slight advantage in terms of computational efficiency because they dispense with the transcription from nodal to modal space and back. The second goal of this paper is to catalogue nodal space formulations of ADER on structured meshes. Modal space formulations have the advantage of yielding a compact, elegant and intuitively obvious set of time evolution equations. They may also be more easily extensible because they permit the inclusion of viscosities and resistivities (Dumbser & Balsara [32]) and stiff source terms. This advantage comes at the cost that one has to be able to efficiently transcribe from modal to nodal space and vice versa, thus making the ADER formulation in modal space quite competitive. If this transcription can be made very efficient, then the more easily understood modal space formulations of ADER become competitive. The third goal of this paper is to show that it is possible to arrive at an efficient strategy for transcribing from modal to nodal space and vice versa. The essential idea is to select one's nodes and to organize one's calculation in such a way that the evaluation of the solution at each nodal point makes if easier to obtain the solution at the subsequent nodal point. Likewise, the evaluation of each mode from the nodal solution makes it simpler to obtain the next mode. Such an exercise has so far only been carried out for structured, hexahedral meshes. A similar simplification should indeed exist for unstructured, tetrahedral meshes. Finding such a simplification is very beneficial because the ADER time-update equations in the modal space of the Dubiner basis is almost as simple, elegant and easy to interpret as the corresponding equations in Hermite basis. We leave that exercise for future work.

The previous paragraph lays out the plan of the paper as it pertains to obtaining a space-time representation of the solution within a zone. An efficient implementation of finite-volume ADER schemes requires a good strategy for obtaining the space-time averaged numerical fluxes at zone boundaries. The traditional method of Cockburn & Shu [19] would require the selection of a large number of quadrature points at each boundary. Riemann problems would have to be evaluated at each of those quadrature points to yield a spatially-averaged representation of the flux at each facial boundary. The evaluation of a large number of Riemann problems at each face makes this approach prohibitively costly for any RK time-stepping scheme. Extension of such a method to ADER's space-time formulation would be even more computationally expensive. As a result, Dumbser *et al.* [29] simplified the process by freezing the wave speeds and eigenstructure at each face. This then permits the resolved, space-time averaged flux from certain Riemann solvers, such as the HLLE Riemann solver (Harten, Lax & van Leer [39], Einfeldt *et al.* [34]) and the linearized Riemann solver (Roe [49], Harten & Hyman [40]), to be written in terms of space-time averages of the conserved variables and fluxes on either side of the zone boundary. The fourth goal of this paper is to show that we can significantly simplify the evaluation of fluxes and electric fields and to provide explicit formulae that are suitable for implementation. If an ADER scheme has been used to obtain a space-time representation of the solution within a zone, then the strategy from Dumbser *et al.* [29] can be used to obtain a space-time averaged numerical flux. If, however, one is using a RK time-stepping strategy then the method can still be very beneficial when evaluating the spatially-averaged numerical flux at a boundary. Thus the description provided here benefits ADER as well as RK time-update strategies. The fifth goal of this paper is, therefore, to catalogue the relative speeds of ADER and RK time-update strategies when all the improvements described in this paragraph are implemented in each of them.

The structure of the compressible MHD eigensystem is well-understood (Jeffrey & Taniuti [43], Brio & Wu [19], Roe and Balsara [50]) making it possible to develop high resolution shock-capturing methods for this system (Dai & Woodward [23], Ryu & Jones [51], Balsara [1], [2], Falle, Komissarov & Joarder [35], Crockett et al [22], Balsara et al. [11], Balsara [10] and Ustyugov *et al.* [63]). An early version of an ADER scheme for MHD was also presented in Taube *et al.* [59]. The mass, momentum and energy equations have a flux form that that parallels Euler flow, with the result that Godunov schemes provide a seamless strategy for their treatment. The magnetic field vector, **B**, in MHD obeys the evolutionary equation

$$\frac{\partial \mathbf{B}}{\partial t} = \nabla \times (\mathbf{v} \times \mathbf{B}) \tag{1}$$

where **v** is the fluid velocity. Eqn. (1) is such that the magnetic field remains divergence-free in its time-evolution, i.e. it satisfies the constraint

$$\nabla \cdot \mathbf{B} = 0 \tag{2}$$

Several early workers (Yee [68], Brackbill & Barnes [16], Brecht *et al.* [17], Evans & Hawley [33] and DeVore [26]) have shown the value of designing time-evolutionary schemes for eqn. (1) that naturally satisfy the constraint in eqn. (2). Such schemes require the normal components of the magnetic field to be defined at zone faces. To be specific, the facially-averaged *x*-component of the magnetic field is defined at the *x*-face of each zone with the *y* and *z*-components of the magnetic field defined at the corresponding *y* and *z* zone faces. In order to evaluate conserved variables and fluxes at any nodal point within each zone, a higher order reconstruction of the magnetic field has to be available within that zone. Such a divergence-free reconstruction strategy for the magnetic field has been designed at second order (Balsara [6], [8]) and it has been extended to higher orders in Balsara [10]. The equations of relativistic MHD (Balsara [7]) also obey eqns. (1) and (2), with the result that all of the advances reported in this and the above papers carry over seamlessly to relativistic MHD. The constraint in eqn. (2) is vitally important in enabling us to take the moments of the magnetic field's components within each zone face and obtain from them a reconstruction of the magnetic field vector over the volume of a zone. The method described in Balsara [10] is, therefore, a watertight method for reconstructing the divergence-free magnetic field. In order to preserve the constraint in eqn. (2), it has to produce more than the minimum requisite number of moments for the magnetic field within each zone. Besides, the moments of the components of the magnetic fields have to be evaluated to the desired order within each zone face. Those are extra steps that are not mandated by the order property but rather by the need to satisfy eqn. (2) at all points within a zone. It would be easier to obtain zone-averaged magnetic fields directly from the facially-averaged magnetic fields. At second order, this is indeed obtained via a simple arithmetic averaging. The zone-averaged magnetic fields can then be directly used to obtain the moments of the magnetic field. The deficiency with a literal implementation of this plan is that the facially-averaged normal components of the magnetic field evaluated from either side of a face would not match up at a face. To avoid this loss of consistency, we design a modified weighted essentially non-oscillatory (WENO) scheme that takes facially averaged magnetic field components and obtains the corresponding zone-averaged magnetic field as well as a few of the moments that are needed for restoring consistency of the normal component of the magnetic field when evaluated from either side of a zone's face. The sixth goal of this paper is to catalogue such a WENO strategy at all orders up to four.

Divergence-free schemes for MHD of the type described in the previous paragraph also require that a multidimensionally upwinded electric field be available at the zone edges. For ideal MHD, the electric field vector $\mathbf{E}$ is related to the velocity and magnetic field vectors by $\mathbf{E} \equiv -\mathbf{v} \times \mathbf{B}$. Several papers have been written on methods for achieving such a multidimensional upwinding within the context of conventional higher order Godunov methodology (Dai & Woodward [24], Ryu *et al.* [52], Balsara & Spicer [3], [4], Balsara [8], Londrillo & DelZanna [46], Gardiner & Stone [37]). All of the authors in the previous sentence try to achieve multidimensional effects while retaining the conventional, one-dimensional Riemann solvers. Genuinely multidimensional Riemann solvers for MHD have also become available (Balsara [12]) and they give us perspective on how multidimensional upwinding might actually be achieved. The most important insights from that work have been integrated into this paper. The seventh goal of the present paper is to catalogue efficient strategies for obtaining the electric fields at zone edges within the context of the conventional approaches while assimilating some of the insights from the multidimensional Riemann solver. The alternative to designing time-evolutionary schemes for eqn. (1) that naturally satisfy the constraint in eqn. (2) consists of modifying the MHD equations (Powell [47], Dedner et al [25]) so as to strictly bring them into a conservation form. Such strategies for numerical MHD should also be benefited by the first five goals of this paper.

Simulations on structured meshes are usually required to satisfy symmetries that usually do not occur on unstructured meshes. For example, when a hydrodynamical or MHD blast is made to propagate through a perfectly static medium on a structured mesh, one might want the velocities in opposite directions to be perfectly symmetrical. Such a perfect symmetry can indeed be achieved in a numerical method, but only if all aspects of the solution methodology are perfectly symmetrical within each zone. There are other applications, like fully developed turbulence, where no such symmetry is desired in the numerical scheme because the application itself does not have any symmetries. When a perfect symmetry is needed, it is best to use symmetrically placed nodes within a zone and to solve the problem in modal space. Such an arrangement is detailed in the body of the text. It carries a small cost penalty stemming from the fact that the number of nodes has to be somewhat larger than the number of modes, resulting in a larger number of flux evaluations at each of those nodes. For all orders described here, the number of nodes used is quite comparable to the number of modes, so the inefficiency is not substantial. For scientific applications where no obvious numerical symmetry needs to be respected, it is acceptable to have an unsymmetrical placement of nodes while only utilizing as many nodes as one has modes. It then becomes more efficient to resort to a nodal ADER scheme that directly carries out the space-time update in nodal space. Such an arrangement of nodes, along with the nodal update equations, is catalogued in the Appendices of this paper.

The schemes catalogued here have been implemented in the first author's RIEMANN code. Section II provides a schematic description of an entire time step using an ADER scheme. Section III catalogues the WENO strategy for obtaining zone-averaged magnetic fields and their moments from the facially-averaged magnetic fields in MHD. This section can be skipped by those who are not interested in MHD. Section IV details the ADER-CG space-time update equations in modal space along with an efficient transcription for going from modal to nodal space and back again from nodal to modal space. Implementation details are explicited for second, third and fourth order. Section V catalogues an efficient strategy for obtaining the resolved, space-time averaged numerical flux from certain Riemann solvers at zone boundaries. Additional extensions for edge-averaged electric fields for MHD are also catalogued. Section VI demonstrates the order of accuracy of the methods developed here. Section VII inter-compares speeds for ADER and RK time stepping. Section VIII catalogues several stringent hydrodynamical tests while Section IX catalogues MHD tests. The Appendices A, B and C catalogue ADER-CG schemes that have been formulated in nodal space at second, third and fourth orders respectively, thereby complementing the three Sub-Sections in Section IV.

## II) Schematic Description of a Time Step Using an ADER Scheme

The MHD equations can be written in conservation form as

$$\frac{\partial}{\partial t}\begin{pmatrix} \rho \\ \rho v_x \\ \rho v_y \\ \rho v_z \\ \varepsilon \\ B_x \\ B_y \\ B_z \end{pmatrix} + \frac{\partial}{\partial x}\begin{pmatrix} \rho v_x \\ \rho v_x^2 + P + \mathbf{B}^2/8\pi - B_x^2/4\pi \\ \rho v_x v_y - B_x B_y/4\pi \\ \rho v_x v_z - B_x B_z/4\pi \\ (\varepsilon + P + \mathbf{B}^2/8\pi)v_x - B_x(\mathbf{v}\cdot\mathbf{B})/4\pi \\ 0 \\ (v_x B_y - v_y B_x) \\ -(v_z B_x - v_x B_z) \end{pmatrix}$$

$$+ \frac{\partial}{\partial y}\begin{pmatrix} \rho v_y \\ \rho v_x v_y - B_x B_y/4\pi \\ \rho v_y^2 + P + \mathbf{B}^2/8\pi - B_y^2/4\pi \\ \rho v_y v_z - B_y B_z/4\pi \\ (\varepsilon + P + \mathbf{B}^2/8\pi)v_y - B_y(\mathbf{v}\cdot\mathbf{B})/4\pi \\ -(v_x B_y - v_y B_x) \\ 0 \\ (v_y B_z - v_z B_y) \end{pmatrix} + \frac{\partial}{\partial z}\begin{pmatrix} \rho v_z \\ \rho v_x v_z - B_x B_z/4\pi \\ \rho v_y v_z - B_y B_z/4\pi \\ \rho v_z^2 + P + \mathbf{B}^2/8\pi - B_z^2/4\pi \\ (\varepsilon + P + \mathbf{B}^2/8\pi)v_z - B_z(\mathbf{v}\cdot\mathbf{B})/4\pi \\ (v_z B_x - v_x B_z) \\ -(v_y B_z - v_z B_y) \\ 0 \end{pmatrix} = 0$$

(3)

where $\rho$ is the density, P is the pressure, $\mathbf{v}$ is the velocity vector, $\mathbf{B}$ is the magnetic field vector, $\varepsilon = \rho \mathbf{v}^2/2 + P/(\gamma-1) + \mathbf{B}^2/8\pi$ is the total energy and $\gamma$ is the ratio of specific heats. The Euler equations can be obtained from eqn. (3) by setting the magnetic fields to zero. We discretize eqn. (3) on a Cartesian mesh with zones of size $\Delta x$, $\Delta y$ and $\Delta z$. Our computational task is to take a time step of size $\Delta t$. We formally write the above conservation law as

$$\frac{\partial U}{\partial t} + \frac{\partial F}{\partial x} + \frac{\partial G}{\partial y} + \frac{\partial H}{\partial z} = S \tag{4}$$

Here *U* is the vector of conserved variables and *F*, *G* and *H* are the fluxes in the three directions. We also extend the utility of this work by including a general, non-stiff source term *S*. We will also find it beneficial to map each zone to a unit reference element in space and time. The coordinates of this unit reference element are given by $(\xi,\eta,\zeta,\tau) \in [-.5,.5]^3 \times [0,1]$. In the unit reference element, eqn. (4) can be written as

$$\frac{\partial u}{\partial \tau} + \frac{\partial f}{\partial \xi} + \frac{\partial g}{\partial \eta} + \frac{\partial h}{\partial \zeta} = s \qquad (5)$$

with the transcription $u = U$, $f = \Delta t\, F/\Delta x$, $g = \Delta t\, G/\Delta y$, $h = \Delta t\, H/\Delta z$ and $s = \Delta t\, S$.

The first five components of eqn. (3) follow a straightforward conservation form and their one-step update from a time $t^n$ to a time $t^{n+1} = t^n + \Delta t$ in a zone labeled by subscripts "$i,j,k$" is given by

$$\bar{U}_{i,j,k}^{n+1} = \bar{U}_{i,j,k}^{n} - \frac{\Delta t}{\Delta x}\left(\bar{F}_{i+1/2,j,k} - \bar{F}_{i-1/2,j,k}\right) - \frac{\Delta t}{\Delta y}\left(\bar{G}_{i,j+1/2,k} - \bar{G}_{i,j-1/2,k}\right) - \frac{\Delta t}{\Delta z}\left(\bar{H}_{i,j,k+1/2} - \bar{H}_{i,j,k-1/2}\right) + \Delta t\, \bar{S}_{i,j,k}.$$

(6)

The overbars on the conserved variables in eqn. (6) denote volumetric averages at times $t^n$ and $t^{n+1}$. For eqn. (6) to be a high order update, the fluxes in eqn. (6) have to be averaged in space and time at the zone faces. The averaging process has to extend to the required order. Likewise, the source term has to be averaged over space and time.

For ideal MHD, the connection between the last three components of eqn. (3) and eqn. (1) can be solidified by making the identification

$$E_x = -G_8 = H_7 \;;\; E_y = F_8 = -H_6 \;;\; E_z = -F_7 = G_6 \;. \qquad (7)$$

Consequently, the last three components of the **F**, **G** and **H** fluxes could also be reinterpreted as electric fields in our dual approach. The electric fields are needed at the edge centers. The discretized, one-step update equations for the facially-averaged magnetic fields in eqn. (1) are then given by

$$\overline{B}_{x;\,i+1/2,j,k}^{n+1} = \overline{B}_{x;\,i+1/2,j,k}^{n} - \frac{\Delta t}{\Delta y \Delta z} \left( \Delta z \overline{E}_{z;\,i+1/2,j+1/2,k} - \Delta z \overline{E}_{z;\,i+1/2,j-1/2,k} + \Delta y \overline{E}_{y;\,i+1/2,j,k-1/2} - \Delta y \overline{E}_{y;\,i+1/2,j,k+1/2} \right)$$

$$\overline{B}_{y;\,i,j-1/2,k}^{n+1} = \overline{B}_{y;\,i,j-1/2,k}^{n} - \frac{\Delta t}{\Delta x \Delta z} \left( \Delta x \overline{E}_{x;\,i,j-1/2,k+1/2} - \Delta x \overline{E}_{x;\,i,j-1/2,k-1/2} + \Delta z \overline{E}_{z;\,i-1/2,j-1/2,k} - \Delta z \overline{E}_{z;\,i+1/2,j-1/2,k} \right)$$

$$\overline{B}_{z;\,i,j,k+1/2}^{n+1} = \overline{B}_{z;\,i,j,k+1/2}^{n} - \frac{\Delta t}{\Delta x \Delta y} \left( \Delta x \overline{E}_{x;\,i,j-1/2,k+1/2} - \Delta x \overline{E}_{x;\,i,j+1/2,k+1/2} + \Delta y \overline{E}_{y;\,i+1/2,j,k+1/2} - \Delta y \overline{E}_{y;\,i-1/2,j,k+1/2} \right)$$

(8)

The overbars on the magnetic field components denote facial averages at times $t^n$ and $t^{n+1}$. The overbars on the electric field components denote appropriate space and time averages at the edges. As before, the averaging process has to extend to the required order.

An entire ADER time step consists of implementing the three sub-steps described in the following three sub-sections. Taken together, these sub-steps yield the numerical fluxes in eqn. (6) and the numerical electric fields in eqn. (8).

**II.1) Reconstruction**

Within each zone we define a space of $L_S$ spatial basis functions. To obtain the order property we need $L_S$ = 4, 10 and 20 for second, third and fourth order schemes respectively. For structured, hexahedral meshes it is simplest to use tensor products of Hermite polynomials. The Hermite polynomials used in this work are given by

$$P_0(\xi) = 1 \;;\quad P_1(\xi) = \xi \;;\quad P_2(\xi) = \xi^2 - \frac{1}{12} \;;\quad P_3(\xi) = \xi^3 - \frac{3}{20}\xi \tag{9}$$

Starting with a vector of zone-centered, conserved variables $\hat{w}_1$ in each zone, we use a reconstruction strategy to obtain the $L_S$ spatial basis functions within that zone. Very efficient implementations of centered WENO reconstruction schemes at orders up to fifth have been described in Balsara et al. [11]. The reconstruction strategies described in that paper are equivalent to, yet much more efficient than, the ones described in Jiang & Shu [44] and Balsara & Shu [5] and they have the added advantage of directly obtaining the moments (weights) in terms of Hermite polynomials. The reconstructed polynomial is denoted by $w(\xi,\eta,\zeta)$ where $(\xi,\eta,\zeta) \in [-.5,.5]^3$ are the coordinates in the reference element, which is the unit cube for our purposes. At second, third and fourth orders the spatially reconstructed polynomials are given by:

$$\begin{aligned}
w(\xi,\eta,\zeta) &= \hat{w}_1 P_0(\xi) P_0(\eta) P_0(\zeta) \\
&+ \hat{w}_2 P_1(\xi) P_0(\eta) P_0(\zeta) + \hat{w}_3 P_0(\xi) P_1(\eta) P_0(\zeta) + \hat{w}_4 P_0(\xi) P_0(\eta) P_1(\zeta)
\end{aligned} \quad (10),$$

$$\begin{aligned}
w(\xi,\eta,\zeta) &= \hat{w}_1 P_0(\xi) P_0(\eta) P_0(\zeta) \\
&+ \hat{w}_2 P_1(\xi) P_0(\eta) P_0(\zeta) + \hat{w}_3 P_0(\xi) P_1(\eta) P_0(\zeta) + \hat{w}_4 P_0(\xi) P_0(\eta) P_1(\zeta) \\
&+ \hat{w}_5 P_2(\xi) P_0(\eta) P_0(\zeta) + \hat{w}_6 P_0(\xi) P_2(\eta) P_0(\zeta) + \hat{w}_7 P_0(\xi) P_0(\eta) P_2(\zeta) \\
&+ \hat{w}_8 P_1(\xi) P_1(\eta) P_0(\zeta) + \hat{w}_9 P_0(\xi) P_1(\eta) P_1(\zeta) + \hat{w}_{10} P_1(\xi) P_0(\eta) P_1(\zeta)
\end{aligned} \quad (11)$$

and

$$\begin{aligned}
w(\xi,\eta,\zeta) &= \hat{w}_1 P_0(\xi) P_0(\eta) P_0(\zeta) \\
&+ \hat{w}_2 P_1(\xi) P_0(\eta) P_0(\zeta) + \hat{w}_3 P_0(\xi) P_1(\eta) P_0(\zeta) + \hat{w}_4 P_0(\xi) P_0(\eta) P_1(\zeta) \\
&+ \hat{w}_5 P_2(\xi) P_0(\eta) P_0(\zeta) + \hat{w}_6 P_0(\xi) P_2(\eta) P_0(\zeta) + \hat{w}_7 P_0(\xi) P_0(\eta) P_2(\zeta) \\
&+ \hat{w}_8 P_1(\xi) P_1(\eta) P_0(\zeta) + \hat{w}_9 P_0(\xi) P_1(\eta) P_1(\zeta) + \hat{w}_{10} P_1(\xi) P_0(\eta) P_1(\zeta) \\
&+ \hat{w}_{11} P_3(\xi) P_0(\eta) P_0(\zeta) + \hat{w}_{12} P_0(\xi) P_3(\eta) P_0(\zeta) + \hat{w}_{13} P_0(\xi) P_0(\eta) P_3(\zeta) \\
&+ \hat{w}_{14} P_2(\xi) P_1(\eta) P_0(\zeta) + \hat{w}_{15} P_2(\xi) P_0(\eta) P_1(\zeta) \\
&+ \hat{w}_{16} P_1(\xi) P_2(\eta) P_0(\zeta) + \hat{w}_{17} P_0(\xi) P_2(\eta) P_1(\zeta) \\
&+ \hat{w}_{18} P_1(\xi) P_0(\eta) P_2(\zeta) + \hat{w}_{19} P_0(\xi) P_1(\eta) P_2(\zeta) \\
&+ \hat{w}_{20} P_1(\xi) P_1(\eta) P_1(\zeta)
\end{aligned} \quad (12)$$

Here eqns. (10), (11) and (12) hold for second, third and fourth order schemes respectively. For magnetic fields, the reconstruction is a little more involved and is described in Section III. The difference between this paper and Balsara [10] is that now even the magnetic fields have the same number of basis functions as the zone centered variables. This provides a conceptual simplicity and computational efficiency to the reconstruction process for numerical MHD.

**II.2) ADER-CG Space-time Representation**

In this step we wish to obtain a space-time representation of the reconstructed polynomials described in eqns. (10), (11) or (12). Our temporal basis functions are taken to be

$$Q_0(\tau)=1 \ , \ Q_1(\tau)=\tau \ , \ Q_2(\tau)=\tau^2 \ , \ Q_3(\tau)=\tau^3. \tag{13}$$

In terms of these temporal basis functions, we seek a space-time extension of eqns. (10), (11) or (12). This is provided by specifying a set of $L$ basis functions in space and time within each zone. To obtain the same order of accuracy in time as in space, we need $L = 5$, 15 and 35 for second, third and fourth order schemes respectively. At second, third and fourth orders the space-time representations that we seek are given by:

$$\begin{aligned} u(\xi,\eta,\zeta,\tau) &= \hat{w}_1 P_0(\xi) P_0(\eta) P_0(\zeta) Q_0(\tau) \\ &+ \hat{w}_2 P_1(\xi) P_0(\eta) P_0(\zeta) Q_0(\tau) + \hat{w}_3 P_0(\xi) P_1(\eta) P_0(\zeta) Q_0(\tau) + \hat{w}_4 P_0(\xi) P_0(\eta) P_1(\zeta) Q_0(\tau) \\ &+ \hat{u}_5 P_0(\xi) P_0(\eta) P_0(\zeta) Q_1(\tau) \end{aligned}$$

$$(14),$$

$$\begin{aligned} u(\xi,\eta,\zeta,\tau) &= \hat{w}_1 P_0(\xi) P_0(\eta) P_0(\zeta) Q_0(\tau) \\ &+ \hat{w}_2 P_1(\xi) P_0(\eta) P_0(\zeta) Q_0(\tau) + \hat{w}_3 P_0(\xi) P_1(\eta) P_0(\zeta) Q_0(\tau) + \hat{w}_4 P_0(\xi) P_0(\eta) P_1(\zeta) Q_0(\tau) \\ &+ \hat{w}_5 P_2(\xi) P_0(\eta) P_0(\zeta) Q_0(\tau) + \hat{w}_6 P_0(\xi) P_2(\eta) P_0(\zeta) Q_0(\tau) + \hat{w}_7 P_0(\xi) P_0(\eta) P_2(\zeta) Q_0(\tau) \\ &+ \hat{w}_8 P_1(\xi) P_1(\eta) P_0(\zeta) Q_0(\tau) + \hat{w}_9 P_0(\xi) P_1(\eta) P_1(\zeta) Q_0(\tau) + \hat{w}_{10} P_1(\xi) P_0(\eta) P_1(\zeta) Q_0(\tau) \\ &+ \hat{u}_{11} P_0(\xi) P_0(\eta) P_0(\zeta) Q_1(\tau) + \hat{u}_{12} P_0(\xi) P_0(\eta) P_0(\zeta) Q_2(\tau) \\ &+ \hat{u}_{13} P_1(\xi) P_0(\eta) P_0(\zeta) Q_1(\tau) + \hat{u}_{14} P_0(\xi) P_1(\eta) P_0(\zeta) Q_1(\tau) + \hat{u}_{15} P_0(\xi) P_0(\eta) P_1(\zeta) Q_1(\tau) \end{aligned}$$

$$(15)$$

and

$$\begin{aligned}
u(\xi,\eta,\zeta,\tau) = {} & \hat{w}_1 P_0(\xi) P_0(\eta) P_0(\zeta) Q_0(\tau) \\
& + \hat{w}_2 P_1(\xi) P_0(\eta) P_0(\zeta) Q_0(\tau) + \hat{w}_3 P_0(\xi) P_1(\eta) P_0(\zeta) Q_0(\tau) + \hat{w}_4 P_0(\xi) P_0(\eta) P_1(\zeta) Q_0(\tau) \\
& + \hat{w}_5 P_2(\xi) P_0(\eta) P_0(\zeta) Q_0(\tau) + \hat{w}_6 P_0(\xi) P_2(\eta) P_0(\zeta) Q_0(\tau) + \hat{w}_7 P_0(\xi) P_0(\eta) P_2(\zeta) Q_0(\tau) \\
& + \hat{w}_8 P_1(\xi) P_1(\eta) P_0(\zeta) Q_0(\tau) + \hat{w}_9 P_0(\xi) P_1(\eta) P_1(\zeta) Q_0(\tau) + \hat{w}_{10} P_1(\xi) P_0(\eta) P_1(\zeta) Q_0(\tau) \\
& + \hat{w}_{11} P_3(\xi) P_0(\eta) P_0(\zeta) Q_0(\tau) + \hat{w}_{12} P_0(\xi) P_3(\eta) P_0(\zeta) Q_0(\tau) + \hat{w}_{13} P_0(\xi) P_0(\eta) P_3(\zeta) Q_0(\tau) \\
& + \hat{w}_{14} P_2(\xi) P_1(\eta) P_0(\zeta) Q_0(\tau) + \hat{w}_{15} P_2(\xi) P_0(\eta) P_1(\zeta) Q_0(\tau) \\
& + \hat{w}_{16} P_1(\xi) P_2(\eta) P_0(\zeta) Q_0(\tau) + \hat{w}_{17} P_0(\xi) P_2(\eta) P_1(\zeta) Q_0(\tau) \\
& + \hat{w}_{18} P_1(\xi) P_0(\eta) P_2(\zeta) Q_0(\tau) + \hat{w}_{19} P_0(\xi) P_1(\eta) P_2(\zeta) Q_0(\tau) \\
& + \hat{w}_{20} P_1(\xi) P_1(\eta) P_1(\zeta) Q_0(\tau) \\
& + \hat{u}_{21} P_0(\xi) P_0(\eta) P_0(\zeta) Q_1(\tau) + \hat{u}_{22} P_0(\xi) P_0(\eta) P_0(\zeta) Q_2(\tau) + \hat{u}_{23} P_0(\xi) P_0(\eta) P_0(\zeta) Q_3(\tau) \\
& + \hat{u}_{24} P_1(\xi) P_0(\eta) P_0(\zeta) Q_1(\tau) + \hat{u}_{25} P_0(\xi) P_1(\eta) P_0(\zeta) Q_1(\tau) + \hat{u}_{26} P_0(\xi) P_0(\eta) P_1(\zeta) Q_1(\tau) \\
& + \hat{u}_{27} P_1(\xi) P_0(\eta) P_0(\zeta) Q_2(\tau) + \hat{u}_{28} P_0(\xi) P_1(\eta) P_0(\zeta) Q_2(\tau) + \hat{u}_{29} P_0(\xi) P_0(\eta) P_1(\zeta) Q_2(\tau) \\
& + \hat{u}_{30} P_2(\xi) P_0(\eta) P_0(\zeta) Q_1(\tau) + \hat{u}_{31} P_0(\xi) P_2(\eta) P_0(\zeta) Q_1(\tau) + \hat{u}_{32} P_0(\xi) P_0(\eta) P_2(\zeta) Q_1(\tau) \\
& + \hat{u}_{33} P_1(\xi) P_1(\eta) P_0(\zeta) Q_1(\tau) + \hat{u}_{34} P_0(\xi) P_1(\eta) P_1(\zeta) Q_1(\tau) + \hat{u}_{35} P_1(\xi) P_0(\eta) P_1(\zeta) Q_1(\tau)
\end{aligned}$$
(16)

Here eqns. (14), (15) and (16) hold for second, third and fourth order schemes respectively. The coefficients of the $L$ basis functions in the above three equations provide the modal representation of the ADER-CG scheme in space-time. In other words, $\hat{u} \equiv \left(\hat{w}_1,..,\hat{w}_{L_S},\hat{u}_{L_S+1},..,\hat{u}_L\right)^T$ is a vector of modes that is stored for each conserved variable within each zone. Observe that $Q_0(\tau)=1$ so that the first $L_S$ of these modes are non-evolutionary. Consequently, at $\tau=0$, eqns. (14), (15) and (16) indeed reduce to eqns. (10), (11) and (12) respectively. The entire ADER-CG time-evolution within a zone is carried by the last $L-L_S$ modes, i.e. only these modes are evolutionary. By comparing eqns. (10) and (14) we realize that the ADER-CG scheme only needs to provide a method for obtaining $\hat{u}_5$ at second order. Similarly, by comparing eqns. (11) and (15) notice that the ADER-CG scheme only needs to provide a method for obtaining $\hat{u}_{11}$ to $\hat{u}_{15}$ at third order. Likewise, compare eqns. (12) and (16) to see that the ADER-CG scheme only needs to provide a method for obtaining $\hat{u}_{21}$ to $\hat{u}_{35}$ at fourth order. Section IV will provide an efficient, iterative strategy for obtaining this space-time representation.

It is worthwhile pointing out that the ADER schemes can also provide a complete space-time representation of the flux vectors $\hat{f} \equiv \left(\hat{f}_1,..,\hat{f}_{L_S},\hat{f}_{L_S+1},..,\hat{f}_L\right)^T$, $\hat{g} \equiv \left(\hat{g}_1,..,\hat{g}_{L_S},\hat{g}_{L_S+1},..,\hat{g}_L\right)^T$ and $\hat{h} \equiv \left(\hat{h}_1,..,\hat{h}_{L_S},\hat{h}_{L_S+1},..,\hat{h}_L\right)^T$ and the source term $\hat{s} \equiv \left(\hat{s}_1,...,\hat{s}_{L_S},\hat{s}_{L_S+1},...,\hat{s}_L\right)^T$. By the end of the ADER iteration, these vectors can be built within each zone. One choice, described here, consists of storing these vectors and using them to simplify the evaluations that are needed in the next step. One may also choose to store only $\hat{u} \equiv \left(\hat{w}_1,..,\hat{w}_{L_S},\hat{u}_{L_S+1},..,\hat{u}_L\right)^T$ at the end of the ADER iteration. In that case, the modal representation of the fluxes will have to be retrieved when evaluating the fluxes and electric fields. This results in a little more programming but yields code that uses memory much more efficiently. We have implemented both strategies and found the latter to be superior.

If non-stiff source terms are present, it would also be advantageous at this stage in the computation to use the modes of $\hat{s}$ within each zone to obtain the space-time averaged expression for $\overline{S}_{i,j,k}$ in eqn. (6). In Section IV we will, therefore, provide closed form expressions for $\overline{S}_{i,j,k}$ in terms of the modes of the source term. The storage associated with saving this bit of information is indeed small.

**II.3) Flux and Electric Field Evaluation**

The strategy presented by Dumbser *et al.* [29] relies on freezing the eigenstructure of the Riemann problem at the space-time barycenter of each face. One then views the flux at a face as being a linear combination of four vector functions. The four vector functions are : a) the conserved variables to the left of the zone boundary given by $U_{L;\,i+1/2,j,k}(y,z,t) \equiv u_{i,j,k}(\xi=1/2,\eta,\zeta,\tau)$, b) the conserved variables to the right of the zone boundary given by $U_{R;\,i+1/2,j,k}(y,z,t) \equiv u_{i+1,j,k}(\xi=-1/2,\eta,\zeta,\tau)$, c) the flux to the left of the zone boundary given by $F_{L;\,i+1/2,j,k}(y,z,t) \equiv f_{i,j,k}(\xi=1/2,\eta,\zeta,\tau)\ \Delta x/\Delta t$ and d) the flux to the right of the zone boundary given by $F_{R;\,i+1/2,j,k}(y,z,t) \equiv f_{i+1,j,k}(\xi=-1/2,\eta,\zeta,\tau)\ \Delta x/\Delta t$. Let us illustrate this for the HLL flux at any general point on the x-boundary that is centered at "$i+1/2,j,k$". Fig. 1 shows the planes where the four vector functions of interest are specified. Consider a situation where the fastest left-going and right-going signal speeds at that boundary are $\lambda_L$ and $\lambda_R$ respectively, see Fig. 1. Following the usual convention for the HLL Riemann solver, we reset $\lambda_L = \min(\lambda_L, 0)$ and $\lambda_R = \max(\lambda_R, 0)$. The HLL flux at any general point on the top x-face of the zone being considered is then given by

$$F_{i+1/2,j,k}(y,z,t) = \left[\frac{\lambda_R}{\lambda_R - \lambda_L}\right] F_{L;\,i+1/2,j,k}(y,z,t) - \left[\frac{\lambda_L}{\lambda_R - \lambda_L}\right] F_{R;\,i+1/2,j,k}(y,z,t)$$
$$+ \left[\frac{\lambda_R \lambda_L}{\lambda_R - \lambda_L}\right]\left(U_{R;\,i+1/2,j,k}(y,z,t) - U_{L;\,i+1/2,j,k}(y,z,t)\right) \quad (17)$$

The space and time averaged flux $\bar{F}_{i+1/2,j,k}$ in eqn. (6) is obtained by averaging the above flux in the y, z and t directions at the x-boundary being considered. But notice that $F_{i+1/2,j,k}(y,z,t)$ depends linearly on the four vector functions $F_{L;\,i+1/2,j,k}(y,z,t)$, $F_{R;\,i+1/2,j,k}(y,z,t)$, $U_{L;\,i+1/2,j,k}(y,z,t)$ and $U_{R;\,i+1/2,j,k}(y,z,t)$ in eqn. (17). Consequently, $\bar{F}_{i+1/2,j,k}$ can be written as space-time averages of the above four vector functions. Analytic forms for these space-time averages can be easily obtained and have been documented in Section V. The linearized Riemann solver of (Roe [49], Harten & Hyman [40]) can also be written in a form that is similar to eqn. (17), see Dumbser *et al.* [29]. As a result, one can even use a more refined numerical flux function for the update in eqn. (6).

For ideal MHD, the electric fields $\bar{E}_{y;\,i+1/2,j,k+1/2}$, $\bar{E}_{y;\,i+1/2,j,k-1/2}$, $\bar{E}_{z;\,i+1/2,j+1/2,k}$ and $\bar{E}_{z;\,i+1/2,j-1/2,k}$ in the first expression of eqn. (8) can be obtained by a space and time averaging of eqn. (17) at the appropriate y and z-edges. A properly upwinded evaluation of $\bar{E}_{y;\,i+1/2,j,k+1/2}$ actually requires contributions from the Riemann problems evaluated in the *x* and *z*-faces that abut the "*i+1/2,j,k+1/2*" edge. Similarly, a properly upwinded evaluation of $\bar{E}_{z;\,i+1/2,j+1/2,k}$ requires contributions from Riemann problems evaluated in the *x* and *y*-faces that abut the "*i+1/2,j+1/2,k*" edge. When the Riemann problem is transonic, it also helps to increase the dissipation (Londrillo & Del Zanna [46], Gardiner & Stone [37], Balsara [12]).

The steps in Sub-sections II.a, II.b and II.c provide a complete description of one ADER timestep.

**II.d) Runge-Kutta Time Stepping**

It is interesting to realize that if the ADER space-time representation from Sub-section II.b is bypassed then Sub-sections II.a and II.c describe all the operations that are needed in a single stage of a multi-stage RK scheme. The result of assembling these stages in a multi-stage time step would indeed be a very efficient RK scheme. For the sake of completeness, we catalogue the RK schemes used in this paper. We formally write eqn. (4) as $\partial U/\partial t = L(U)$. The second and third order accurate strong stability preserving RK schemes from Shu & Osher [54] are stable for a CFL number of unity for one-dimensional problems, though the CFL number is inversely proportional to the dimensionality of the problem. They are given by

$$U^{(1)} = U^n + \Delta t\, L(U^n)$$
$$U^{n+1} = \frac{1}{2}U^n + \frac{1}{2}U^{(1)} + \frac{1}{2}\Delta t\, L(U^{(1)})$$
(18)

and

$$U^{(1)} = U^n + \Delta t\, L(U^n)$$
$$U^{(2)} = \frac{3}{4}U^n + \frac{1}{4}U^{(1)} + \frac{1}{4}\Delta t\, L(U^{(1)}) \quad . \tag{19}$$
$$U^{n+1} = \frac{1}{3}U^n + \frac{2}{3}U^{(2)} + \frac{2}{3}\Delta t\, L(U^{(2)})$$

The optimality of the above schemes has been proved by Gottlieb & Shu [38]. Fourth order accurate RK schemes with more than four sub-stages have been constructed by Spiteri & Ruuth [56], [57]. Their optimal five stage RK scheme retains fourth order of accuracy and is stable up to a CFL number of ~ 1.5, though it too is inversely proportional to the dimensionality of the problem. It is given by

$$U^{(1)} = U^n + 0.39175222700392\, \Delta t\, L(U^n)$$
$$U^{(2)} = 0.44437049406734\, U^n + 0.55562950593266\, U^{(1)} + 0.36841059262959\, \Delta t\, L(U^{(1)})$$
$$U^{(3)} = 0.62010185138540\, U^n + 0.37989814861460\, U^{(2)} + 0.25189177424738\, \Delta t\, L(U^{(2)})$$
$$U^{(4)} = 0.17807995410773\, U^n + 0.82192004589227\, U^{(3)} + 0.54497475021237\, \Delta t\, L(U^{(3)}) \quad .$$
$$\begin{aligned}U^{n+1} = &\ 0.00683325884039\, U^n + 0.51723167208978\, U^{(2)} \\ &+ 0.12759831133288\, U^{(3)} + 0.34833675773694\, U^{(4)} \\ &+ 0.08460416338212\, \Delta t\, L(U^{(3)}) + 0.22600748319395\, \Delta t\, L(U^{(4)})\end{aligned}$$

(20)

The methods described in this paper can also, therefore, be used to design optimal RK schemes of high order. This paper presents speed comparisons of optimal RK schemes vis-à-vis ADER-CG schemes.

## III) WENO Strategy for Obtaining Zone-averaged Magnetic Fields

The problem of obtaining zone-averaged magnetic fields and their moments is best illustrated by starting with the second order case. We wish to reconstruct the magnetic field in such a fashion as to retain consistency at zone boundaries. Fig. 2 shows the facially-averaged $x$-components of the magnetic field on a two dimensional mesh.

Because we only need to focus on the x-directional variation, we show one row of zones in Fig. 1. The zone-averaged *x*-magnetic field at second order in zone "*i,j,k*" is most easily obtained by averaging the facial magnetic fields at either *x*-boundary of that zone. Thus we have $\overline{B}_{x;i,j,k} = 0.5\left(\overline{B}_{x;i+1/2,j,k} + \overline{B}_{x;i-1/2,j,k}\right)$. Suppose a TVD-preserving slope limiter is applied to the above zone-averaged magnetic field in order to obtain its *x*-directional variation. The variation of the magnetic field in each zone would then be such that the facially averaged *x*-magnetic field from zones "*i,j,k*" and "*i+1,j,k*" would not match up with $\overline{B}_{x;i+1/2,j,k}$ at the *x*-face "*i+1/2,j,k*". This is the source of the inconsistency. However, if the variation of $\overline{B}_{x;i,j,k}$ in the *x*-direction is given by $\left(\overline{B}_{x;i+1/2,j,k} - \overline{B}_{x;i-1/2,j,k}\right)$, then the inconsistency would be resolved. The *y* and *z*-variations of $\overline{B}_{x;i,j,k}$ can still be obtained via slope limiters applied in the two transverse directions. Thus we have a procedure for obtaining all the variations of $\overline{B}_{x;i,j,k}$ that are required in eqn. (10).

The above discussion shows us that at third order we wish to find $\overline{B}_{x;i,j,k}$ as well as its linear and quadratic variations in the x-direction through a process that differs from the traditional zone-centered, non-oscillatory limiting. Such a process is described in Sub-section III.1. The remaining moments of $\overline{B}_{x;i,j,k}$, as are required for eqn. (11), can be obtained by applying a traditional reconstruction strategy to the zone-averaged x-magnetic field. At fourth order we will again need to find $\overline{B}_{x;i,j,k}$ as well as its linear, quadratic and cubic variations in the x-direction through a process that differs from the traditional one. Such a process is described in Sub-section III.2. The remaining moments of $\overline{B}_{x;i,j,k}$, as are required for eqn. (12), can be obtained by any traditional reconstruction strategy.

**III.1) Third Order Reconstruction of the Normal Component of the Magnetic Field**

The WENO reconstruction calls for assembling all the stencils that cover the zone of interest. At third order we have a left-biased stencil, $S_1$, that relies on the variables $\{\overline{B}_{x;i-3/2,j,k}, \overline{B}_{x;i-1/2,j,k}, \overline{B}_{x;i+1/2,j,k}\}$ and a right-biased stencil, $S_2$, that relies on the variables $\{\overline{B}_{x;i-1/2,j,k}, \overline{B}_{x;i+1/2,j,k}, \overline{B}_{x;i+3/2,j,k}\}$. Notice that both stencils cover the zone "$i,j,k$", consequently, either stencil will produce an interpolating function in the $x$-direction that matches the facial averages $\overline{B}_{x;i-1/2,j,k}$ and $\overline{B}_{x;i+1/2,j,k}$. The reconstructed polynomial with the moments that are required at third order can generally be written as

$$b(\xi) = b_0 + b_x P_1(\xi) + b_{xx} P_2(\xi). \tag{21}$$

For the stencil $S_1$ we have

$$\begin{aligned} b_0 &= \left(8\,\overline{B}_{x;i-1/2,j,k} - \overline{B}_{x;i-3/2,j,k} + 5\,\overline{B}_{x;i+1/2,j,k}\right)/12 \\ b_x &= \overline{B}_{x;i+1/2,j,k} - \overline{B}_{x;i-1/2,j,k} \\ b_{xx} &= \left(\overline{B}_{x;i+1/2,j,k} + \overline{B}_{x;i-3/2,j,k} - 2\,\overline{B}_{x;i-1/2,j,k}\right)/2 \end{aligned} \tag{22}$$

and for stencil $S_2$ we have

$$\begin{aligned} b_0 &= \left(8\,\overline{B}_{x;i+1/2,j,k} - \overline{B}_{x;i+3/2,j,k} + 5\,\overline{B}_{x;i-1/2,j,k}\right)/12 \\ b_x &= \overline{B}_{x;i+1/2,j,k} - \overline{B}_{x;i-1/2,j,k} \\ b_{xx} &= \left(\overline{B}_{x;i-1/2,j,k} + \overline{B}_{x;i+3/2,j,k} - 2\,\overline{B}_{x;i+1/2,j,k}\right)/2 \end{aligned} \tag{23}$$

The smoothness measures for this form of WENO scheme are given by

$$\text{IS} = b_x^2 + 13\,b_{xx}^2/3. \tag{24}$$

The smoothness measures enable us to make a non-linear, convex combination of the moments from each of the competing stencils. The method presented here is so designed that the zeroth moment yields the zone-averaged value of the $x$-magnetic field. This completes our description of the WENO reconstruction of the normal magnetic field at third order. The remaining moments of the *x*-component of the magnetic field can be obtained via any suitable reconstruction strategy. The same formulae can be applied in the other two directions for the *y* and *z*-components of the magnetic field.

**III.2) Fourth Order Reconstruction of the Normal Component of the Magnetic Field**

At fourth order, the WENO reconstruction procedure requires us to consider a left-biased stencil, $S_1$, that relies on the variables $\{\bar{B}_{x;i-5/2,j,k}, \bar{B}_{x;i-3/2,j,k}, \bar{B}_{x;i-1/2,j,k}, \bar{B}_{x;i+1/2,j,k}\}$, a centered stencil, $S_2$, that relies on the variables $\{\bar{B}_{x;i-3/2,j,k}, \bar{B}_{x;i-1/2,j,k}, \bar{B}_{x;i+1/2,j,k}, \bar{B}_{x;i+3/2,j,k}\}$ and a right-biased stencil, $S_3$, that relies on the variables $\{\bar{B}_{x;i-1/2,j,k}, \bar{B}_{x;i+1/2,j,k}, \bar{B}_{x;i+3/2,j,k}, \bar{B}_{x;i+5/2,j,k}\}$. Notice that all three stencils cover the zone "*i,j,k*", consequently, each of them will produce an interpolating function that matches the facial averages $\bar{B}_{x;i-1/2,j,k}$ and $\bar{B}_{x;i+1/2,j,k}$. The reconstructed polynomial with the moments that are required at fourth order can generally be written as

$$b(\xi) = b_0 + b_x P_1(\xi) + b_{xx} P_2(\xi) + b_{xxx} P_3(\xi). \tag{25}$$

For the stencil $S_1$ we have

$$\begin{aligned}
b_0 &= \left(9\,\bar{B}_{x;i+1/2,j,k} + 19\,\bar{B}_{x;i-1/2,j,k} - 5\,\bar{B}_{x;i-3/2,j,k} + \bar{B}_{x;i-5/2,j,k}\right)/24 \\
b_x &= \left(59\,\bar{B}_{x;i+1/2,j,k} - 57\,\bar{B}_{x;i-1/2,j,k} - 3\,\bar{B}_{x;i-3/2,j,k} + \bar{B}_{x;i-5/2,j,k}\right)/60 \\
b_{xx} &= \left(3\,\bar{B}_{x;i+1/2,j,k} - 7\,\bar{B}_{x;i-1/2,j,k} + 5\,\bar{B}_{x;i-3/2,j,k} - \bar{B}_{x;i-5/2,j,k}\right)/4 \\
b_{xxx} &= \left(\bar{B}_{x;i+1/2,j,k} - 3\,\bar{B}_{x;i-1/2,j,k} + 3\,\bar{B}_{x;i-3/2,j,k} - \bar{B}_{x;i-5/2,j,k}\right)/6
\end{aligned} \tag{26),}$$

for the stencil $S_2$ we have

$$b_0 = \left(-\overline{B}_{x;i+3/2,j,k} + 13\,\overline{B}_{x;i+1/2,j,k} + 13\,\overline{B}_{x;i-1/2,j,k} - \overline{B}_{x;i-3/2,j,k}\right)/24$$
$$b_x = \left(-\overline{B}_{x;i+3/2,j,k} + 63\,\overline{B}_{x;i+1/2,j,k} - 63\,\overline{B}_{x;i-1/2,j,k} + \overline{B}_{x;i-3/2,j,k}\right)/60$$
$$b_{xx} = \left(\overline{B}_{x;i+3/2,j,k} - \overline{B}_{x;i+1/2,j,k} - \overline{B}_{x;i-1/2,j,k} + \overline{B}_{x;i-3/2,j,k}\right)/4$$
$$b_{xxx} = \left(\overline{B}_{x;i+3/2,j,k} - 3\,\overline{B}_{x;i+1/2,j,k} + 3\,\overline{B}_{x;i-1/2,j,k} - \overline{B}_{x;i-3/2,j,k}\right)/6$$
(27)

and for the stencil $S_3$ we have

$$b_0 = \left(\overline{B}_{x;i+5/2,j,k} - 5\,\overline{B}_{x;i+3/2,j,k} + 19\,\overline{B}_{x;i+1/2,j,k} + 9\,\overline{B}_{x;i-1/2,j,k}\right)/24$$
$$b_x = \left(-\overline{B}_{x;i+5/2,j,k} + 3\,\overline{B}_{x;i+3/2,j,k} + 57\,\overline{B}_{x;i+1/2,j,k} - 59\,\overline{B}_{x;i-1/2,j,k}\right)/60$$
$$b_{xx} = \left(-\overline{B}_{x;i+5/2,j,k} + 5\,\overline{B}_{x;i+3/2,j,k} - 7\,\overline{B}_{x;i+1/2,j,k} + 3\,\overline{B}_{x;i-1/2,j,k}\right)/4$$
$$b_{xxx} = \left(\overline{B}_{x;i+5/2,j,k} - 3\,\overline{B}_{x;i+3/2,j,k} + 3\,\overline{B}_{x;i+1/2,j,k} - \overline{B}_{x;i-1/2,j,k}\right)/6$$
(28)

The smoothness measures for this form of WENO scheme are given by

$$\text{IS} = \left(b_x + 0.1\,b_{xxx}\right)^2 + 13\,b_{xx}^2/3 + 781\,b_{xxx}^2/20.$$
(29)

As in the previous Sub-section, the smoothness measures enable us to make a non-linear, convex combination of the moments, with the zeroth moment yielding the zone-averaged *x*-magnetic field. This completes our description of the WENO reconstruction of the normal magnetic field at fourth order. As before, the remaining moments of the *x*-component of the magnetic field can be obtained via any suitable reconstruction strategy.

## IV) ADER-CG Space-time Update in Modal Space

In this Section we describe ADER-CG schemes in modal space at orders ranging from two to four. We focus exclusively on formulae that would be implemented in numerical schemes. The derivations of these schemes are provided in Section 3 of Balsara *et al.* [11] and are not repeated here. Instead, in the ensuing three Sub-sections we catalogue recent innovations that facilitate efficient computation with ADER-CG schemes.

**IV.1) ADER-CG Scheme in Modal Space at Second Order**

We describe the ADER-CG scheme in modal space at second order in this Subsection. In other words, we describe the process of starting with eqn. (10) and arriving at eqn. (14) by carrying out one iteration of the formula described in the next equation. The evolutionary equation in modal space is given by

$$\hat{u}_5 = -\hat{f}_2 - \hat{g}_3 - \hat{h}_4 + \hat{s}_1 + \frac{2}{3}\hat{s}_5 . \tag{30}$$

Notice that eqn. (30) does not require all the modes of all the fluxes. We can simply get by with the modes $\hat{f}_2$, $\hat{g}_3$ and $\hat{h}_4$ for the $x$, $y$ and $z$-fluxes.

To obtain the above-mentioned modes for the fluxes, we first define an ordered set of seven symmetrically-placed nodes in the reference space-time element given by

$$\{(1/2,0,0,0); (-1/2,0,0,0); (0,1/2,0,0);(0,-1/2,0,0); \\ (0,0,1/2,0); (0,0,-1/2,0); (0,0,0,1)\} \tag{31}$$

Let nodal variables be denoted by overbars and modal variables be denoted by carets. We now obtain the conserved variables at each of the above nodal points by using the following formulae

$$\bar{u}_1 = \hat{w}_1 + 0.5\,\hat{w}_2 ; \quad \bar{u}_2 = \hat{w}_1 - 0.5\,\hat{w}_2 ; \quad \bar{u}_3 = \hat{w}_1 + 0.5\,\hat{w}_3 ; \quad \bar{u}_4 = \hat{w}_1 - 0.5\,\hat{w}_3 ; \\ \bar{u}_5 = \hat{w}_1 + 0.5\,\hat{w}_4 ; \quad \bar{u}_6 = \hat{w}_1 - 0.5\,\hat{w}_4 ; \quad \bar{u}_7 = \hat{w}_1 + \hat{u}_5 \tag{32}$$

The conserved variables from eqn. (32) can then be used to evaluate the fluxes at the nodal points. However, not all the fluxes need to be evaluated because we only need the fluxes at specific nodes in order to evaluate the requisite modes of the fluxes. This becomes clear when the formulae for the modes $\hat{f}_2$, $\hat{g}_3$ and $\hat{h}_4$ are catalogued as

$$\hat{f}_2 = \overline{f}_1 - \overline{f}_2 \; ; \quad \hat{g}_3 = \overline{g}_3 - \overline{g}_4 \; ; \quad \hat{h}_4 = \overline{h}_5 - \overline{h}_6 \; . \tag{33}$$

The sources, if present, have to be evaluated at all the nodal points. The equations that provide the source terms in modal space are

$$\hat{s}_1 = \left( \overline{s}_1 + \overline{s}_2 + \overline{s}_3 + \overline{s}_4 + \overline{s}_5 + \overline{s}_6 \right) / 6 \; ; \quad \hat{s}_5 = \overline{s}_7 - \hat{s}_1 \; . \tag{34}$$

After $\hat{u}_5$ is built, the complete modal space representation of $\hat{f}$, $\hat{g}$ and $\hat{h}$ may also be built and stored.

The space-time integral of the source term, which is needed in eqn. (6) for each zone "$i,j,k$", is given by

$$\overline{S}_{i,j,k} = \left( \hat{s}_1 + 0.5 \, \hat{s}_5 \right) / \Delta t \tag{35}$$

This completes our description of the second order ADER-CG scheme in modal space. It is comforting to notice that it is simply a second order predictor-corrector type scheme. A nodal space formulation of ADER-CG schemes at second order is given in Appendix A.

**IV.2) ADER-CG Scheme in Modal Space at Third Order**

We describe the ADER-CG scheme in modal space at third order in this Sub-section. In other words, we describe the process of starting with eqn. (11) and arriving at eqn. (15) by carrying out two iterations of the formulae described in the next equation. (There exists formal theory based on the Picard iteration which supports the claim that "M−1" iterations are adequate for an $M^{th}$ order scheme.) Notice that a majority of the fluxes and source terms need to be evaluated at $\tau = 0$ in the unit reference element. As a result, they do not need to be evaluated again in the second iteration. In this and the next Sub-sections we continue our convention that carets denote modes while overbars denote nodal points. The third order evolutionary equations in modal space are given by

$$\hat{u}_{11} = -\hat{f}_2 - \hat{g}_3 - \hat{h}_4 + \hat{s}_1 - \frac{3}{10}\hat{s}_{12} \ ; \ \hat{u}_{12} = -\frac{\hat{f}_{13}}{2} - \frac{\hat{g}_{14}}{2} - \frac{\hat{h}_{15}}{2} + \frac{\hat{s}_{11}}{2} + \frac{3}{5}\hat{s}_{12} \ ;$$

$$\hat{u}_{13} = -2\hat{f}_5 - \hat{g}_8 - \hat{h}_{10} + \hat{s}_2 + \frac{2}{3}\hat{s}_{13} \ ; \ \hat{u}_{14} = -\hat{f}_8 - 2\hat{g}_6 - \hat{h}_9 + \hat{s}_3 + \frac{2}{3}\hat{s}_{14} \ ; \quad (36)$$

$$\hat{u}_{15} = -\hat{f}_{10} - \hat{g}_9 - 2\hat{h}_7 + \hat{s}_4 + \frac{2}{3}\hat{s}_{15}$$

As in the second order case, we only need the $\hat{f}_2$, $\hat{f}_5$, $\hat{f}_8$, $\hat{f}_{10}$ and $\hat{f}_{13}$ modes for the *x*-flux in eqn. (36). Of these, only $\hat{f}_{13}$ will have to be rebuilt in the second iteration since the remaining are only built once at $\tau = 0$. Similarly, we only need the $\hat{g}_3$, $\hat{g}_6$, $\hat{g}_8$, $\hat{g}_9$ and $\hat{g}_{14}$ modes for the *y*-flux in eqn. (36), of which only $\hat{g}_{14}$ needs to be rebuilt in the second iteration. Likewise, we only need the $\hat{h}_4$, $\hat{h}_7$, $\hat{h}_9$, $\hat{h}_{10}$ and $\hat{h}_{15}$ modes for the *z*-flux in eqn. (36), of which only $\hat{h}_{15}$ needs to be rebuilt in the second iteration.

To obtain the above-mentioned modes for the fluxes, we first define an ordered set of twenty symmetrically-placed nodes in the reference space-time element given by

$$\begin{aligned} \{ & (0,0,0,0); (1/2,1/2,0,0); (1/2,-1/2,0,0); (1/2,0,1/2,0); (1/2,0,-1/2,0); \\ & (-1/2,1/2,0,0); (-1/2,-1/2,0,0); (-1/2,0,1/2,0); (-1/2,0,-1/2,0); \\ & (0,1/2,1/2,0); (0,-1/2,1/2,0); (0,1/2,-1/2,0); (0,-1/2,-1/2,0); \quad . \quad (37) \\ & (1/2,0,0,1/2); (-1/2,0,0,1/2); (0,1/2,0,1/2); (0,-1/2,0,1/2); \\ & (0,0,1/2,1/2); (0,0,-1/2,1/2); (0,0,0,1) \} \end{aligned}$$

We now obtain the conserved variables at each of the above nodal points by using the following formulae

$$\bar{u}_1 = \hat{w}_1 - (\hat{w}_5 + \hat{w}_6 + \hat{w}_7)/12 \;;$$
$$\bar{u}_2 = \bar{u}_1 + 0.5\,(\hat{w}_2 + \hat{w}_3) + 0.25\,(\hat{w}_5 + \hat{w}_6 + \hat{w}_8) \;;\; \bar{u}_3 = \bar{u}_2 - \hat{w}_3 - 0.5\,\hat{w}_8 \;;$$
$$\bar{u}_4 = \bar{u}_1 + 0.5\,(\hat{w}_2 + \hat{w}_4) + 0.25\,(\hat{w}_5 + \hat{w}_7 + \hat{w}_{10}) \;;\; \bar{u}_5 = \bar{u}_4 - \hat{w}_4 - 0.5\,\hat{w}_{10} \;;$$
$$\bar{u}_6 = \bar{u}_3 - \hat{w}_2 + \hat{w}_3 \;;\; \bar{u}_7 = \bar{u}_2 - \hat{w}_2 - \hat{w}_3 \;;\; \bar{u}_8 = \bar{u}_5 - \hat{w}_2 + \hat{w}_4 \;;\; \bar{u}_9 = \bar{u}_4 - \hat{w}_2 - \hat{w}_4 \;;$$
$$\bar{u}_{10} = \bar{u}_1 + 0.5\,(\hat{w}_3 + \hat{w}_4) + 0.25\,(\hat{w}_6 + \hat{w}_7 + \hat{w}_9) \;;\; \bar{u}_{11} = \bar{u}_{10} - \hat{w}_3 - 0.5\,\hat{w}_9 \;;$$
$$\bar{u}_{12} = \bar{u}_{11} + \hat{w}_3 - \hat{w}_4 \;;\; \bar{u}_{13} = \bar{u}_{10} - \hat{w}_3 - \hat{w}_4 \tag{38}$$

and

$$\bar{u}_{14} = 0.25\,(\bar{u}_2 + \bar{u}_3 + \bar{u}_4 + \bar{u}_5 + \hat{u}_{12} + \hat{u}_{13}) + 0.5\,\hat{u}_{11} - 0.125\,(\hat{w}_6 + \hat{w}_7) \;;$$
$$\bar{u}_{16} = 0.25\,(\bar{u}_2 + \bar{u}_6 + \bar{u}_{10} + \bar{u}_{12} + \hat{u}_{12} + \hat{u}_{14}) + 0.5\,\hat{u}_{11} - 0.125\,(\hat{w}_5 + \hat{w}_7) \;;$$
$$\bar{u}_{18} = 0.25\,(\bar{u}_4 + \bar{u}_8 + \bar{u}_{10} + \bar{u}_{11} + \hat{u}_{12} + \hat{u}_{15}) + 0.5\,\hat{u}_{11} - 0.125\,(\hat{w}_5 + \hat{w}_6) \;; \tag{39}$$
$$\bar{u}_{15} = \bar{u}_{14} - \hat{w}_2 - 0.5\,\hat{u}_{13} \;;\; \bar{u}_{17} = \bar{u}_{16} - \hat{w}_3 - 0.5\,\hat{u}_{14} \;;\; \bar{u}_{19} = \bar{u}_{18} - \hat{w}_4 - 0.5\,\hat{u}_{15} \;;$$
$$\bar{u}_{20} = \bar{u}_1 + \hat{u}_{11} + \hat{u}_{12}$$

Notice that each nodal value utilizes the ones that were built before it, making the computation very efficient. Notice too that the set of conserved variables at the first thirteen nodal points in eqn. (38) only need to be built once, i.e. in the first iteration. The conserved variables at the last seven nodal points, i.e. the ones in eqn. (39), will need to be built once in the first iteration and then rebuilt in the second iteration. This is because only the nodal values in eqn. (39) depend on $\hat{u}_{11}$, $\hat{u}_{12}$, $\hat{u}_{13}$, $\hat{u}_{14}$, and $\hat{u}_{15}$. The conserved variables from eqns. (38) and (39) can then be used to evaluate the fluxes at the nodes. The fluxes at the first thirteen nodal points only need to be built once; the fluxes at the remaining seven nodal points need to be rebuilt for each iteration. Once the fluxes are built at the nodal points, we can evaluate the desired modes of the fluxes. This is most easily done by having a set of temporary variables, which we denote by a "$q$". The requisite modes of the $x$-flux are given as follows

$$q_1 = \bar{f}_2 + \bar{f}_3 + \bar{f}_4 + \bar{f}_5 \ ; \ q_2 = \bar{f}_6 + \bar{f}_7 + \bar{f}_8 + \bar{f}_9 \ ; \ q_3 = \bar{f}_{10} + \bar{f}_{11} + \bar{f}_{12} + \bar{f}_{13} \ ;$$
$$\hat{f}_2 = 0.25(q_1 - q_2) \ ; \ \hat{f}_5 = 0.5(q_1 + q_2 - q_3) - 2\bar{f}_1 \ ;$$
$$\hat{f}_8 = \bar{f}_2 - \bar{f}_3 - \bar{f}_6 + \bar{f}_7 \ ; \ \hat{f}_{10} = \bar{f}_4 - \bar{f}_5 - \bar{f}_8 + \bar{f}_9 \ ; \ \hat{f}_{13} = 2\left(\bar{f}_{14} - \bar{f}_{15} - \hat{f}_2\right)$$
(40)

the modes of the *y*-flux are given as follows

$$q_4 = \bar{g}_2 + \bar{g}_6 + \bar{g}_{10} + \bar{g}_{12} \ ; \ q_5 = \bar{g}_3 + \bar{g}_7 + \bar{g}_{11} + \bar{g}_{13} \ ; \ q_6 = \bar{g}_4 + \bar{g}_5 + \bar{g}_8 + \bar{g}_9 \ ;$$
$$\hat{g}_3 = 0.25(q_4 - q_5) \ ; \ \hat{g}_6 = 0.5(q_4 + q_5 - q_6) - 2\bar{g}_1 \ ;$$
$$\hat{g}_8 = \bar{g}_2 - \bar{g}_3 - \bar{g}_6 + \bar{g}_7 \ ; \ \hat{g}_9 = \bar{g}_{10} - \bar{g}_{11} - \bar{g}_{12} + \bar{g}_{13} \ ; \ \hat{g}_{14} = 2\left(\bar{g}_{16} - \bar{g}_{17} - \hat{g}_3\right)$$
(41)

and the modes of the *z*-flux are given as follows

$$q_7 = \bar{h}_4 + \bar{h}_8 + \bar{h}_{10} + \bar{h}_{11} \ ; \ q_8 = \bar{h}_5 + \bar{h}_9 + \bar{h}_{12} + \bar{h}_{13} \ ; \ q_9 = \bar{h}_2 + \bar{h}_3 + \bar{h}_6 + \bar{h}_7 \ ;$$
$$\hat{h}_4 = 0.25(q_7 - q_8) \ ; \ \hat{h}_7 = 0.5(q_7 + q_8 - q_9) - 2\bar{h}_1 \ ;$$
$$\hat{h}_9 = \bar{h}_{10} - \bar{h}_{11} - \bar{h}_{12} + \bar{h}_{13} \ ; \ \hat{h}_{10} = \bar{h}_4 - \bar{h}_5 - \bar{h}_8 + \bar{h}_9 \ ; \ \hat{h}_{15} = 2\left(\bar{h}_{18} - \bar{h}_{19} - \hat{h}_4\right)$$
(42)

The sources, if present, have to be evaluated at all the nodal points. Many of the modes for the source terms are obtained by setting $f \to s$ in eqn. (40), $g \to s$ in eqn. (41) and $h \to s$ in eqn. (42). The remaining modes that are not contained in eqns. (40) to (42) are given below

$$\hat{s}_1 = \bar{s}_1 + (\hat{s}_5 + \hat{s}_6 + \hat{s}_7)/12 \ ; \ \hat{s}_{11} = 2(\bar{s}_{14} + \bar{s}_{15} + \bar{s}_{16} + \bar{s}_{17} + \bar{s}_{18} + \bar{s}_{19} - 6\,\hat{s}_1)/3 - \bar{s}_{20} + \bar{s}_1 \ ;$$
$$\hat{s}_{12} = \bar{s}_{20} - \bar{s}_1 - \hat{s}_{11}$$
.(43)

Once $\hat{u}_{11}$ to $\hat{u}_{15}$ are obtained at the end of the second iteration, the complete modal space representation of $\hat{f}$, $\hat{g}$ and $\hat{h}$ may also be rebuilt and stored.

The space-time integral of the source term, which is needed in eqn. (6) for each zone "i,j,k", is then given by

$$\overline{S}_{i,j,k} = \left( \hat{s}_1 + 0.5\, \hat{s}_{11} + \hat{s}_{12}/3 \right)/\Delta t \,. \tag{44}$$

This completes our description of the third order ADER-CG scheme in modal space. A nodal space formulation of ADER-CG schemes at third order is given in Appendix B.

**IV.3) ADER-CG Scheme in Modal Space at Fourth Order**

We describe the ADER-CG scheme in modal space at fourth order in this Subsection. In other words, we describe the process of starting with eqn. (12) and arriving at eqn. (16) by carrying out three iterations of the formulae described in the next equation. As with the third order case, the majority of the fluxes and source terms need to be evaluated only at $\tau = 0$ in the unit reference element. As a result, they do not need to be evaluated again in the second and third iterations. The fourth order evolutionary equations in modal space are given by

$$\hat{u}_{21} = -\frac{\hat{f}_{11}}{10} - \hat{f}_2 - \frac{\hat{g}_{12}}{10} - \hat{g}_3 - \frac{\hat{h}_{13}}{10} - \hat{h}_4 + \hat{s}_1 + \frac{8}{70}\hat{s}_{23} \,;$$

$$\hat{u}_{22} = -\frac{\hat{f}_{24}}{2} - \frac{\hat{g}_{25}}{2} - \frac{\hat{h}_{26}}{2} + \frac{\hat{s}_{21}}{2} - \frac{3}{7}\hat{s}_{23} \,; \quad \hat{u}_{23} = -\frac{\hat{f}_{27}}{3} - \frac{\hat{g}_{28}}{3} - \frac{\hat{h}_{29}}{3} + \frac{\hat{s}_{22}}{3} + \frac{4}{7}\hat{s}_{23} \,;$$

$$\hat{u}_{24} = -2\,\hat{f}_5 - \hat{g}_8 - \hat{h}_{10} + \hat{s}_2 - \frac{3}{10}\hat{s}_{27} \,; \quad \hat{u}_{25} = -\hat{f}_8 - 2\,\hat{g}_6 - \hat{h}_9 + \hat{s}_3 - \frac{3}{10}\hat{s}_{28} \,;$$

$$\hat{u}_{26} = -\hat{f}_{10} - \hat{g}_9 - 2\,\hat{h}_7 + \hat{s}_4 - \frac{3}{10}\hat{s}_{29} \,; \quad \hat{u}_{27} = -\hat{f}_{30} - \frac{\hat{g}_{33}}{2} - \frac{\hat{h}_{35}}{2} + \frac{\hat{s}_{24}}{2} + \frac{3}{5}\hat{s}_{27} \,;$$

$$\hat{u}_{28} = -\frac{\hat{f}_{33}}{2} - \hat{g}_{31} - \frac{\hat{h}_{34}}{2} + \frac{\hat{s}_{25}}{2} + \frac{3}{5}\hat{s}_{28} \,; \quad \hat{u}_{29} = -\frac{\hat{f}_{35}}{2} - \frac{\hat{g}_{34}}{2} - \hat{h}_{32} + \frac{\hat{s}_{26}}{2} + \frac{3}{5}\hat{s}_{29} \,;$$

$$\hat{u}_{30} = -3\,\hat{f}_{11} - \hat{g}_{14} - \hat{h}_{15} + \hat{s}_5 + \frac{2}{3}\hat{s}_{30} \,; \quad \hat{u}_{31} = -\hat{f}_{16} - 3\,\hat{g}_{12} - \hat{h}_{17} + \hat{s}_6 + \frac{2}{3}\hat{s}_{31} \,;$$

$$\hat{u}_{32} = -\hat{f}_{18} - \hat{g}_{19} - 3\,\hat{h}_{13} + \hat{s}_7 + \frac{2}{3}\hat{s}_{32} \,; \quad \hat{u}_{33} = -2\,\hat{f}_{14} - 2\,\hat{g}_{16} - \hat{h}_{20} + \hat{s}_8 + \frac{2}{3}\hat{s}_{33} \,;$$

$$\hat{u}_{34} = -\hat{f}_{20} - 2\,\hat{g}_{17} - 2\,\hat{h}_{19} + \hat{s}_9 + \frac{2}{3}\hat{s}_{34} \,; \quad \hat{u}_{35} = -2\,\hat{f}_{15} - \hat{g}_{20} - 2\,\hat{h}_{18} + \hat{s}_{10} + \frac{2}{3}\hat{s}_{35}$$
$$. \tag{45}$$

As before, we only need the $\hat{f}_2$, $\hat{f}_5$, $\hat{f}_8$, $\hat{f}_{10}$, $\hat{f}_{11}$, $\hat{f}_{14}$, $\hat{f}_{15}$, $\hat{f}_{16}$, $\hat{f}_{18}$, $\hat{f}_{20}$, $\hat{f}_{24}$, $\hat{f}_{27}$, $\hat{f}_{30}$, $\hat{f}_{33}$ and $\hat{f}_{35}$ modes for the *x*-flux in eqn. (45). Of these, only $\hat{f}_{14}$, $\hat{f}_{15}$, $\hat{f}_{16}$, $\hat{f}_{18}$, $\hat{f}_{24}$, $\hat{f}_{27}$, $\hat{f}_{30}$, $\hat{f}_{33}$ and $\hat{f}_{35}$ will have to be rebuilt in the second and third iterations since the remaining only need to be built once at $\tau = 0$. Likewise, we only need the $\hat{g}_3$, $\hat{g}_6$, $\hat{g}_8$, $\hat{g}_9$, $\hat{g}_{12}$, $\hat{g}_{14}$, $\hat{g}_{16}$, $\hat{g}_{17}$, $\hat{g}_{19}$, $\hat{g}_{20}$, $\hat{g}_{25}$, $\hat{g}_{28}$, $\hat{g}_{31}$, $\hat{g}_{33}$ and $\hat{g}_{34}$ modes for the *y*-flux in eqn. (45). Of these, only $\hat{g}_{14}$, $\hat{g}_{16}$, $\hat{g}_{17}$, $\hat{g}_{19}$, $\hat{g}_{25}$, $\hat{g}_{28}$, $\hat{g}_{31}$, $\hat{g}_{33}$ and $\hat{g}_{34}$ will have to be rebuilt in the second and third iterations since the remaining are only built once at $\tau = 0$. Similarly, we only need the $\hat{h}_4$, $\hat{h}_7$, $\hat{h}_9$, $\hat{h}_{10}$, $\hat{h}_{13}$, $\hat{h}_{15}$, $\hat{h}_{17}$, $\hat{h}_{18}$, $\hat{h}_{19}$, $\hat{h}_{20}$, $\hat{h}_{26}$, $\hat{h}_{29}$, $\hat{h}_{32}$, $\hat{h}_{34}$ and $\hat{h}_{35}$ modes for the *z*-flux in eqn. (45). Of these, only $\hat{h}_{15}$, $\hat{h}_{17}$, $\hat{h}_{18}$, $\hat{h}_{19}$, $\hat{h}_{26}$, $\hat{h}_{29}$, $\hat{h}_{32}$, $\hat{h}_{34}$ and $\hat{h}_{35}$ have to be rebuilt in the second and third iterations since the remaining only need to be built once at $\tau = 0$.

To obtain the above-mentioned modes for the fluxes, we first define an ordered set of forty-two symmetrically-placed nodes in the reference space-time element given by

$\{ (0,0,0,0); (1/2,0,0,0); (1/4,0,0,0); (-1/4,0,0,0); (-1/2,0,0,0); (0,1/2,0,0);$
$(0,1/4,0,0); (0,-1/4,0,0); (0,-1/2,0,0); (0,0,1/2,0); (0,0,1/4,0); (0,0,-1/4,0);$
$(0,0,-1/2,0); (1/2,1/2,1/2,0); (1/2,-1/2,1/2,0); (1/2,1/2,-1/2,0); (1/2,-1/2,-1/2,0);$
$(-1/2,1/2,1/2,0); (-1/2,-1/2,1/2,0); (-1/2,1/2,-1/2,0); (-1/2,-1/2,-1/2,0);$
$(0,0,0,1/3); (1/2,1/2,0,1/3); (1/2,-1/2,0,1/3); (1/2,0,1/2,1/3); (1/2,0,-1/2,1/3);$
$(-1/2,1/2,0,1/3); (-1/2,-1/2,0,1/3); (-1/2,0,1/2,1/3); (-1/2,0,-1/2,1/3);$
$(0,1/2,1/2,1/3); (0,-1/2,1/2,1/3); (0,1/2,-1/2,1/3); (0,-1/2,-1/2,1/3);$
$(0,0,0,2/3); (1/2,0,0,2/3); (-1/2,0,0,2/3); (0,1/2,0,2/3); (0,-1/2,0,2/3);$
$(0,0,1/2,2/3); (0,0,-1/2,2/3); (0,0,0,1)\}$

(46)

We now obtain the conserved variables at each of the above nodal points by using the following formulae

$$\bar{u}_1 = \hat{w}_1 - (\hat{w}_5 + \hat{w}_6 + \hat{w}_7)/12 \;\; ; \;\; \bar{u}_2 = \bar{u}_1 + (60\hat{w}_2 + 30\hat{w}_5 + 6\hat{w}_{11} - 5\hat{w}_{16} - 5\hat{w}_{18})/120 \;\; ;$$

$$\bar{u}_3 = 0.5\,(\bar{u}_1 + \bar{u}_2) - 0.015625(4\hat{w}_5 + 3\hat{w}_{11}) \;\; ; \;\; \bar{u}_4 = 3\,\bar{u}_1 + \bar{u}_2 - 3\,\bar{u}_3 - 0.09375\,\hat{w}_{11} \;\; ;$$

$$\bar{u}_5 = 6\,\bar{u}_1 + 3\,\bar{u}_2 - 8\,\bar{u}_3 - 0.375\,\hat{w}_{11} \;\; ; \;\; \bar{u}_6 = \bar{u}_1 + (60\,\hat{w}_3 + 30\,\hat{w}_6 + 6\,\hat{w}_{12} - 5\,\hat{w}_{14} - 5\,\hat{w}_{19})/120 \;\; ;$$

$$\bar{u}_7 = 0.5\,(\bar{u}_1 + \bar{u}_6) - 0.015625\,(4\,\hat{w}_6 + 3\,\hat{w}_{12}) \;\; ; \;\; \bar{u}_8 = 3\,\bar{u}_1 + \bar{u}_6 - 3\,\bar{u}_7 - 0.09375\,\hat{w}_{12} \;\; ;$$

$$\bar{u}_9 = 6\,\bar{u}_1 + 3\,\bar{u}_6 - 8\,\bar{u}_7 - 0.375\,\hat{w}_{12} \;\; ; \;\; \bar{u}_{10} = \bar{u}_1 + (60\,\hat{w}_4 + 30\,\hat{w}_7 + 6\,\hat{w}_{13} - 5\,\hat{w}_{15} - 5\,\hat{w}_{17})/120 \;\; ;$$

$$\bar{u}_{11} = 0.5\,(\bar{u}_1 + \bar{u}_{10}) - 0.015625\,(4\,\hat{w}_7 + 3\,\hat{w}_{13}) \;\; ; \;\; \bar{u}_{12} = 3\,\bar{u}_1 + \bar{u}_{10} - 3\,\bar{u}_{11} - 0.09375\,\hat{w}_{13} \;\; ;$$

$$\bar{u}_{13} = 6\,\bar{u}_1 + 3\,\bar{u}_{10} - 8\,\bar{u}_{11} - 0.375\,\hat{w}_{13} \;\; ;$$

$$\bar{u}_{14} = \bar{u}_2 + \bar{u}_7 + \bar{u}_{10} - 2\,\bar{u}_1 + 0.25\,(\hat{w}_3 + \hat{w}_8 + \hat{w}_9 + \hat{w}_{10}) + 5\,(\hat{w}_{14} + \hat{w}_{19})/48 + 23\,\hat{w}_{12}/320$$
$$+ 0.125\,(\hat{w}_{15} + \hat{w}_{16} + \hat{w}_{17} + \hat{w}_{18} + \hat{w}_{20}) + 0.1875\,\hat{w}_6 \;\; ;$$

$$\bar{u}_{15} = \bar{u}_{14} + \bar{u}_9 - \bar{u}_6 - 0.5\,(\hat{w}_8 + \hat{w}_9) - 0.25\,(\hat{w}_{14} + \hat{w}_{19} + \hat{w}_{20}) \;\; ;$$

$$\bar{u}_{16} = \bar{u}_{14} + \bar{u}_{13} - \bar{u}_{10} - 0.5\,(\hat{w}_9 + \hat{w}_{10}) - 0.25\,(\hat{w}_{15} + \hat{w}_{17} + \hat{w}_{20}) \;\; ;$$

$$\bar{u}_{18} = \bar{u}_{14} + \bar{u}_5 - \bar{u}_2 - 0.5\,(\hat{w}_8 + \hat{w}_{10}) - 0.25\,(\hat{w}_{16} + \hat{w}_{18} + \hat{w}_{20}) \;\; ;$$

$$\bar{u}_{19} = \bar{u}_{18} + \bar{u}_9 - \bar{u}_6 + 0.5\,(\hat{w}_8 - \hat{w}_9) - 0.25\,(\hat{w}_{14} + \hat{w}_{19} - \hat{w}_{20}) \;\; ;$$

$$\bar{u}_{20} = \bar{u}_{18} + \bar{u}_{13} - \bar{u}_{10} - 0.5\,(\hat{w}_9 - \hat{w}_{10}) - 0.25\,(\hat{w}_{15} + \hat{w}_{17} - \hat{w}_{20}) \;\; ;$$

$$\bar{u}_{17} = \bar{u}_{15} + \bar{u}_{16} - \bar{u}_{14} + \hat{w}_9 + 0.5\,\hat{w}_{20} \;\; ; \;\; \bar{u}_{21} = \bar{u}_{19} + \bar{u}_{20} - \bar{u}_{18} + \hat{w}_9 - 0.5\,\hat{w}_{20}$$

(47)

and

$$\bar{u}_{22} = \bar{u}_1 + (36\hat{u}_{21} + 12\hat{u}_{22} + 4\hat{u}_{23} - 3\hat{u}_{30} - 3\hat{u}_{31} - 3\hat{u}_{32})/108 \ ;$$

$$\bar{u}_{23} = \bar{u}_{22} + \bar{u}_2 + \bar{u}_6 - 2\bar{u}_1$$
$$+ (18\hat{w}_8 + 9\hat{w}_{14} + 9\hat{w}_{16} + 12\hat{u}_{24} + 12\hat{u}_{25} + 4\hat{u}_{27} + 4\hat{u}_{28} + 6\hat{u}_{30} + 6\hat{u}_{31} + 6\hat{u}_{33})/72 \ ;$$

$$\bar{u}_{24} = \bar{u}_{23} - \bar{u}_6 + \bar{u}_9 - 0.5\hat{w}_8 - 0.25\hat{w}_{14} - (6\hat{u}_{25} + 2\hat{u}_{28} + 3\hat{u}_{33})/18 \ ;$$

$$\bar{u}_{25} = 0.5(\bar{u}_{14} + \bar{u}_{15} + \bar{u}_{23} + \bar{u}_{24}) - \bar{u}_2 - 0.5\hat{w}_6 - 0.25\hat{w}_{16} - 0.125\hat{w}_{17}$$
$$+ (6\hat{u}_{26} + 2\hat{u}_{29} - 3\hat{u}_{31} + 3\hat{u}_{32} + 3\hat{u}_{35})/36 \ ;$$

$$\bar{u}_{26} = \bar{u}_{23} + \bar{u}_{24} - \bar{u}_{25} - 0.5(\hat{w}_6 - \hat{w}_7) - 0.25(\hat{w}_{16} - \hat{w}_{18}) - (\hat{u}_{31} - \hat{u}_{32})/6 \ ;$$

$$\bar{u}_{27} = \bar{u}_5 + \bar{u}_{23} - \bar{u}_2 - 0.5\hat{w}_8 - 0.25\hat{w}_{16} - (6\hat{u}_{24} + 2\hat{u}_{27} + 3\hat{u}_{33})/18 \ ;$$

$$\bar{u}_{28} = \bar{u}_9 + \bar{u}_{27} - \bar{u}_6 + 0.5\hat{w}_8 - 0.25\hat{w}_{14} - (6\hat{u}_{25} + 2\hat{u}_{28} - 3\hat{u}_{33})/18 \ ;$$

$$\bar{u}_{29} = 0.5(\bar{u}_{18} + \bar{u}_{19} + \bar{u}_{27} + \bar{u}_{28}) - \bar{u}_5 - 0.5\hat{w}_6 + 0.25\hat{w}_{16} - 0.125\hat{w}_{17}$$
$$+ (6\hat{u}_{26} + 2\hat{u}_{29} - 3\hat{u}_{31} + 3\hat{u}_{32} - 3\hat{u}_{35})/36 \ ;$$

$$\bar{u}_{30} = \bar{u}_{27} + \bar{u}_{28} - \bar{u}_{29} - 0.5(\hat{w}_6 - \hat{w}_7) + 0.25(\hat{w}_{16} - \hat{w}_{18}) - (\hat{u}_{31} - \hat{u}_{32})/6 \ ;$$

$$\bar{u}_{31} = 0.5(\bar{u}_{14} + \bar{u}_{18} + \bar{u}_{25} + \bar{u}_{29}) - \bar{u}_{10} - 0.5\hat{w}_5 - 0.25\hat{w}_{15} - 0.125\hat{w}_{14}$$
$$+ (6\hat{u}_{25} + 2\hat{u}_{28} - 3\hat{u}_{30} + 3\hat{u}_{31} + 3\hat{u}_{34})/36 \ ;$$

$$\bar{u}_{32} = \bar{u}_9 + \bar{u}_{31} - \bar{u}_6 - 0.5\hat{w}_9 - 0.25\hat{w}_{19} - (6\hat{u}_{25} + 2\hat{u}_{28} + 3\hat{u}_{34})/18 \ ;$$

$$\bar{u}_{33} = \bar{u}_{23} + \bar{u}_{27} - \bar{u}_{31} - 0.5(\hat{w}_5 - \hat{w}_7) - 0.25(\hat{w}_{14} - \hat{w}_{19}) - (\hat{u}_{30} - \hat{u}_{32})/6 \ ;$$

$$\bar{u}_{34} = \bar{u}_{24} + \bar{u}_{28} - \bar{u}_{32} - 0.5(\hat{w}_5 - \hat{w}_7) + 0.25(\hat{w}_{14} - \hat{w}_{19}) - (\hat{u}_{30} - \hat{u}_{32})/6 \ ;$$

$$q_1 = (36\hat{u}_{21} + 24\hat{u}_{22} + 16\hat{u}_{23})/54 \ ; \quad q_2 = (18\hat{u}_{24} + 12\hat{u}_{27})/54 \ ; \quad q_3 = (6\hat{u}_{30} - 3\hat{u}_{31} - 3\hat{u}_{32})/54 \ ;$$

$$q_4 = (18\hat{u}_{25} + 12\hat{u}_{28})/54 \ ; \quad q_5 = (-3\hat{u}_{30} + 6\hat{u}_{31} - 3\hat{u}_{32})/54 \ ; \quad q_6 = (18\hat{u}_{26} + 12\hat{u}_{29})/54 \ ;$$

$$q_7 = (-3\hat{u}_{30} - 3\hat{u}_{31} + 6\hat{u}_{32})/54 \ ;$$

$$\bar{u}_{35} = 2\bar{u}_{22} - \bar{u}_1 + 2(\hat{u}_{22} + \hat{u}_{23})/9 \ ; \quad \bar{u}_{36} = \bar{u}_2 + q_1 + q_2 + q_3 \ ; \quad \bar{u}_{37} = \bar{u}_5 + q_1 - q_2 + q_3 \ ;$$

$$\bar{u}_{38} = \bar{u}_6 + q_1 + q_4 + q_5 \ ; \quad \bar{u}_{39} = \bar{u}_9 + q_1 - q_4 + q_5 \ ; \quad \bar{u}_{40} = \bar{u}_{10} + q_1 + q_6 + q_7 \ ;$$

$$\bar{u}_{41} = \bar{u}_{13} + q_1 - q_6 + q_7 \ ; \quad \bar{u}_{42} = \bar{u}_1 + \hat{u}_{21} + \hat{u}_{22} + \hat{u}_{23} - (\hat{u}_{30} + \hat{u}_{31} + \hat{u}_{32})/12$$

. (48)

As in the third order case, each nodal value utilizes the ones that were built before it, making the computation very efficient. Notice too that the set of conserved variables at the first twenty-one nodal points in eqn. (47) only need to be built once, i.e. in the first iteration. The conserved variables at the next twenty-one nodal points, i.e. the ones in eqn. (48), will need to be built once in the first iteration and then rebuilt in the subsequent iterations. The conserved variables from eqns. (47) and (48) can then be used to evaluate the fluxes at the nodes. The fluxes at the first twenty-one nodal points only need to be built once; the fluxes at the remaining twenty-one nodal points need to be rebuilt for each iteration. Once the fluxes are built at the nodal points, we can evaluate the desired modes of the fluxes. This is most easily done by having a set of temporary variables, which we denote by a "$q$". The requisite modes of the $x$-flux are given as follows

$$q_1 = \bar{f}_{14} + \bar{f}_{15} + \bar{f}_{16} + \bar{f}_{17} \; ; \; q_2 = \bar{f}_{18} + \bar{f}_{19} + \bar{f}_{20} + \bar{f}_{21} \; ; \; q_3 = \bar{f}_{14} + \bar{f}_{16} + \bar{f}_{18} + \bar{f}_{20} \; ;$$
$$q_4 = \bar{f}_{15} + \bar{f}_{17} + \bar{f}_{19} + \bar{f}_{21} \; ; \; q_5 = \bar{f}_{14} + \bar{f}_{15} + \bar{f}_{18} + \bar{f}_{19} \; ; \; q_6 = \bar{f}_{16} + \bar{f}_{17} + \bar{f}_{20} + \bar{f}_{21} \; ;$$
$$q_7 = \bar{f}_{23} + \bar{f}_{24} - \bar{f}_{25} - \bar{f}_{26} - \bar{f}_{27} - \bar{f}_{28} + \bar{f}_{29} + \bar{f}_{30} \; ; \; q_8 = \bar{f}_{23} + \bar{f}_{27} - \bar{f}_{24} - \bar{f}_{28} - \bar{f}_{31} - \bar{f}_{33} + \bar{f}_{32} + \bar{f}_{34} \; ;$$
$$q_9 = \bar{f}_{25} + \bar{f}_{29} - \bar{f}_{26} - \bar{f}_{30} - \bar{f}_{31} - \bar{f}_{32} + \bar{f}_{33} + \bar{f}_{34} \; ; \; q_{10} = \bar{f}_{23} + \bar{f}_{24} + \bar{f}_{25} + \bar{f}_{26} - \bar{f}_{27} - \bar{f}_{28} - \bar{f}_{29} - \bar{f}_{30} \; ;$$
$$\hat{f}_5 = 2(\bar{f}_2 - 2\bar{f}_1 + \bar{f}_5) \; ; \; \hat{f}_8 = 0.5(\bar{f}_{14} - \bar{f}_{15} + \bar{f}_{16} - \bar{f}_{17} - \bar{f}_{18} + \bar{f}_{19} - \bar{f}_{20} + \bar{f}_{21}) \; ;$$
$$\hat{f}_{10} = 0.5(\bar{f}_{14} + \bar{f}_{15} - \bar{f}_{16} - \bar{f}_{17} - \bar{f}_{18} - \bar{f}_{19} + \bar{f}_{20} + \bar{f}_{21}) \; ; \; \hat{f}_{11} = 16(\bar{f}_2 - 2\bar{f}_3 + 2\bar{f}_4 - \bar{f}_5)/3 \; ;$$
$$\hat{f}_2 = 2(\bar{f}_2 - \bar{f}_5)/3 - 0.1\hat{f}_{11} + (q_1 - q_2)/12 \; ; \; \hat{f}_{20} = \bar{f}_{14} - \bar{f}_{15} - \bar{f}_{16} + \bar{f}_{17} - \bar{f}_{18} + \bar{f}_{19} + \bar{f}_{20} - \bar{f}_{21} \; ;$$
$$\hat{f}_{14} = q_8 + 0.5(q_3 - q_4) - 2(\bar{f}_6 - \bar{f}_9) \; ; \; \hat{f}_{15} = q_9 + 0.5(q_5 - q_6) - 2(\bar{f}_{10} - \bar{f}_{13}) \; ;$$
$$\hat{f}_{16} = q_7 + 0.5(q_1 - q_2) - 2(\bar{f}_2 - \bar{f}_5) \; ; \; \hat{f}_{17} = -2q_9 + \hat{f}_{15} \; ; \; \hat{f}_{18} = -2q_7 + \hat{f}_{16} \; ;$$
$$\hat{f}_{19} = -2q_8 + \hat{f}_{14} \; ; \; \hat{f}_{27} = 1.125(q_1 - q_2) + 4.5(\bar{f}_{36} - \bar{f}_{37}) - 2.25q_{10} \; ;$$
$$\hat{f}_{24} = 3(\bar{f}_{36} - \bar{f}_{37}) - 0.75q_{10} + 0.375(\hat{f}_{16} + \hat{f}_{18}) - \hat{f}_{27} \; ; \; \hat{f}_{30} = 3(\bar{f}_{36} - 2\bar{f}_{35} + \bar{f}_{37}) - 1.5\hat{f}_5 \; ;$$
$$\hat{f}_{33} = 3(\bar{f}_{23} - \bar{f}_{27} - \bar{f}_{24} + \bar{f}_{28}) - 3\hat{f}_8 \; ; \; \hat{f}_{35} = 3(\bar{f}_{25} - \bar{f}_{29} - \bar{f}_{26} + \bar{f}_{30}) - 3\hat{f}_{10}$$

, (49)

the modes of the $y$-flux are given as follows

$$q_1 = \overline{g}_{14} + \overline{g}_{15} + \overline{g}_{16} + \overline{g}_{17} \; ; \; q_2 = \overline{g}_{18} + \overline{g}_{19} + \overline{g}_{20} + \overline{g}_{21} \; ; \; q_3 = \overline{g}_{14} + \overline{g}_{16} + \overline{g}_{18} + \overline{g}_{20} \; ;$$

$$q_4 = \overline{g}_{15} + \overline{g}_{17} + \overline{g}_{19} + \overline{g}_{21} \; ; \; q_5 = \overline{g}_{14} + \overline{g}_{15} + \overline{g}_{18} + \overline{g}_{19} \; ; \; q_6 = \overline{g}_{16} + \overline{g}_{17} + \overline{g}_{20} + \overline{g}_{21} \; ;$$

$$q_7 = \overline{g}_{23} + \overline{g}_{24} - \overline{g}_{25} - \overline{g}_{26} - \overline{g}_{27} - \overline{g}_{28} + \overline{g}_{29} + \overline{g}_{30} \; ; \; q_8 = \overline{g}_{23} + \overline{g}_{27} - \overline{g}_{24} - \overline{g}_{28} - \overline{g}_{31} - \overline{g}_{33} + \overline{g}_{32} + \overline{g}_{34} \; ;$$

$$q_9 = \overline{g}_{25} + \overline{g}_{29} - \overline{g}_{26} - \overline{g}_{30} - \overline{g}_{31} - \overline{g}_{32} + \overline{g}_{33} + \overline{g}_{34} \; ; \; q_{11} = \overline{g}_{23} - \overline{g}_{24} + \overline{g}_{27} - \overline{g}_{28} + \overline{g}_{31} - \overline{g}_{32} + \overline{g}_{33} - \overline{g}_{34} \; ;$$

$$\hat{g}_6 = 2(\overline{g}_6 - 2\overline{g}_1 + \overline{g}_9) \; ; \; \hat{g}_8 = 0.5(\overline{g}_{14} - \overline{g}_{15} + \overline{g}_{16} - \overline{g}_{17} - \overline{g}_{18} + \overline{g}_{19} - \overline{g}_{20} + \overline{g}_{21}) \; ;$$

$$\hat{g}_9 = 0.5(\overline{g}_{14} - \overline{g}_{15} - \overline{g}_{16} + \overline{g}_{17} + \overline{g}_{18} - \overline{g}_{19} - \overline{g}_{20} + \overline{g}_{21}) \; ; \; \hat{g}_{12} = 16(\overline{g}_6 - 2\overline{g}_7 + 2\overline{g}_8 - \overline{g}_9)/3 \; ;$$

$$\hat{g}_3 = 2(\overline{g}_6 - \overline{g}_9)/3 - 0.1\hat{g}_{12} + (q_3 - q_4)/12 \; ; \; \hat{g}_{20} = \overline{g}_{14} - \overline{g}_{15} - \overline{g}_{16} + \overline{g}_{17} - \overline{g}_{18} + \overline{g}_{19} + \overline{g}_{20} - \overline{g}_{21} \; ;$$

$$\hat{g}_{14} = q_8 + 0.5(q_3 - q_4) - 2(\overline{g}_6 - \overline{g}_9) \; ; \; \hat{g}_{15} = q_9 + 0.5(q_5 - q_6) - 2(\overline{g}_{10} - \overline{g}_{13}) \; ;$$

$$\hat{g}_{16} = q_7 + 0.5(q_1 - q_2) - 2(\overline{g}_2 - \overline{g}_5) \; ; \; \hat{g}_{17} = -2q_9 + \hat{g}_{15} \; ; \; \hat{g}_{18} = -2q_7 + \hat{g}_{16} \; ;$$

$$\hat{g}_{19} = -2q_8 + \hat{g}_{14} \; ; \; \hat{g}_{28} = 1.125(q_3 - q_4) + 4.5(\overline{g}_{38} - \overline{g}_{39}) - 2.25q_{11} \; ;$$

$$\hat{g}_{25} = 3(\overline{g}_{38} - \overline{g}_{39}) - 0.75q_{11} + 0.375(\hat{g}_{14} + \hat{g}_{19}) - \hat{g}_{28} \; ; \; \hat{g}_{31} = 3(\overline{g}_{38} - 2\overline{g}_{35} + \overline{g}_{39}) - 1.5\hat{g}_6 \; ;$$

$$\hat{g}_{33} = 3(\overline{g}_{23} - \overline{g}_{27} - \overline{g}_{24} + \overline{g}_{28}) - 3\hat{g}_8 \; ; \; \hat{g}_{34} = 3(\overline{g}_{31} - \overline{g}_{32} - \overline{g}_{33} + \overline{g}_{34}) - 3\hat{g}_9$$

, (50)

and the modes of the $z$-flux are given as follows

$$q_1 = \overline{h}_{14} + \overline{h}_{15} + \overline{h}_{16} + \overline{h}_{17} \; ; \; q_2 = \overline{h}_{18} + \overline{h}_{19} + \overline{h}_{20} + \overline{h}_{21} \; ; \; q_3 = \overline{h}_{14} + \overline{h}_{16} + \overline{h}_{18} + \overline{h}_{20} \; ;$$

$$q_4 = \overline{h}_{15} + \overline{h}_{17} + \overline{h}_{19} + \overline{h}_{21} \; ; \; q_5 = \overline{h}_{14} + \overline{h}_{15} + \overline{h}_{18} + \overline{h}_{19} \; ; \; q_6 = \overline{h}_{16} + \overline{h}_{17} + \overline{h}_{20} + \overline{h}_{21} \; ;$$

$$q_7 = \overline{h}_{23} + \overline{h}_{24} - \overline{h}_{25} - \overline{h}_{26} - \overline{h}_{27} - \overline{h}_{28} + \overline{h}_{29} + \overline{h}_{30} \; ; \; q_8 = \overline{h}_{23} + \overline{h}_{27} - \overline{h}_{24} - \overline{h}_{28} - \overline{h}_{31} - \overline{h}_{33} + \overline{h}_{32} + \overline{h}_{34} \; ;$$

$$q_9 = \overline{h}_{25} + \overline{h}_{29} - \overline{h}_{26} - \overline{h}_{30} - \overline{h}_{31} - \overline{h}_{32} + \overline{h}_{33} + \overline{h}_{34} \; ; \; q_{12} = \overline{h}_{25} - \overline{h}_{26} + \overline{h}_{29} - \overline{h}_{30} + \overline{h}_{31} + \overline{h}_{32} - \overline{h}_{33} - \overline{h}_{34} \; ;$$

$$\hat{h}_7 = 2(\overline{h}_{10} - 2\overline{h}_1 + \overline{h}_{13}) \; ; \; \hat{h}_9 = 0.5(\overline{h}_{14} - \overline{h}_{15} - \overline{h}_{16} + \overline{h}_{17} + \overline{h}_{18} - \overline{h}_{19} - \overline{h}_{20} + \overline{h}_{21}) \; ;$$

$$\hat{h}_{10} = 0.5(\overline{h}_{14} + \overline{h}_{15} - \overline{h}_{16} - \overline{h}_{17} - \overline{h}_{18} - \overline{h}_{19} + \overline{h}_{20} + \overline{h}_{21}) \; ; \; \hat{h}_{13} = 16(\overline{h}_{10} - 2\overline{h}_{11} + 2\overline{h}_{12} - \overline{h}_{13})/3 \; ;$$

$$\hat{h}_4 = 2(\overline{h}_{10} - \overline{h}_{13})/3 - 0.1\hat{h}_{13} + (q_5 - q_6)/12 \; ; \; \hat{h}_{20} = \overline{h}_{14} - \overline{h}_{15} - \overline{h}_{16} + \overline{h}_{17} - \overline{h}_{18} + \overline{h}_{19} + \overline{h}_{20} - \overline{h}_{21} \; ;$$

$$\hat{h}_{14} = q_8 + 0.5(q_3 - q_4) - 2(\overline{h}_6 - \overline{h}_9) \; ; \; \hat{h}_{15} = q_9 + 0.5(q_5 - q_6) - 2(\overline{h}_{10} - \overline{h}_{13}) \; ;$$

$$\hat{h}_{16} = q_7 + 0.5(q_1 - q_2) - 2(\overline{h}_2 - \overline{h}_5) \; ; \; \hat{h}_{17} = -2q_9 + \hat{h}_{15} \; ; \; \hat{h}_{18} = -2q_7 + \hat{h}_{16} \; ;$$

$$\hat{h}_{19} = -2q_8 + \hat{h}_{14} \; ; \; \hat{h}_{29} = 1.125(q_5 - q_6) + 4.5(\overline{h}_{40} - \overline{h}_{41}) - 2.25q_{12} \; ;$$

$$\hat{h}_{26} = 3(\overline{h}_{40} - \overline{h}_{41}) - 0.75q_{12} + 0.375(\hat{h}_{15} + \hat{h}_{17}) - \hat{h}_{29} \; ; \; \hat{h}_{32} = 3(\overline{h}_{40} - 2\overline{h}_{35} + \overline{h}_{41}) - 1.5\hat{h}_7 \; ;$$

$$\hat{h}_{34} = 3(\overline{h}_{31} - \overline{h}_{32} - \overline{h}_{33} + \overline{h}_{34}) - 3\hat{h}_9 \; ; \; \hat{h}_{35} = 3(\overline{h}_{25} - \overline{h}_{26} - \overline{h}_{29} + \overline{h}_{30}) - 3\hat{h}_{10}$$

. (51)

The sources, if present, have to be evaluated at all the nodal points. Many of the modes for the source terms are obtained by setting $f \to s$ in eqn. (49), $g \to s$ in eqn. (50) and $h \to s$ in eqn. (51). The remaining modes that are not contained in eqns. (49) to (51) are given below

$$\hat{s}_1 = 0.125 \left(\bar{s}_{14} + \bar{s}_{15} + \bar{s}_{16} + \bar{s}_{17} + \bar{s}_{18} + \bar{s}_{19} + \bar{s}_{20} + \bar{s}_{21}\right) - \left(\hat{s}_5 + \hat{s}_6 + \hat{s}_7\right)/6 \;;$$
$$\hat{s}_{23} = 4.5 \left(\bar{s}_{42} - \bar{s}_1\right) - 13.5\left(\bar{s}_{35} - \bar{s}_{22}\right) \;;\; \hat{s}_{22} = 2.25 \left(\bar{s}_1 - \bar{s}_{22} - \bar{s}_{35} + \bar{s}_{42}\right) - 1.5\hat{s}_{23} \;;. \quad (52)$$
$$\hat{s}_{21} = \bar{s}_{42} - \bar{s}_1 - \hat{s}_{22} - \hat{s}_{23} + \left(\hat{s}_{30} + \hat{s}_{31} + \hat{s}_{32}\right)/12$$

Once $\hat{u}_{21}$ to $\hat{u}_{35}$ are obtained at the end of the third iteration, the complete modal space representation of $\hat{f}$, $\hat{g}$ and $\hat{h}$ may also be rebuilt and stored.

The space-time integral of the source term, which is needed in eqn. (6) for each zone "$i,j,k$", is then given by

$$\bar{S}_{i,j,k} = \left(\hat{s}_1 + 0.5 \hat{s}_{21} + \hat{s}_{22}/3 + 0.25 \hat{s}_{23}\right)/\Delta t \quad . \quad (53)$$

This completes our description of the fourth order ADER-CG scheme in modal space. A nodal space formulation of ADER-CG schemes at fourth order is given in Appendix C.

## V) Obtaining the Resolved Numerical Flux at Zone Boundaries

Eqns. (6), (8) and (17) have shown us that the numerical flux (and electric fields for MHD) can be obtained by: a) freezing the wave structure at the facial barycenter and b) obtaining suitable space-time averages of the conserved variables and fluxes at either side of a zone boundary. The first step consists of obtaining a physical set of conserved variables at either side of a face that are centered in space and time. These are the variables that will be used to construct the frozen wave speeds that are needed in the HLLE Riemann solver in eqn. (17). If a linearized Riemann solver is desired, the above-mentioned variables can also be used to obtain the frozen eigenvectors. The conserved variables at the space-time barycenters of top and bottom *x*-faces of the reference element at zone "*i,j,k*" are given by

$$u_{bx+;i,j,k} \equiv u_{i,j,k}\left(\xi = 1/2, \eta = 0, \zeta = 0, \tau = 1/2\right)$$
$$u_{bx-;i,j,k} \equiv u_{i,j,k}\left(\xi = -1/2, \eta = 0, \zeta = 0, \tau = 1/2\right) \quad . \tag{54}$$

The upper panel in Fig. 1 shows the placement of the conserved variables from eqn. (54) at the x-faces of a one-dimensional mesh. Since cataloguing explicit formulae at all six faces of the reference element is very repetitive, we simply demonstrate the process of obtaining the conserved variables and *x*-fluxes at the top and bottom *x*-faces of the reference element at zone "*i,j,k*". Within that zone, we assume that the descriptions from the previous Section have enabled us to construct eqns. (14), (15) or (16) at second, third or fourth orders.

Our primary task in this Section is to obtain closed-form expressions for eqn. (54) and the following space-time averages:

$$\langle u \rangle_{x+;i,j,k} \equiv \iiint_{(\eta,\zeta,\tau)\in[-.5,.5]^2 \times [0,1]} u_{i,j,k}\left(\xi = 1/2, \eta, \zeta, \tau\right) d\eta \, d\zeta \, d\tau$$
$$\langle u \rangle_{x-;i,j,k} \equiv \iiint_{(\eta,\zeta,\tau)\in[-.5,.5]^2 \times [0,1]} u_{i,j,k}\left(\xi = -1/2, \eta, \zeta, \tau\right) d\eta \, d\zeta \, d\tau \quad . \tag{55}$$

The dashed lines in the lower panel of Fig. 1 show the locations where the integrals from eqn. (55) are evaluated at the *x*-faces of a mesh. Notice that $\langle u \rangle_{x+;i,j,k}$ is a space-time average of the conserved variables at the lower side of the top *x*-face of zone "*i,j,k*" while $\langle u \rangle_{x-;i,j,k}$ is a similar average at the upper side of the bottom *x*-face of the same zone. If the space-time representations $\hat{f}$, $\hat{g}$ and $\hat{h}$ are also retained, expressions analogous to the ones above can be asserted for the fluxes. The terms in eqn. (55) can be used in eqn. (17) to yield the numerical flux. We illustrate this for the construction of the *x*-flux $\overline{F}_{i+1/2,j,k}$ for the HLL Riemann solver. Notice from Fig. 1 that $u_{bx+;i,j,k}$ and $u_{bx-;i+1,j,k}$ are physical states at the space-time barycenter of the *x*-face at "*i+1/2,j,k*" and can, therefore, be used to build the frozen eigenvalues at the zone boundary. (For the linearized Riemann solver, they can also be used to build the frozen eigenvectors at the zone boundary.) From these eigenvalues, we can build $\lambda_L$ and $\lambda_R$ that are needed in eqn. (17), see the upper panel in Fig. 1. We can then use $\langle u \rangle_{x+;i,j,k}$ and $\langle u \rangle_{x-;i+1,j,k}$ as well as analogous expressions for the *x*-fluxes to obtain a space-time averaged version of eqn. (17) as

$$\overline{F}_{i+1/2,j,k} = \left[\frac{\lambda_R}{\lambda_R - \lambda_L}\right] \frac{\Delta x}{\Delta t} \langle f \rangle_{x+;i,j,k} - \left[\frac{\lambda_L}{\lambda_R - \lambda_L}\right] \frac{\Delta x}{\Delta t} \langle f \rangle_{x-;i+1,j,k} + \left[\frac{\lambda_R \lambda_L}{\lambda_R - \lambda_L}\right] \left( \langle u \rangle_{x-;i+1,j,k} - \langle u \rangle_{x+;i,j,k} \right) \tag{56}$$

Please see both panels of Fig. 1 in order to fully appreciate why eqn. (56) is a space-time averaged version of eqn. (17). This illustrates the process of obtaining numerical fluxes from eqns. (54) and (55).

For MHD, we also need the following space-time averages at the zone edges:

$$\langle u \rangle_{x+;y+;i,j,k} \equiv \iint_{(\zeta,\tau)\in[-.5,.5]\times[0,1]} u_{i,j,k}\left(\xi=1/2,\eta=1/2,\zeta,\tau\right) d\zeta\, d\tau$$

$$\langle u \rangle_{x+;y-;i,j,k} \equiv \iint_{(\zeta,\tau)\in[-.5,.5]\times[0,1]} u_{i,j,k}\left(\xi=1/2,\eta=-1/2,\zeta,\tau\right) d\zeta\, d\tau$$

$$\langle u \rangle_{x+;z+;i,j,k} \equiv \iint_{(\eta,\tau)\in[-.5,.5]\times[0,1]} u_{i,j,k}\left(\xi=1/2,\eta,\zeta=1/2,\tau\right) d\eta\, d\tau$$

$$\langle u \rangle_{x+;z-;i,j,k} \equiv \iint_{(\eta,\tau)\in[-.5,.5]\times[0,1]} u_{i,j,k}\left(\xi=1/2,\eta,\zeta=-1/2,\tau\right) d\eta\, d\tau$$

$$\langle u \rangle_{x-;y+;i,j,k} \equiv \iint_{(\zeta,\tau)\in[-.5,.5]\times[0,1]} u_{i,j,k}\left(\xi=-1/2,\eta=1/2,\zeta,\tau\right) d\zeta\, d\tau$$

$$\langle u \rangle_{x-;y-;i,j,k} \equiv \iint_{(\zeta,\tau)\in[-.5,.5]\times[0,1]} u_{i,j,k}\left(\xi=-1/2,\eta=-1/2,\zeta,\tau\right) d\zeta\, d\tau$$

$$\langle u \rangle_{x-;z+;i,j,k} \equiv \iint_{(\eta,\tau)\in[-.5,.5]\times[0,1]} u_{i,j,k}\left(\xi=-1/2,\eta,\zeta=1/2,\tau\right) d\eta\, d\tau$$

$$\langle u \rangle_{x-;z-;i,j,k} \equiv \iint_{(\eta,\tau)\in[-.5,.5]\times[0,1]} u_{i,j,k}\left(\xi=-1/2,\eta,\zeta=-1/2,\tau\right) d\eta\, d\tau \qquad (57)$$

When the HLL flux is used, the electric fields are just averages of appropriate components of eqn. (56), see eqn. (7). In the transonic case, the dissipation can be doubled by doubling the third term in eqn. (17).

For the rest of this Section, we drop the subscripts "i,j,k" since we will only be cataloguing the procedure for taking the modal space-time representation of the conserved variables and fluxes within a zone and using them to obtain closed-form expressions for the terms in eqns. (54), (55) and (57). The next three Sub-sections catalogue closed-form formulae for these terms at second, third and fourth order.

If the modal representation of the fluxes is not retained, one has to identify a minimal set of space-time quadrature points within each face, derive a modal representation from those quadrature points and then use the modal representation to obtain the above integrals. This results in a memory efficient formulation that is also computationally efficient. This is also the method of choice when designing RK schemes using the methods described here. We have carried out such procedures and implemented them in code. In order to retain a compact structure for this paper, we detail the procedure that is sketched in this paragraph in Appendix D.

**IV.1) Flux and Electric Field Evaluation at Second Order**

The second order expressions are easy to verify. They are reminiscent of a predictor-corrector update. As in the previous Section, we use temporary variables, which we denote by a "$q$". For the terms that are defined in eqn. (54) we have the following second order expressions

$$u_{bx+} = \hat{w}_1 + 0.5\left(\hat{w}_2 + \hat{u}_5\right) \ ; \ u_{bx-} = \hat{w}_1 + 0.5\left(-\hat{w}_2 + \hat{u}_5\right). \tag{58}$$

For the terms that are defined in eqn. (55) we have the following second order expressions

$$\langle u \rangle_{x+} = u_{bx+} \ ; \ \langle u \rangle_{x-} = u_{bx-}. \tag{59}$$

For the terms in eqn. (57) we have the following second order expressions

$$\begin{aligned}
\langle u \rangle_{x+;y+} &= u_{bx+} + 0.5\,\hat{w}_3 \ ; \ \langle u \rangle_{x+;y-} = u_{bx+} - 0.5\,\hat{w}_3 \ ; \\
\langle u \rangle_{x+;z+} &= u_{bx+} + 0.5\,\hat{w}_4 \ ; \ \langle u \rangle_{x+;z-} = u_{bx+} - 0.5\,\hat{w}_4 \ ; \\
\langle u \rangle_{x-;y+} &= u_{bx-} + 0.5\,\hat{w}_3 \ ; \ \langle u \rangle_{x-;y-} = u_{bx-} - 0.5\,\hat{w}_3 \ ; \\
\langle u \rangle_{x-;z+} &= u_{bx-} + 0.5\,\hat{w}_4 \ ; \ \langle u \rangle_{x-;z-} = u_{bx-} - 0.5\,\hat{w}_4
\end{aligned} \tag{60}$$

If the modal representation of the fluxes is also saved, analogues to eqns. (59) and (60) can also be obtained for the *x*-flux. If the modal representation of the fluxes is not saved, Appendix D provides a computationally efficient strategy for obtaining the space-time modes of the *x*-flux in the *x*-face at second order. Similar equations can be written at the *y* and *z*-faces. This completes our description of the flux and electric field evaluation at second order.

### IV.2) Flux and Electric Field Evaluation at Third Order

As in the previous Section, we use temporary variables, which we denote by a "$q$". For the terms that are defined in eqn. (54) we have the following third order expressions

$$u_{bx+} = \hat{w}_1 + 0.5\,(\hat{w}_2 + \hat{u}_{11}) + 0.25\,(\hat{u}_{12} + \hat{u}_{13}) + (2\,\hat{w}_5 - \hat{w}_6 - \hat{w}_7)/12$$
$$u_{bx-} = u_{bx+} - \hat{w}_2 - 0.5\,\hat{u}_{13}$$
(61)

For the terms that are defined in eqn. (55) we have the following third order expressions

$$q_1 = (\hat{w}_6 + \hat{w}_7 + \hat{u}_{12})/12 \;;\; \langle u \rangle_{x+} = u_{bx+} + q_1 \;;\; \langle u \rangle_{x-} = u_{bx-} + q_1.$$
(62)

For the terms that are defined in eqn. (57) we have the following third order expressions

$$\begin{aligned}
&q_2 = 0.5\,\hat{w}_3 + 0.25\,(\hat{w}_8 + \hat{u}_{14}) \;;\; q_3 = 0.5\,\hat{w}_4 + 0.25\,(\hat{w}_{10} + \hat{u}_{15}) \;; \\
&\langle u \rangle_{x+;y+} = \langle u \rangle_{x+} + \hat{w}_6/6 + q_2 \;;\; \langle u \rangle_{x+;y-} = \langle u \rangle_{x+;y+} - 2\,q_2 \;; \\
&\langle u \rangle_{x+;z+} = \langle u \rangle_{x+} + \hat{w}_7/6 + q_3 \;;\; \langle u \rangle_{x+;z-} = \langle u \rangle_{x+;z+} - 2\,q_3 \;; \\
&q_4 = 0.5\,\hat{w}_3 - 0.25\,(\hat{w}_8 - \hat{u}_{14}) \;;\; q_5 = 0.5\,\hat{w}_4 - 0.25\,(\hat{w}_{10} - \hat{u}_{15}) \;; \\
&\langle u \rangle_{x-;y+} = \langle u \rangle_{x-} + \hat{w}_6/6 + q_4 \;;\; \langle u \rangle_{x-;y-} = \langle u \rangle_{x-;y+} - 2\,q_4 \;; \\
&\langle u \rangle_{x-;z+} = \langle u \rangle_{x-} + \hat{w}_7/6 + q_5 \;;\; \langle u \rangle_{x-;z-} = \langle u \rangle_{x-;z+} - 2\,q_5
\end{aligned}$$
(63)

As in the previous Section, we observe that once a few terms have been obtained, the remaining evaluations in the three preceding equations can be dramatically simplified. If the modal representation of the fluxes is also saved, analogues to eqns. (62) and (63) can also be obtained for the $x$-flux. If the modal representation of the fluxes is not saved, Appendix D provides a computationally efficient strategy for obtaining the space-time modes of the $x$-flux in the $x$-face at third order. Similar equations can be written at the $y$ and z-faces. This completes our description of the flux and electric field evaluation at third order.

**IV.3) Flux and Electric Field Evaluation at Fourth Order**

For the terms that are defined in eqn. (54) we have the following fourth order expressions

$$q_1 = \hat{w}_1 + 0.5\,\hat{u}_{21} + 0.25\,\hat{u}_{22} + 0.125\,\hat{u}_{23} + (4\,\hat{w}_5 - 2\,\hat{w}_6 - 2\,\hat{w}_7 + 2\,\hat{u}_{30} - \hat{u}_{31} - \hat{u}_{32})/24\;;$$
$$q_2 = 0.5\,\hat{w}_2 + 0.05\,\hat{w}_{11} + 0.25\,\hat{u}_{24} + 0.125\,\hat{u}_{27} - (\hat{w}_{16} + \hat{w}_{18})/24\;;$$
$$u_{bx+} = q_1 + q_2\;;\quad u_{bx-} = q_1 - q_2$$

(64)

For the terms that are defined in eqn. (55) we have the following fourth order expressions

$$\langle u \rangle_{x+} = u_{bx+} + (2\,\hat{w}_6 + 2\,\hat{w}_7 + \hat{w}_{16} + \hat{w}_{18} + 2\,\hat{u}_{22} + 3\,\hat{u}_{23} + \hat{u}_{27} + \hat{u}_{31} + \hat{u}_{32})/24$$
$$\langle u \rangle_{x-} = u_{bx-} + (2\,\hat{w}_6 + 2\,\hat{w}_7 - \hat{w}_{16} - \hat{w}_{18} + 2\,\hat{u}_{22} + 3\,\hat{u}_{23} - \hat{u}_{27} + \hat{u}_{31} + \hat{u}_{32})/24$$

(65)

For the terms that are defined in eqn. (57) we have the following fourth order expressions

$$q_3 = \hat{w}_3 + 0.5\,(\hat{w}_8 + \hat{u}_{25}) + 0.25\,\hat{u}_{33} + 0.1\,\hat{w}_{12} + (\hat{w}_{14} + 2\,\hat{u}_{28})/6\;;$$
$$q_4 = \hat{w}_4 + 0.5\,(\hat{w}_{10} + \hat{u}_{26}) + 0.25\,\hat{u}_{35} + 0.1\,\hat{w}_{13} + (\hat{w}_{15} + 2\,\hat{u}_{29})/6\;;$$
$$\langle u \rangle_{x+;y+} = \langle u \rangle_{x+} + (2\,\hat{w}_6 + \hat{w}_{16} + \hat{u}_{31})/12 + 0.5\,q_3\;;\quad \langle u \rangle_{x+;y-} = \langle u \rangle_{x+;y+} - q_3\;;$$
$$\langle u \rangle_{x+;z+} = \langle u \rangle_{x+} + (2\,\hat{w}_7 + \hat{w}_{18} + \hat{u}_{32})/12 + 0.5\,q_4\;;\quad \langle u \rangle_{x+;z-} = \langle u \rangle_{x+;z+} - q_4\;;$$
$$q_5 = \hat{w}_3 - 0.5\,(\hat{w}_8 - \hat{u}_{25}) - 0.25\,\hat{u}_{33} + 0.1\,\hat{w}_{12} + (\hat{w}_{14} + 2\,\hat{u}_{28})/6\;;$$
$$q_6 = \hat{w}_4 - 0.5\,(\hat{w}_{10} - \hat{u}_{26}) - 0.25\,\hat{u}_{35} + 0.1\,\hat{w}_{13} + (\hat{w}_{15} + 2\,\hat{u}_{29})/6\;;$$
$$\langle u \rangle_{x-;y+} = \langle u \rangle_{x-} + (2\,\hat{w}_6 - \hat{w}_{16} + \hat{u}_{31})/12 + 0.5\,q_5\;;\quad \langle u \rangle_{x-;y-} = \langle u \rangle_{x-;y+} - q_5\;;$$
$$\langle u \rangle_{x-;z+} = \langle u \rangle_{x-} + (2\,\hat{w}_7 - \hat{w}_{18} + \hat{u}_{32})/12 + 0.5\,q_6\;;\quad \langle u \rangle_{x-;z-} = \langle u \rangle_{x-;z+} - q_6$$

(66)

As in the previous Section, we observe that once a few terms have been obtained, the remaining evaluations in the three preceding equations can be dramatically simplified. If the modal representation of the fluxes is also saved, analogues to eqns. (65) and (66) can also be obtained for the *x*-flux. If the modal representation of the fluxes is not saved, Appendix D provides a computationally efficient strategy for obtaining the space-time modes of the *x*-flux in the *x*-face at fourth order. Similar equations can be written at the *y* and *z*-faces. This completes our description of the flux and electric field evaluation at fourth order.

## VI) Order of Accuracy of the ADER-CG Schemes

We present here a number of two and three dimensional tests to demonstrate the order of accuracy of these schemes. Each of the two dimensional tests was run with a CFL number of 0.45 and the three dimensional tests were run with a CFL number of 0.3. Each of the tests presented in this section utilize an HLLE Riemann solver. The entire set of test problems was also run with the linearized Riemann solver, and consistent orders of accuracy were observed in each case. Please note that our second order scheme actually uses slopes that are constructed from the r=3 WENO interpolants in Jiang & Shu [44]. Consequently, the second order results shown here are somewhat superior to those that would be obtained from a second order TVD scheme. The results presented in this section are from the ADER-CG modal space representation presented in this paper. We have run the same problems using the nodal space representation and found nearly identical values of the errors for each case.

### VI.1) Isentropic unmagnetized vortex in two dimensions

The first test we present is an isentropic vortex, as presented by Balsara and Shu [5]. In this problem, a hydrodynamic vortex is added to a mean flow oblique to the mesh at an angle of 45°. The problem is solved on a grid in the x-y plane extending over a domain of [-5, 5] × [-5, 5] with periodic boundary conditions. The mean flow is given by $(\rho, P, v_x, v_y, B_x, B_y, B_z) = (1, 1, 1, 1, 0, 0, 0)$, and the variation from the mean flow due to the vortex by

$$(\delta v_x, \delta v_y) = \frac{\varepsilon}{2\pi} e^{0.5(1-r^2)}(-y, x) \; ; \; \delta T = -\frac{(\gamma-1)\varepsilon^2}{8\gamma\pi^2} e^{(1-r^2)} \; ; \; \delta S = 0,$$

where the temperature and entropy are given by $T = P/\rho$ and $S = P/\rho^\gamma$ respectively. The ratio of specific heats is taken to be $\gamma = 1.4$. The simulation is run to time t=10, so that the vortex has completely traversed the grid once. The parameter $\varepsilon$ controls the strength of the vortex, which we set to $\varepsilon = 5$. The problem is initialized using numerical quadrature with accuracy equal to that of the scheme. Note that in the fourth order problem, we double the size of the domain to [-10, 10] × [-10, 10] and double the simulation stopping time to t=20 because the nonzero values of the exponential function at the boundaries are picked up by the fourth order scheme on the smaller domain.

Table I shows the results of the accuracy tests at second, third and fourth orders, where the errors are measured in the density variable. Observe that the schemes meet their expected order of accuracy in the $L_1$ and $L_\infty$ norms in each of the cases we have tested. It is worth pointing out that the errors of the 256×256 simulation at third order are comparable to the errors of the 512×512 simulation at second order, demonstrating the value of a higher-order scheme.

**Table I**: The accuracy analysis for the two-dimensional isentropic vortex problem using the second, third and fourth order ADER-CG schemes in the modal representation with an HLLE Riemann solver. The errors are computed from the density variable.

| Order | Number of zones | $L_1$ error | $L_1$ order | $L_\infty$ error | $L_\infty$ order |
|---|---|---|---|---|---|
| 2nd | 64×64 | 9.9560×10⁻⁴ | | 2.1303×10⁻² | |
| | 128×128 | 2.0912×10⁻⁴ | 2.25 | 4.5133×10⁻³ | 2.24 |
| | 256×256 | 4.7517×10⁻⁵ | 2.14 | 9.2646×10⁻⁴ | 2.28 |
| | 512×512 | 1.1614×10⁻⁵ | 2.03 | 2.0240×10⁻⁴ | 2.20 |
| 3rd | 64×64 | $6.8025 \times 10^{-4}$ | | $1.1245 \times 10^{-2}$ | |
| | 128×128 | $1.0221 \times 10^{-4}$ | 2.74 | $1.6252 \times 10^{-3}$ | 2.79 |
| | 256×256 | $1.3103 \times 10^{-5}$ | 2.96 | $2.1295 \times 10^{-4}$ | 2.93 |
| | 512×512 | $1.6432 \times 10^{-6}$ | 2.99 | $2.6799 \times 10^{-5}$ | 3.00 |
| 4th | 64×64 | $4.4035 \times 10^{-4}$ | | $2.8125 \times 10^{-2}$ | |
| | 128×128 | $2.4834 \times 10^{-5}$ | 4.15 | $2.0067 \times 10^{-3}$ | 3.81 |
| | 256×256 | $1.0118 \times 10^{-6}$ | 4.62 | $7.9187 \times 10^{-5}$ | 4.66 |
| | 512×512 | $3.7938 \times 10^{-8}$ | 4.74 | $3.0047 \times 10^{-6}$ | 4.72 |

**VI.2) Magnetized isodensity vortex in two dimensions**

We now turn our attention to the magnetized vortex, which is described in Balsara [8]. We use the same periodic problem domain as in the previous test, and the same values for the mean flow. We double the domain and stopping time for the fourth order scheme in this case as well. In this problem, however, we take the ratio of specific heats to be $\gamma = 5/3$. The variation from the mean flow for the magnetized vortex is given by

$$(\delta v_x, \delta v_y) = \frac{\kappa}{2\pi} e^{0.5(1-r^2)}(-y, x) \ ; \ (\delta B_x, \delta B_y) = \frac{\mu}{2\pi} e^{0.5(1-r^2)}(-y, x) \ ;$$

$$\delta P = \frac{1}{8\pi}(\frac{\mu}{2\pi})^2 (1-r^2) e^{(1-r^2)} - \frac{1}{2}(\frac{\kappa}{2\pi})^2 e^{(1-r^2)}$$

and we take the density to be constant across the entire domain.

Table II shows the error measured in the x-directional magnetic field at the end of the simulation. See again that the scheme meets design accuracy at second, third, and fourth orders. Here too, the third order ADER-CG scheme produces errors at third order which are comparable to the second order scheme at twice the resolution. The benefit of going to higher orders is evident, therefore, because one can achieve similar or better numerical accuracies on computational meshes of half the resolution.

**Table II**: The accuracy analysis for the two-dimensional magnetized vortex problem using the second, third and fourth order ADER-CG schemes in the modal representation with an HLLE Riemann solver. The errors are computed from the *x*-magnetic field variable.

| Order | Number of zones | $L_1$ error | $L_1$ order | $L_\infty$ error | $L_\infty$ order |
|---|---|---|---|---|---|
| 2nd | 64×64 | $2.5329 \times 10^{-3}$ | | $3.2997 \times 10^{-2}$ | |
| | 128×128 | $5.9004 \times 10^{-4}$ | 2.10 | $6.4889 \times 10^{-3}$ | 2.34 |
| | 256×256 | $1.4406 \times 10^{-4}$ | 2.03 | $1.4320 \times 10^{-3}$ | 2.18 |
| | 512×512 | $3.5885 \times 10^{-5}$ | 2.01 | $3.5155 \times 10^{-4}$ | 2.03 |
| 3rd | 64×64 | $8.5732 \times 10^{-4}$ | | $1.4232 \times 10^{-2}$ | |
| | 128×128 | $1.1041 \times 10^{-4}$ | 2.96 | $1.6385 \times 10^{-3}$ | 3.12 |
| | 256×256 | $1.3834 \times 10^{-5}$ | 2.99 | $2.1258 \times 10^{-4}$ | 2.94 |
| | 512×512 | $1.7657 \times 10^{-6}$ | 2.97 | $2.6809 \times 10^{-5}$ | 2.98 |
| 4th | 64×64 | $5.1853 \times 10^{-4}$ | | $3.7648 \times 10^{-2}$ | |
| | 128×128 | $2.0506 \times 10^{-5}$ | 4.66 | $1.5512 \times 10^{-3}$ | 4.60 |
| | 256×256 | $9.3604 \times 10^{-7}$ | 4.45 | $5.4514 \times 10^{-5}$ | 4.83 |
| | 512×512 | $5.2602 \times 10^{-8}$ | 4.15 | $2.5203 \times 10^{-6}$ | 4.43 |

**VI.3) Torsional Alfven wave in three dimensions**

We now turn our attention to three-dimensional accuracy tests. First we consider the propagation of a torsional Alfven wave as described by Balsara, *et al.* [11]. This problem is initialized with the Alfven wave propagating in the $x'$ direction of an ($x'$, $y'$, $z'$) coordinate system, which is then rotated into the (x, y, z) coordinate system, where $x'$ points along the diagonal of the mesh. The initial variables in the ($x'$, $y'$, $z'$) coordinate system are given by

$$\rho = 1 \; ; \; P = 10 \; ; \; \gamma = 1.4 \; ; \; \Phi = \frac{2\pi}{\lambda}\left(x' - 2\,t\right) \; ;$$

$$v_{x'} = 1 \; ; \; v_{y'} = \varepsilon \cos \Phi \; ; \; v_{z'} = \varepsilon \sin \Phi \; ;$$

$$B_{x'} = \sqrt{4\pi\rho} \; ; \; B_{y'} = -\varepsilon\sqrt{4\pi\rho} \cos \Phi \; ; \; B_{z'} = -\varepsilon\sqrt{4\pi\rho} \sin \Phi$$

where we set the constants $\varepsilon = 0.02$ and $\lambda = \sqrt{3}$. When initializing the problem, we ensure the magnetic field is divergence-free by utilizing the vector potential

$$A_{x'} = 0 \; ; \; A_{y'} = \varepsilon\lambda\sqrt{\rho/\pi}\cos\Phi \; ; \; A_{z'} = \sqrt{4\pi\rho}\; y' + \varepsilon\lambda\sqrt{\rho/\pi}\sin\Phi.$$

The vector potential is evaluated by numerical quadrature at each edge, which yields a value for the magnetic field in each face. The problem is simulated on a unit cube with periodic boundaries and run to a time of $t = \sqrt{3}/2$, by which time the Alfven wave has traversed the cube's diagonal.

The actual problem is solved on a unit cube with periodic boundary conditions in the (x,y,z) coordinate frame which is rotated relative to the (x', y', z') coordinate system. The rotation matrix is given by **A** so that we have

$$\mathbf{A} = \begin{bmatrix} \cos\psi\cos\phi - \cos\theta\sin\phi\sin\psi & \cos\psi\sin\phi + \cos\theta\cos\phi\sin\psi & \sin\psi\sin\theta \\ -\sin\psi\cos\phi - \cos\theta\sin\phi\cos\psi & -\sin\psi\sin\phi + \cos\theta\cos\phi\cos\psi & \cos\psi\sin\theta \\ \sin\theta\sin\phi & -\sin\theta\cos\phi & \cos\theta \end{bmatrix}$$

where $\phi = -\pi/4$, $\theta = \sin^{-1}\left(-\sqrt{2/3}\right)$ and $\psi = \sin^{-1}\left(\left(\sqrt{2}-\sqrt{6}\right)/4\right)$. As a result, the position vector $\mathbf{r}'$ in the primed frame transforms to the position vector $\mathbf{r}$ in the unprimed frame as $\mathbf{r} = \mathbf{A}\,\mathbf{r}'$. Other vectors transform similarly. The effect of the rotation is to make the wave propagate along the diagonal of the unit cube.

Table III presents the errors of this test for second, third, and fourth orders. Each scheme meets its designed order property. The third order scheme on the 64×64×64 mesh has errors which are already a factor of two smaller than the second order scheme on the 96×96×96 mesh. Also, notice that the fourth order simulation on the 32×32×32 mesh, has errors which are smaller than any of the second or third order simulations.

**Table III**: The accuracy analysis for the three-dimensional torsional Alfven wave using the second, third and fourth order ADER-CG schemes in the modal representation with an HLLE Riemann solver. The errors are computed from the *x*-magnetic field variable.

| Order | Number of zones | $L_1$ error | $L_1$ order | $L_\infty$ error | $L_\infty$ order |
|---|---|---|---|---|---|
| 2nd | 16×16 | $1.8020 \times 10^{-2}$ | | $2.8617 \times 10^{-2}$ | |
| | 32×32 | $3.6368 \times 10^{-3}$ | 2.30 | $5.7164 \times 10^{-3}$ | 2.32 |
| | 64×64 | $6.9888 \times 10^{-4}$ | 2.38 | $1.0975 \times 10^{-3}$ | 2.39 |
| | 96×96 | $2.8698 \times 10^{-4}$ | 2.20 | $4.5086 \times 10^{-4}$ | 2.19 |
| 3rd | 16×16 | $8.3352 \times 10^{-3}$ | | $1.2990 \times 10^{-2}$ | |
| | 32×32 | $1.0277 \times 10^{-3}$ | 3.03 | $1.6721 \times 10^{-3}$ | 2.96 |
| | 64×64 | $1.3007 \times 10^{-4}$ | 2.98 | $2.3739 \times 10^{-4}$ | 2.82 |
| | 96×96 | $3.9276 \times 10^{-5}$ | 2.96 | $7.3238 \times 10^{-5}$ | 2.90 |
| 4th | 16×16 | $7.6961 \times 10^{-4}$ | | $1.1966 \times 10^{-3}$ | |
| | 32×32 | $3.0104 \times 10^{-5}$ | 4.67 | $4.6556 \times 10^{-5}$ | 4.69 |
| | 64×64 | $1.9767 \times 10^{-6}$ | 3.93 | $3.8464 \times 10^{-6}$ | 3.60 |
| | 96×96 | $3.9630 \times 10^{-7}$ | 3.96 | $7.0009 \times 10^{-7}$ | 4.20 |

**VI.4) Density wave in three dimensions**

Finally, we consider a purely hydrodynamic density wave propagating in three dimensions, as described in Balsara, *et al.* [11]. We solve this problem on the same periodic domain as the torsional Alfven wave above. The flow variables are given initially in the $(x', y', z')$ coordinate system by

$$\rho = 1 + \varepsilon \sin \Phi \; ; \; P = 1 \; ; \; \gamma = 1.4 \; ; \; \Phi = \frac{2\pi}{\lambda}(x' - t) \; ;$$
$$v_{x'} = 1 \; ; \; v_{y'} = 0 \; ; \; v_{z'} = 0 \; ; \; B_{x'} = 0 \; ; \; B_{y'} = 0 \; ; \; B_{z'} = 0$$

and rotated into the unit cube, as above. We have chosen $\varepsilon = 0.2$ and $\lambda = \sqrt{3}$. The simulation is run to a time of $t = \sqrt{3}$ by which time the density wave has propagated the length of the cube's diagonal. The actual problem is solved in the (x,y,z) coordinate frame which is rotated relative to the $(x', y', z')$ coordinate system using the same rotation matrix that was detailed in the previous Sub-section.

The errors in density for this problem are presented in Table IV. For this problem, each of the schemes demonstrates its expected order of accuracy. We see again that the third order scheme on the 64×64×64 zone mesh has errors comparable to the second order scheme with 96×96×96 resolution. Again, too, the fourth order scheme at a resolution of 32×32×32 zones outpreforms all of the second and third order simulations we have conducted.

**Table IV**: The accuracy analysis for the three-dimensional density wave using the second, third and fourth order ADER-CG schemes in the modal representation with an HLLE Riemann solver. The errors are computed from the density variable.

| Order | Number of zones | $L_1$ error | $L_1$ order | $L_\infty$ error | $L_\infty$ order |
|---|---|---|---|---|---|
| 2nd | 16×16 | $2.8009\times10^{-2}$ | | $4.5285\times10^{-2}$ | |
| | 32×32 | $4.4752\times10^{-3}$ | 2.65 | $8.3212\times10^{-3}$ | 2.44 |
| | 64×64 | $7.7461\times10^{-4}$ | 2.53 | $1.5860\times10^{-4}$ | 2.40 |
| | 96×96 | $3.0293\times10^{-4}$ | 2.31 | $6.3249\times10^{-4}$ | 2.26 |
| 3rd | 16×16 | $1.9211\times10^{-2}$ | | $3.0296\times10^{-2}$ | |
| | 32×32 | $2.7334\times10^{-3}$ | 2.81 | $4.5948\times10^{-3}$ | 2.72 |
| | 64×64 | $3.5766\times10^{-4}$ | 2.93 | $5.7929\times10^{-4}$ | 2.99 |
| | 96×96 | $1.0737\times10^{-4}$ | 2.97 | $1.7158\times10^{-4}$ | 3.01 |
| 4th | 16×16 | $1.3409\times10^{-3}$ | | $2.0625\times10^{-3}$ | |
| | 32×32 | $6.5790\times10^{-5}$ | 4.35 | $1.0433\times10^{-4}$ | 4.30 |
| | 64×64 | $3.8174\times10^{-6}$ | 4.11 | $6.2961\times10^{-6}$ | 4.05 |
| | 96×96 | $7.5004\times10^{-7}$ | 4.01 | $1.2170\times10^{-7}$ | 4.07 |

## VII) Comparing Speeds for ADER and RK Time Stepping

In this Section we cross-compare the speeds of various ADER-CG schemes as applied to three-dimensional MHD and Euler flow blast wave problems on a $97^3$ zone mesh. The blast wave problems will themselves be described in the ensuing two sections. Details of the physical problem, however, have no bearing on the speed. A centered WENO reconstruction of the appropriate order was applied to the conserved variables in all instances. The schemes of interest, along with their speeds in zones updated per second are catalogued in Table V. All test problems were run on a dedicated Intel Nehalem processor with a clock speed of 2.53 GHz. Only one core of the multi-core CPU was used so as to best bring out the serial performance.

We first compare the nodal and modal formulations of ADER-CG which were coupled to an HLL Riemann solver. We see that for all the orders tested, the nodal and modal formulations of ADER perform at roughly the same speed. The modal formulations do evaluate the fluxes at a slightly larger number of nodal points than the nodal formulations. They however compensate for this by having a very efficient transcription from nodal to modal space as well as a very compact formulation of the ADER update equations in modal space, as was shown in Section IV. This is equally true for the MHD and Euler problems tested. As a result, we conclude that the modal and nodal formulations of ADER-CG perform quite comparably in all instances.

We now consider the modal formulation of ADER-CG along with a linearized Riemann solver. We see that the linearized Riemann solver makes a considerable difference in the speed at lower orders, specifically at $2^{nd}$ order. This is also the order at which the quality of the Riemann solver makes a substantial change in the quality of the solution. Thus, if one wants a better quality solution at $2^{nd}$ order, it would behoove one to use a linearized Riemann solver. However, we see that at $3^{rd}$ and $4^{th}$ orders, the linearized Riemann solver does not suffer from a substantial speed penalty relative to the HLL Riemann solver. This can be understood by realizing that the strategy that was sketched in Sub-Section II.3 and detailed in Section V is very efficient at obtaining higher order numerical fluxes and electric fields. Thus, as the cost of the ADER-CG algorithm increases with increasing order, the cost of the Riemann solver is very efficiently amortized by the formulations presented in this paper. When a linearized Riemann solver is used, the third order ADER schemes are half as fast as the second order ADER schemes and the fourth order ADER schemes are a third as fast as the third order ADER schemes.

Let us now compare ADER schemes to RK schemes. It is worth observing that the third order ADER scheme with a linearized Riemann solver is practially as fast as the second order RK scheme with a linearized Riemann solver. Similarly, the fourth order ADER schemes with a linearized Riemann solver cost only forty percent more than the third order RK scheme with a similar Riemann solver. We also see that ADER schemes make it beneficial to use more sophisticated Riemann solvers with their reduced dissipation, because the Riemann solver is called only once per time step and the cost of the Riemann solver does not factor significantly in the overall cost of the scheme, especially at higher orders. The RK schemes invoke the reconstruction strategy and Riemann solver multiple times per time step; consequently, they do not share this advantage. Observe too that the third order ADER scheme with the HLL Riemann solver is only 40% slower than the second order RK scheme with the same Riemann solver. A similar trend is observed when comparing the fourth order ADER scheme to the third order RK scheme when both schemes use the HLLE Riemann solver. The fourth order RK schemes can, in principle, operate with a CFL of 1.5. We see, however, that this increase in the CFL number does not compensate for the decrease in the speed per time step. In summary, this paragraph shows that when we are willing to pay a fixed cost per simulation (as is often the case in engineering applications), the ADER schemes can yield almost an order of magnitude improvement relative to RK schemes.

It is also interesting to briefly examine the memory footprint of ADER schemes. For the implementation catalogued here, the second order ADER scheme has the same memory footprint as a typical second order TVD scheme. The third order ADER scheme requires 1.62 times the memory as the second order scheme. The fourth order ADER scheme requires 1.78 times the memory as the corresponding third order scheme. We see, therefore, that the memory usage of ADER schemes increases in a well-contained fashion with increasing order. The modal and nodal formulations of ADER have the same memory footprint at all orders. At the risk of expressing a viewpoint, we therefore conclude that the third order ADER-WENO scheme, either with an HLL or linearized Riemann solver, represents an excellent upgrade path for scientists and engineers who are working with a second order TVD scheme. The third order ADER-WENO scheme in modal space is easy to implement, operates with a speed that is comparable to the corresponding second order RK-TVD scheme and has a memory footprint that is only moderately larger than the second order RK-TVD scheme. It is as robust as the second order scheme (a property that is also shared by the fourth order scheme) and has dissipation characteristics that are substantially better, as we shall show in the subsequent two Sub-sections.

The ADER schemes have one further advantage relative to the RK schemes. The strong stability preserving RK schemes do not have a dense output (Horn [42]). As a result, it is not possible to use the solution at all of the sub-stages to interpolate in time. Eqns. (14), (15) and (16) show that the ADER schemes can indeed be used to interpolate the solution to intermediate times. The ability to interpolate the solution to intermediate times is very useful in adaptive mesh refinement (AMR) simulations. Schemes for AMR-hydrodynamics and AMR-MHD indeed do rely on this ability to interpolate the solution in time at fine-coarse mesh interfaces (Berger & Colella [15], Balsara [6]). ADER schemes should, therefore, prove very useful in AMR simulations.

**Table V**: Speed (in zones updated per second) of various ADER-WENO and RK-WENO schemes for MHD and Euler test problems run on a dedicated Intel Nehalem processor with a clock speed of 2.53 GHz. Only one core was used.

| Scheme | Riemann Solver | $2^{nd}$ Order Zones/sec | $3^{rd}$ Order Zones/sec | $4^{th}$ Order Zones/sec |
|---|---|---|---|---|
| ADER-WENO-modal-MHD | HLL | 118,280 | 46,418 | 12,076 |
| ADER-WENO-nodal-MHD | HLL | 121,362 | 48,880 | 13,001 |
| ADER-WENO-modal-MHD | Linearized | 60,186 | 33,424 | 10,396 |
| RK-WENO-MHD | HLL | 72,047 | 21,791 | 3,778 |
| RK-WENO-MHD | Linearized | 33,475 | 14,057 | 3,051 |
| ADER-WENO-modal-Euler | HLL | 198,817 | 75,105 | 21,065 |
| ADER-WENO-nodal-Euler | HLL | 207,198 | 78,994 | 21,115 |
| ADER-WENO-modal-Euler | Linearized | 154,674 | 67,025 | 20,348 |
| RK-WENO-Euler | HLL | 126,935 | 37,049 | 8,623 |
| RK-WENO-Euler | Linearized | 92,448 | 31,598 | 8,119 |

## VIII) Hydrodynamical Tests with ADER Time Stepping

In this Section we present several hydrodynamical tests in two and three dimensions to demonstrate the accuracy and stability of the ADER-CG scheme presented in this paper. These will serve to demonstrate the utility and robustness of the ADER method for simulating hydrodynamic flows.

**VIII.1) Forward facing step problem in two dimensions**

This problem was first proposed by Woodward and Colella [65]. It consists of a two-dimensional wind tunnel with a domain of [0, 3]×[0, 1]. An ideal gas with a polytropic index of 1.4 flows in at the left boundary at a speed of Mach 3, a density of 1.4 , a pressure of 1 and a ratio of specific heats of 1.4. A forward-facing step is located with its upper corner at the position (0.6, 0.2). Outflow boundary conditions are applied on the right boundary. The top and bottom walls have reflective boundary conditions. We treat the singularity at the corner of the box in the same manner as Woodward and Colella [65]. The simulation is run to a time of 4.0 with a CFL number of 0.45.

Fig. 3 shows the final density from simulations that have been run on 240×80, 480×160 and 960×320 zone meshes. All the simulations reported here were run with a fourth order ADER-WENO scheme with a linearized Riemann solver. All the shocks have sharp profiles and the vortex sheet shows little or no spreading along the length of the computational domain. The 480×160 zone simulation shows the onset of the vortex sheet roll up and the 960×320 zone simulation shows a very pronounced roll up. A very good second order scheme would need at least 960×320 zones to show any evidence of vortex sheet roll up, whereas the fourth order scheme displays this at half the resolution. This demonstrates the effectiveness of high order schemes.

**VIII.2) Double Mach reflection in two dimensions**

This problem was proposed by Woodward and Colella [65], and we use the same parameters as those authors. The problem simulates the similarity solution that develops in multi-dimensions when an angled wedge is put in a supersonic flow. The computational domain of this problem is [0, 4]×[0, 1]. Initially, a Mach 10 shock is positioned at an angle of 60° to the bottom boundary, meeting that boundary at $x = 1/6$. For values of $x < 1/6$, the post-shock conditions are used at the boundary, which mimics the start of the wedge. For $x > 1/6$, the boundary is reflective, which mimics the windward face of the wedge. The upper boundary is set to exactly track the motion of the shock. The left and right edges have inflow and outflow boundary conditions, respectively. The unshocked material is initialized with a density of 1.4, a pressure of unity and a ratio of specific heats of 1.4. The problem was run to a time of 0.2 using a CFL number of 0.45. As is customary, we only image the domain [0, 3]×[0, 1].

Fig. 4 shows the density variable at the final time at 960×240 and 1920×480 zone resolution. The two smaller panels at the bottom of the figure show a blow-up of the region around the Mach stem for both computations. The fourth order ADER-WENO scheme was used with a linearized Riemann solver. Notice that our scheme resolves all the structures, including the instabilities that develop around the Mach stem, that were shown in Cockburn & Shu [19] in their fourth order simulation of this problem at the same resolution. The ADER schemes do, however, offer the advantage that they are substantially more efficient at higher orders than the corresponding RK time stepping that has been the traditional method for obtaining higher order accuracy in time.

**VIII.3) Hydrodynamic blast wave in three dimensions**

This problem consists of a centrally located high pressure region surrounded by an initially low pressure fluid. It was solved on a computational domain of [-0.5, 0.5] ×[-0.5, 0.5] ×[-0.5, 0.5] using a mesh of $96^3$ zones. The initial density is set to unity everywhere, and the fluid is initially at rest. The pressure is set to 0.1 for r > 0.1, and to 1000 for r < 0.1. The ratio of specific heats is given by $\gamma$ =1.4. The problem was run to a time of 0.1 time units with a CFL number of 0.3 using all the ADER schemes catalogued here.

Fig. 5 shows the density, pressure and magnitude of the velocity in the midplane of the simulation when the fourth order ADER-WENO scheme was used with an HLL Riemann solver. We show the results from the modal formulation. We see that the blast wave is perfectly symmetrical. The oppositely oriented velocities were found to be equal in magnitude up to numerical round-off at all orders when the modal ADER formulation was used. The nodal ADER formulation produced an asymmetry in the final velocities of 0.04% at second order, 0.0002% at third order and 0.007% at fourth order. We see, therefore, that the asymmetries are very small, but they do not decrease with increasing order. The loss of symmetry stems from our asymmetric choice of nodal points in the nodal formulation of the ADER scheme and is not governed by the order property.

## IX) MHD Tests with ADER Time Stepping

In this Section, we extend the tests presented above to include a number of MHD tests in two and three dimensions. These tests will evaluate the accuracy and stability of the ADER-CG scheme for MHD flows. They will demonstrate especially the robustness of the magnetic field reconstruction provided in Section III.

### IX.1) The decay of Alfven waves in two dimensions

This problem studies the numerical dissipation of the scheme by examining the long term amplitude decay of torsional Alfven waves, and was first presented in Balsara [8]. The problem consists of the propagation of Alfven waves at a small angle to the mesh. The length of the box is $\ell = 6$, and the computational domain extends from $[-\ell/2, \ell/2] \times [-\ell/2, \ell/2]$. The Alfven waves propagate along an angle of $\tan^{-1}(1/\ell) \approx 9.462°$ to the y-axis. The unit normal vector along the axis of propagation can thus be written as

$$\hat{n} = n_x \hat{i} + n_y \hat{j} = \frac{1}{\sqrt{\ell^2 + 1}} \hat{i} + \frac{\ell}{\sqrt{\ell^2 + 1}} \hat{j}.$$

The phase of the wave is given by

$$\phi = \frac{2\pi}{n_y}(n_x x + n_y y - V_A t), \text{ where } V_A = \frac{B_0}{\sqrt{4\pi\rho_0}}.$$

The velocity is given by

$$\mathbf{v} = (v_0 n_x - \varepsilon n_y \cos\phi)\hat{i} + (v_0 n_y - \varepsilon n_x \cos\phi)\hat{j} + \varepsilon \sin\phi \hat{k}.$$

The magnetic field is

$$\mathbf{B} = (B_0 n_x + \varepsilon n_y \sqrt{4\pi\rho_0} \cos\phi)\hat{i} + (B_0 n_y - \varepsilon n_x \sqrt{4\pi\rho_0} \cos\phi)\hat{j} - \varepsilon\sqrt{4\pi\rho_0} \sin\phi \hat{k},$$

which corresponds to a vector potential

$$\mathbf{A} = -\frac{\varepsilon\sqrt{4\pi\rho_0}}{2\pi} \cos\phi \hat{i} + (-B_0 n_y x + B_0 n_x y + \frac{\varepsilon n_y \sqrt{4\pi\rho_0}}{2\pi} \sin\phi)\hat{k}$$

which is used to initialize the magnetic field in a divergence-free manner. In the test problem presented here, we set the mean velocity $v_0 = 0$, the mean magnetic field $B_0 = 1$ and the parameter $\varepsilon = 0.2$. The pressure and density are set to unity everywhere. The simulation was run to a stopping time of 129, by which time the waves had crossed the computational grid a number of times.

Fig. 6 compares the time history of the maximal values of the velocities and the magnetic fields in the *z* direction for the ADER-CG scheme at different orders. The decay in the velocity and field variables for the second, third and fourth ADER-WENO schemes with an HLLE Riemann solver are shown in the upper two panels of Fig. 6. The lower two panels show the velocity and magnetic field decay as a function of time for the same ADER-WENO schemes, but this time using a linearized Riemann solver. Notice that the upper and lower panels don't have the same vertical scales. We see that the HLL Riemann solver causes a considerable decay in the amplitude of the Alfven wave for the second and third order schemes. The difference between the two schemes would be even greater but for the fact that our second order WENO reconstruction actually uses the slopes from the r=3 WENO interpolation from Jiang & Shu [44]. The fourth order ADER-WENO scheme, even with an HLL Riemann solver, shows an almost minimal decay in the Alfven wave amplitude. When the linearized Riemann solver is used with the ADER-WENO schemes, the wave dissipation is much improved at second and third orders. This is our motivation for claiming that the third order ADER-WENO scheme with a linearized Riemann solver as a very worthy replacement for a second order RK scheme. By comparing the upper and lower panels in Fig. 6 we also see that the reconstruction scheme along with the ADER time update strategy at fourth order are so good that the quality of the Riemann solver makes no significant difference in the overall dissipation characteristics of the numerical scheme.

**IX.2) The two dimensional rotor problem**

This problem was presented in Balsara and Spicer [4], and we follow the version of the problem as described in Balsara *et al.* [11]. A high density, rapidly rotating cylinder of fluid is placed in a low-density fluid initially at rest, with a moagnetic field threading the entire domain. The initial density of the cylinder was 10 and of the ambient fluid was 1. The magnetic field was set to 2.5, and the initial pressure was set to unity in both fluids. The computational domain was [-0.5, 0.5]×[-0.5, 0.5], solved on a mesh of 1000×1000 zones. The angular velocity of the cylinder was constant, and the magnitude of the velocity at its edge was 1. The radius of the cylinder was 0.1 and the density and velocity was linearly tapered to match the ambient fluid over 6 zones (that is, out to a radius of 0.106). The ratio of specific heats is given by $\gamma$=5/3. The simulation is run to a time of 0.29 with a CFL number of 0.45.

Fig. 7 shows the density, pressure, Mach number and magnitude of the magnetic field for the rotor's simulation when the fourth order nodal ADER-WENO scheme was used with a linearized Riemann solver. The reason for carrying out this simulation at such a high resolution stems from the claim in Fuchs, Mishra & Risebro [36] that this problem produces negative pressures at high resolutions. 20 contours are shown for each figure along with the minimum and maximum values of the variable in Fig. 7. We find that the pressure remains positive all through the simulation, showing that even at high orders the problem remains well-behaved if everything is done right. The density variable does not show any of the degradation that was reported in Londrillo and Del Zanna [46], neither was any degradation observed in the density variable in the original paper that first presented this problem.

**IX.3) Field loop advection in two dimensions**

This problem is set up on a 128x64 zone domain that spans [-1,1]×[-0.5,0.5]. The problem consists of a two-dimensional loop of magnetic field with a very low magnetic pressure compared to the gas pressure. The magnetic pressure is constant inside the loop and falls abruptly to zero at the loop's boundary which is initially set up at a radius of 0.3 units. The details of the set-up are described in Gardiner & Stone [37] and are not repeated here. The problem was run to a stopping time of unity with a CFL number of 0.45 using second, third and fourth order ADER-WENO schemes and an HLL Riemann solver. The second order scheme in just this simulation utilized an MC limiter.

Fig. 8 shows the magnitude of the magnetic field for the field loop advection problem. The panels show the ADER-WENO schemes with HLL Riemann solvers at second, third and fourth orders. The plot shows the field loop after it has executed one complete orbit around the computational domain. We see that there is no numerical diffusion of the loop's boundaries and there are no oscillations in the magnetic pressure within the loop for all of the simulations shown. We see, however, that the second order scheme shows a slight directional bias while the third and fourth order schemes are free of such a bias. We also see that the fourth order scheme captures the loop's profile much more crisply than the third order scheme. We see, therefore, that the higher order schemes have adequate multidimensional dissipation for the advection of the magnetic field while simultaneously yielding a superior solution with increasing order.

**IX.4) MHD blast wave in three dimensions**

This problem is an MHD extension of the problem proposed in Sub-section VIII.3, and parallels similar problems from Balsara & Spicer [4] and Balsara, *et al.* [11]. The computational domain remains the unit cube, and the MHD blast problem is solved on a computational mesh of $97^3$ zones. The gas has unit density and $\gamma =1.4$. The interior and exterior gas pressures remain 1000 and 0.1, respectively, but in this case, we include a magnetic field threading the entire domain which points along the diagonal of the box, given by $(B_x, B_y, B_z) = (20/\sqrt{3},\ 20/\sqrt{3},\ 20/\sqrt{3})$. Note that this magnetic field corresponds to a plasma-$\beta$ of 0.00628 in the ambient medium. The problem was run to a time of 0.014 with a CFL number of 0.3 using all the ADER schemes catalogued here.

Fig. 9 shows the density, pressure, magnitude of the velocity and magnitude of the magnetic field in the midplane of the simulation when the fourth order ADER-WENO scheme was used with a linearized Riemann solver. We show the results from the modal formulation. We see that the blast wave is symmetrical about the magnetic field's orientation. The oppositely oriented velocities were found to be equal in magnitude up to numerical round-off at all orders when the modal ADER formulation was used. The nodal ADER formulation produced an asymmetry in the final velocities of 0.0021% at second order, 0.14% at third order and 0.0071% at fourth order. As in the hydrodynamic blast problem, we see that the asymmetries are very small and they do not decrease with increasing order. As before, the loss of symmetry stems from our asymmetric choice of nodal points and is not governed by the order property.

## X) Conclusions

In this paper we have presented a very efficient formulation of the ADER-CG time stepping algorithim and shown that it performs very well. All of the schemes catalogued here have been implemented in the first author's RIEMANN code. We catalogue the advanced reported in this paper in pointwise form below:

1) We hope to have made the ADER schemes much more accessible by providing two very useful formulations in modal as well as nodal space. Sufficient amount of implementation-related detail has been provided to make it easy for readers to use these schemes in their application codes.

2) We have catalogued an efficient strategy for transcribing from modal space to nodal space and vice versa. The essential idea consists of using each node or mode that has been evaluated to help us in the evaluation of the subsequent one. When this transcription is used, the modal and nodal ADER-CG schemes perform with speeds that are within 10% of each other.

3) We have also provided a fast and efficient strategy for the evaluation of the numerical flux for conservation laws in general and electric fields for the particular case of divergence-free MHD.

4) An efficient WENO-based strategy for obtaining zone-averaged magnetic fields from face-centered magnetic fields has also been presented. We also show that our strategy helps in simplifying the problem of reconstructing the magnetic fields at higher orders.

5) The ADER schemes are shown to be almost twice as fast as Runge-Kutta schemes at the same order of accuracy. The third order ADER-WENO scheme, either with an HLL or linearized Riemann solver, represents an excellent upgrade path for scientists and engineers who are working with a second order Runge-Kutta based total variation diminishing (TVD) scheme.

6) We have also presented several demonstrations of the ADER scheme's order of accuracy for hydrodynamical and MHD problems. Several stringent test problems have also been catalogued.

## Acknowledgements

DSB and CM acknowledge support via NSF grant NSF-AST-0947765. DSB also acknowledges NASA grants NASA-NNX07AG93G and NASA-NNX08AG69G. The majority of simulations were performed on a cluster at UND that is run by the Center for Research Computing.

## Appendix A: ADER-CG Scheme in Nodal Space at Second Order

The reconstruction problem at second order yields eqn. (10). Starting from this equation, we wish to obtain eqn. (14) using an ADER-CG scheme that is formulated in nodal space. Because eqn. (14) has five modes, we define a set of five nodes in the reference space-time element. They are given by

$$\{(0,0,0,0);\ (1/2,0,0,0);\ (0,1/2,0,0);\ (0,0,1/2,0);\ (0,0,0,1)\} \tag{A.1}$$

A set of five polynomial basis functions can then be defined with the special property that each of them is unity at just one of the nodes and zero at the rest of the nodes. We call these basis functions the nodal basis set. The ADER-CG scheme is then formulated in that basis using Section 3 of Balsara *et al.* (2009). The nodal basis functions are not very important for making an implementation of ADER-CG. However, just for the sake of showing the reader how such a scheme is formulated, we catalogue them below. They are given by

$$P_1[\xi,\eta,\zeta,\tau]=1-2\xi-2\eta-2\zeta-\tau \;;\; P_2[\xi,\eta,\zeta,\tau]=2\xi \;;$$
$$P_3[\xi,\eta,\zeta,\tau]=2\eta \;;\; P_4[\xi,\eta,\zeta,\tau]=2\zeta \;;\; P_5[\xi,\eta,\zeta,\tau]=\tau \tag{A.2}$$

The above nodal basis functions then permit us to write the space-time variation of eqn. (14) in an equivalent way as follows

$$u(\xi,\eta,\zeta,\tau) = \bar{u}_1\, P_1[\xi,\eta,\zeta,\tau] + \bar{u}_2\, P_2[\xi,\eta,\zeta,\tau] + \bar{u}_3\, P_3[\xi,\eta,\zeta,\tau]$$
$$+ \bar{u}_4\, P_4[\xi,\eta,\zeta,\tau] + \bar{u}_5\, P_5[\xi,\eta,\zeta,\tau] \tag{A.3}$$

Notice that eqn. (A.3) retrieves the nodal values because the bases in eqn. (A.2) are indeed nodal bases. Consequently, $\{\bar{u}_1, \bar{u}_2, \bar{u}_3, \bar{u}_4, \bar{u}_5\}$ are indeed the values of the conserved variables at the nodes in eqn. (A.1). We will detail an efficient transcription strategy that enables us to obtain the coefficients of eqn. (A.3) from the coefficients of eqn. (14) and vice versa in the next paragraph. The basis functions in eqn. (A.2) can now be used to formulate an ADER-CG scheme in nodal space. The second order ADER-CG formulation in nodal space requires one iteration of the following evolutionary equation

$$\bar{u}_5 = \bar{u}_1 - 2\left(\bar{f}_2 - \bar{f}_1\right) - 2\left(\bar{g}_3 - \bar{g}_1\right) - 2\left(\bar{h}_3 - \bar{h}_1\right) + \bar{s}_1/3 + 2\,\bar{s}_5/3 \tag{A.4}$$

We use the overbars in eqn. (A.4) denote nodal values, just as we did in Section IV. By looking at the location of the nodes we easily recognize that eqn. (A.4) is a finite difference approximation for the conservation law within the reference element. At higher orders we cannot make such an obvious identification, yet it is comforting to see that the second order ADER-CG yields such a familiar simplification.

To complete our description we need two further things. First, we need a computationally efficient strategy for obtaining the first four nodal values, i.e. the first four coefficients of eqn. (A.3), from the first four modal values in eqn. (10). This is done as follows

$$\bar{u}_1 = \hat{w}_1 \ ; \ \bar{u}_2 = \hat{w}_1 + 0.5 \, \hat{w}_2 \ ; \ \bar{u}_3 = \hat{w}_1 + 0.5 \, \hat{w}_3 \ ; \ \bar{u}_4 = \hat{w}_1 + 0.5 \, \hat{w}_4 \tag{A.5}$$

Notice that eqn. (A.4) provides the fifth nodal value. Second, we need a computationally efficient strategy for obtaining the space-time representation in eqn. (14) from the nodal values that we have built. The reconstruction has already yielded the first four modes in eqn. (14). The fifth mode in eqn. (14) is given by

$$\hat{u}_5 = \bar{u}_5 - \bar{u}_1 \tag{A.6}$$

The space-time integral of the source term, which is needed in eqn. (6) for each zone "$i,j,k$", is then given by

$$\bar{S}_{i,j,k} = 0.5 \, (\bar{s}_1 + \bar{s}_5)/\Delta t \tag{A.7}$$

This completes our description of the ADER-CG scheme in nodal space at second order.

## Appendix B: ADER-CG Scheme in Nodal Space at Third Order

The reconstruction problem at third order yields eqn. (11). Starting from this equation, we wish to obtain eqn. (15) using an ADER-CG scheme that is formulated in nodal space. Because eqn. (15) has fifteen modes, we define a set of fifteen nodes in the reference space-time element. They are given by

$$\{(0,0,0,0);\ (1/2,0,0,0);\ (-1/2,0,0,0);\ (0,1/2,0,0);\ (0,-1/2,0,0);$$
$$(0,0,1/2,0);\ (0,0,-1/2,0);\ (0,1/2,1/2,0);\ (1/2,0,1/2,0);\ (1/2,1/2,0,0);\quad \text{(B.1)}$$
$$(0,0,0,1/2);\ (1/2,0,0,1/2);\ (0,1/2,0,1/2);\ (0,0,1/2,1/2);\ (0,0,0,1)\}$$

It is most efficient to build the following temporary variables once at $\tau = 0$

$$\begin{aligned}
q_{11} &= \bar{u}_1 + 0.5\left(\bar{f}_3 - \bar{f}_1\right) + 0.5\left(\bar{g}_5 - \bar{g}_1\right) + 0.5\left(\bar{h}_7 - \bar{h}_1\right) \\
&\quad + \left(-9\,\bar{s}_1 + 2\,\bar{s}_2 + 2\,\bar{s}_3 + 2\,\bar{s}_4 + 2\,\bar{s}_5 + 2\,\bar{s}_6 + 2\,\bar{s}_7\right)/24 \\
q_{12} &= \bar{u}_2 + (\,36\,\bar{f}_1 - 24\,\bar{f}_2 - 12\,\bar{f}_3 - 36\,\bar{g}_1 + 24\,\bar{g}_2 + 24\,\bar{g}_4 + 12\,\bar{g}_5 - 24\,\bar{g}_{10} \\
&\quad - 36\,\bar{h}_1 + 24\,\bar{h}_2 + 24\,\bar{h}_6 + 12\,\bar{h}_7 - 24\,\bar{h}_9 + 7\,\bar{s}_1 - 8\,\bar{s}_2 - 4\,\bar{s}_3 \\
&\quad + 2\,\bar{s}_4 + 2\,\bar{s}_5 + 2\,\bar{s}_6 + 2\,\bar{s}_7\,)/24 \\
q_{13} &= \bar{u}_4 + (\,-36\,\bar{f}_1 + 24\,\bar{f}_2 + 12\,\bar{f}_3 + 24\,\bar{f}_4 - 24\,\bar{f}_{10} + 36\,\bar{g}_1 - 24\,\bar{g}_4 - 12\,\bar{g}_5 \\
&\quad - 36\,\bar{h}_1 + 24\,\bar{h}_4 + 24\,\bar{h}_6 + 12\,\bar{h}_7 - 24\,\bar{h}_8 + 7\,\bar{s}_1 + 2\,\bar{s}_2 + 2\,\bar{s}_3 - 8\,\bar{s}_4 \qquad \text{(B.2)} \\
&\quad - 4\,\bar{s}_5 + 2\,\bar{s}_6 + 2\,\bar{s}_7\,)/24 \\
q_{14} &= \bar{u}_6 + (\,-36\,\bar{f}_1 + 24\,\bar{f}_2 + 12\,\bar{f}_3 + 24\,\bar{f}_6 - 24\,\bar{f}_9 - 36\,\bar{g}_1 + 24\,\bar{g}_4 + 12\,\bar{g}_5 \\
&\quad + 24\,\bar{g}_6 - 24\,\bar{g}_8 + 36\,\bar{h}_1 - 24\,\bar{h}_6 - 12\,\bar{h}_7 + 7\,\bar{s}_1 + 2\,\bar{s}_2 + 2\,\bar{s}_3 + 2\,\bar{s}_4 \\
&\quad + 2\,\bar{s}_5 - 8\,\bar{s}_6 - 4\,\bar{s}_7\,)/24 \\
q_{15} &= \bar{u}_1 + (\,-60\,\bar{f}_1 + 30\,\bar{f}_2 + 30\,\bar{f}_3 - 60\,\bar{g}_1 + 30\,\bar{g}_4 + 30\,\bar{g}_5 - 60\,\bar{h}_1 + 30\,\bar{h}_6 \\
&\quad + 30\,\bar{h}_7 - 27\,\bar{s}_1 + 5\,\bar{s}_2 + 5\,\bar{s}_3 + 5\,\bar{s}_4 + 5\,\bar{s}_5 + 5\,\bar{s}_6 + 5\,\bar{s}_7\,)/30
\end{aligned}$$

The third order ADER-CG formulation in nodal space requires two iterations of the following evolutionary equations

$$q_1 = 0.5\left(\bar{f}_{11} - \bar{f}_{12}\right) + 0.5\left(\bar{g}_{11} - \bar{g}_{13}\right) + 0.5\left(\bar{h}_{11} - \bar{h}_{14}\right)\ ;\ q_2 = q_1 + \left(-4\,\bar{s}_{11} - 3\,\bar{s}_{15}\right)/24\ ;$$
$$\bar{u}_{11} = q_{11} + q_1 + 0.125\left(4\,\bar{s}_{11} - \bar{s}_{15}\right)\ ;\ \bar{u}_{12} = q_{12} + q_2 + 2\,\bar{s}_{12}/3\ ;\ \bar{u}_{13} = q_{13} + q_2 + 2\,\bar{s}_{13}/3\ ;$$
$$\bar{u}_{14} = q_{14} + q_2 + 2\,\bar{s}_{14}/3\ ;\ \bar{u}_{15} = q_{15} + 4\,q_1 + 0.1\left(8\,\bar{s}_{11} + \bar{s}_{15}\right)$$
$$\text{(B.3)}$$

To complete our description we need two further things. First, we need a computationally efficient strategy for obtaining the first ten nodal values from the first ten modal values in eqn. (11). This is done as follows

$$\bar{u}_1 = \hat{w}_1 - (\hat{w}_5 + \hat{w}_6 + \hat{w}_7)/12 \; ; \; \bar{u}_2 = \bar{u}_1 + 0.5\,\hat{w}_2 + 0.25\,\hat{w}_5 \; ; \; \bar{u}_3 = 2\,\bar{u}_1 - \bar{u}_2 + 0.5\,\hat{w}_5 \; ;$$
$$\bar{u}_4 = \bar{u}_1 + 0.5\,\hat{w}_3 + 0.25\,\hat{w}_6 \; ; \; \bar{u}_5 = 2\,\bar{u}_1 - \bar{u}_4 + 0.5\,\hat{w}_6 \; ; \; \bar{u}_6 = \bar{u}_1 + 0.5\,\hat{w}_4 + 0.25\,\hat{w}_7 \; ;$$
$$\bar{u}_7 = 2\,\bar{u}_1 - \bar{u}_6 + 0.5\,\hat{w}_7 \; ; \; \bar{u}_8 = \bar{u}_4 + \bar{u}_6 - \bar{u}_1 + 0.25\,\hat{w}_9 \; ; \; \bar{u}_9 = \bar{u}_2 + \bar{u}_6 - \bar{u}_1 + 0.25\,\hat{w}_{10} \; ;$$
$$\bar{u}_{10} = \bar{u}_2 + \bar{u}_4 - \bar{u}_1 + 0.25\,\hat{w}_8$$
(B.4)

Notice that eqn. (B.3) provides the last five nodal values. Second, we need a computationally efficient strategy for obtaining the space-time representation in eqn. (15) from the nodal values that we have built. The reconstruction has already yielded the first ten modes in eqn. (15). The eleventh through fifteenth modes in eqn. (15) are given by

$$\hat{u}_{12} = 2\,(\bar{u}_{15} - 2\,\bar{u}_{11} + \bar{u}_1) \; ; \; \hat{u}_{11} = 2\,(\bar{u}_{11} - \bar{u}_1 - 0.25\,\hat{u}_{12}) \; ; \; \hat{u}_{13} = 4\,(\bar{u}_{12} - \bar{u}_{11} - \bar{u}_2 + \bar{u}_1) \; ;$$
$$\hat{u}_{14} = 4\,(\bar{u}_{13} - \bar{u}_{11} - \bar{u}_4 + \bar{u}_1) \; ; \; \hat{u}_{15} = 4\,(\bar{u}_{14} - \bar{u}_{11} - \bar{u}_6 + \bar{u}_1)$$
(B.5)

The space-time integral of the source term, which is needed in eqn. (6) for each zone "i,j,k", is then given by

$$\bar{S}_{i,j,k} = (-5\,\bar{s}_1 + \bar{s}_2 + \bar{s}_3 + \bar{s}_4 + \bar{s}_5 + \bar{s}_6 + \bar{s}_7 + 4\,\bar{s}_{11} + \bar{s}_{15})/(6\,\Delta t) \tag{B.6}$$

This completes our description of the ADER-CG scheme in nodal space at third order.

## Appendix C: ADER-CG Scheme in Nodal Space at Fourth Order

The reconstruction problem at fourth order yields eqn. (12). Starting from this equation, we wish to obtain eqn. (16) using an ADER-CG scheme that is formulated in nodal space. Because eqn. (16) has thirty-five modes, we define a set of thirty-five nodes in the reference space-time element. They are given by

$$\{(0,0,0,0); (1/2,1/2,1/2,0); (1/2,0,0,0); (-1/2,0,0,0); (0,1/2,0,0); (0,-1/2,0,0);$$
$$(0,0,1/2,0); (0,0,-1/2,0); (0,1/2,1/2,0); (-1/2,0,1/2,0); (1/2,0,1/2,0); (1/2,-1/2,0,0);$$
$$(1/2,1/2,0,0); (0,1/2,-1/2,0); (-1/4,-1/2,-1/2,0); (1/4,-1/2,-1/2,0);$$
$$(-1/2,-1/4,-1/2,0); (-1/2,1/4,-1/2,0); (-1/2,-1/2,-1/4,0); (-1/2,-1/2,1/4,0);$$
$$(0,0,0,1/3); (1/2,0,0,1/3); (-1/2,0,0,1/3); (0,1/2,0,1/3); (0,-1/2,0,1/3);$$
$$(0,0,1/2,1/3); (0,0,-1/2,1/3); (0,1/2,1/2,1/3); (1/2,0,1/2,1/3); (1/2,1/2,0,1/3);$$
$$(0,0,0,2/3); (1/2,0,0,2/3); (0,1/2,0,2/3); (0,0,1/2,2/3); (0,0,0,1)\}$$

(C.1)

It is most efficient to build fifteen temporary variables once at $\tau = 0$ for each of the nodal update terms. The terms correspond to the contributions to the nodal update equations from fluxes that are evaluated at $\tau = 0$. It is, however, cumbersome to list all the terms. Consequently, we list only the first two terms in the journal version of the paper, cataloguing the rest in the web version of the paper (arXiv:1006.2146).

$$q_{21} = \bar{u}_1 +$$
$$(-650370\bar{f}_1 + 191520\bar{f}_2 + 305340\bar{f}_3 + 295750\bar{f}_4 + 238560\bar{f}_5 + 28000\bar{f}_6 + 389760\bar{f}_7$$
$$+ 159040\bar{f}_8 - 203840\bar{f}_9 - 80640\bar{f}_{10} - 284480\bar{f}_{11} + 70560\bar{f}_{12} - 213920\bar{f}_{13}$$
$$+ 10080\bar{f}_{14} + 202720\bar{f}_{15} - 180320\bar{f}_{16} - 89600\bar{f}_{17} - 89600\bar{f}_{18} - 49280\bar{f}_{19}$$
$$- 49280\bar{f}_{20} - 650370\bar{g}_1 + 191520\bar{g}_2 + 389760\bar{g}_3 + 159040\bar{g}_4 + 305340\bar{g}_5$$
$$+ 295750\bar{g}_6 + 238560\bar{g}_7 + 28000\bar{g}_8 - 213920\bar{g}_9 + 10080\bar{g}_{10} - 203840\bar{g}_{11}$$
$$- 80640\bar{g}_{12} - 284480\bar{g}_{13} + 70560\bar{g}_{14} - 49280\bar{g}_{15} - 49280\bar{g}_{16} + 202720\bar{g}_{17}$$
$$- 180320\bar{g}_{18} - 89600\bar{g}_{19} - 89600\bar{g}_{20} - 650370\bar{h}_1 + 191520\bar{h}_2 + 238560\bar{h}_3$$
$$+ 28000\bar{h}_4 + 389760\bar{h}_5 + 159040\bar{h}_6 + 305340\bar{h}_7 + 295750\bar{h}_8 - 284480\bar{h}_9 + 70560\bar{h}_{10}$$
$$- 213920\bar{h}_{11} + 10080\bar{h}_{12} - 203840\bar{h}_{13} - 80640\bar{h}_{14} - 89600\bar{h}_{15} - 89600\bar{h}_{16}$$
$$- 49280\bar{h}_{17} - 49280\bar{h}_{18} + 202720\bar{h}_{19} - 180320\bar{h}_{20} - 89091\bar{s}_1 + 17955\bar{s}_3 + 17955\bar{s}_4$$
$$+ 17955\bar{s}_5 + 17955\bar{s}_6 + 17955\bar{s}_7 + 17955\bar{s}_8 \,)/215460$$

$q_{22} = \bar{u}_3 +$

$(4225410\bar{f}_1 - 1149120\bar{f}_2 - 2262960\bar{f}_3 - 1666770\bar{f}_4 - 1431360\bar{f}_5 - 168000\bar{f}_6$
$- 2338560\bar{f}_7 - 954240\bar{f}_8 + 1223040\bar{f}_9 + 483840\bar{f}_{10} + 1706880\bar{f}_{11} - 423360\bar{f}_{12}$
$+ 1283520\bar{f}_{13} - 60480\bar{f}_{14} - 1216320\bar{f}_{15} + 1081920\bar{f}_{16} + 537600\bar{f}_{17} + 537600\bar{f}_{18}$
$+ 295680\bar{f}_{19} + 295680\bar{f}_{20} - 1735650\bar{g}_1 + 574560\bar{g}_2 + 953820\bar{g}_3 + 477120\bar{g}_4$
$+ 916020\bar{g}_5 + 671790\bar{g}_6 + 715680\bar{g}_7 + 84000\bar{g}_8 - 641760\bar{g}_9 + 30240\bar{g}_{10}$
$- 611520\bar{g}_{11} - 26460\bar{g}_{12} - 853440\bar{g}_{13} + 211680\bar{g}_{14} - 147840\bar{g}_{15} - 147840\bar{g}_{16}$
$+ 608160\bar{g}_{17} - 540960\bar{g}_{18} - 268800\bar{g}_{19} - 268800\bar{g}_{20} - 2645370\bar{h}_1 + 574560\bar{h}_2$
$+ 981540\bar{h}_3 + 510720\bar{h}_4 + 1197000\bar{h}_5 + 510720\bar{h}_6 + 1340640\bar{h}_7 + 1049370\bar{h}_8$
$- 853860\bar{h}_9 - 853860\bar{h}_{11} - 638400\bar{h}_{13} - 215460\bar{h}_{14} - 255360\bar{h}_{15} - 255360\bar{h}_{16}$
$- 255360\bar{h}_{17} - 255360\bar{h}_{18} + 606480\bar{h}_{19} - 542640\bar{h}_{20} + 474069\bar{s}_1 - 114912\bar{s}_2$
$- 411810\bar{s}_3 - 192514\bar{s}_4 - 120561\bar{s}_5 + 6755\bar{s}_6 - 145131\bar{s}_7 - 14539\bar{s}_8 + 122234\bar{s}_9$
$+ 13104\bar{s}_{10} + 171248\bar{s}_{11} - 11466\bar{s}_{12} + 159782\bar{s}_{13} - 1638\bar{s}_{14} - 119392\bar{s}_{15} + 110432\bar{s}_{16}$
$+ 35840\bar{s}_{17} + 35840\bar{s}_{18} + 29288\bar{s}_{19} + 29288\bar{s}_{20}) / 646380$

$q_{23} = \bar{u}_4 +$

$(3363570\bar{f}_1 - 1149120\bar{f}_2 - 1832040\bar{f}_3 - 1235850\bar{f}_4 - 1431360\bar{f}_5 - 168000\bar{f}_6$
$- 2338560\bar{f}_7 - 954240\bar{f}_8 + 1223040\bar{f}_9 + 483840\bar{f}_{10} + 1706880\bar{f}_{11} - 423360\bar{f}_{12}$
$+ 1283520\bar{f}_{13} - 60480\bar{f}_{14} - 1216320\bar{f}_{15} + 1081920\bar{f}_{16} + 537600\bar{f}_{17} + 537600\bar{f}_{18}$
$+ 295680\bar{f}_{19} + 295680\bar{f}_{20} - 1208970\bar{g}_1 + 574560\bar{g}_2 + 966420\bar{g}_3 + 152880\bar{g}_4$
$+ 815220\bar{g}_5 + 249270\bar{g}_6 + 660240\bar{g}_7 + 16800\bar{g}_8 - 588000\bar{g}_9 - 22680\bar{g}_{10} - 610680\bar{g}_{11}$
$- 34020\bar{g}_{12} - 860160\bar{g}_{13} + 272160\bar{g}_{14} - 144480\bar{g}_{15} - 144480\bar{g}_{16} + 581280\bar{g}_{17}$
$- 567840\bar{g}_{18} - 53760\bar{g}_{19} - 53760\bar{g}_{20} - 2118690\bar{h}_1 + 574560\bar{h}_2 + 880740\bar{h}_3$
$+ 88200\bar{h}_4 + 1141560\bar{h}_5 + 443520\bar{h}_6 + 1353240\bar{h}_7 + 725130\bar{h}_8 - 853020\bar{h}_9 - 7560\bar{h}_{10}$
$- 860580\bar{h}_{11} + 60480\bar{h}_{12} - 584640\bar{h}_{13} - 268380\bar{h}_{14} - 282240\bar{h}_{15} - 282240\bar{h}_{16}$
$- 40320\bar{h}_{17} - 40320\bar{h}_{18} + 609840\bar{h}_{19} - 539280\bar{h}_{20} - 577695\bar{s}_1 + 114912\bar{s}_2 + 304080\bar{s}_3$
$+ 84784\bar{s}_4 + 228291\bar{s}_5 + 100975\bar{s}_6 + 252861\bar{s}_7 + 122269\bar{s}_8 - 122234\bar{s}_9 - 13104\bar{s}_{10}$
$- 171248\bar{s}_{11} + 11466\bar{s}_{12} - 159782\bar{s}_{13} + 1638\bar{s}_{14} + 119392\bar{s}_{15} - 110432\bar{s}_{16} - 35840\bar{s}_{17}$
$- 35840\bar{s}_{18} - 29288\bar{s}_{19} - 29288\bar{s}_{20}) / 646380$

$q_{24} = \bar{u}_5 +$

$(-2645370\bar{f}_1 + 574560\bar{f}_2 + 1340640\bar{f}_3 + 1049370\bar{f}_4 + 981540\bar{f}_5 + 510720\bar{f}_6$
$+ 1197000\bar{f}_7 + 510720\bar{f}_8 - 638400\bar{f}_9 - 215460\bar{f}_{10} - 853860\bar{f}_{11} - 853860\bar{f}_{13}$
$+ 606480\bar{f}_{15} - 542640\bar{f}_{16} - 255360\bar{f}_{17} - 255360\bar{f}_{18} - 255360\bar{f}_{19} - 255360\bar{f}_{20}$
$+ 4225410\bar{g}_1 - 1149120\bar{g}_2 - 2338560\bar{g}_3 - 954240\bar{g}_4 - 2262960\bar{g}_5 - 1666770\bar{g}_6$
$- 1431360\bar{g}_7 - 168000\bar{g}_8 + 1283520\bar{g}_9 - 60480\bar{g}_{10} + 1223040\bar{g}_{11} + 483840\bar{g}_{12}$
$+ 1706880\bar{g}_{13} - 423360\bar{g}_{14} + 295680\bar{g}_{15} + 295680\bar{g}_{16} - 1216320\bar{g}_{17} + 1081920\bar{g}_{18}$
$+ 537600\bar{g}_{19} + 537600\bar{g}_{20} - 1735650\bar{h}_1 + 574560\bar{h}_2 + 715680\bar{h}_3 + 84000\bar{h}_4 + 953820\bar{h}_5$
$+ 477120\bar{h}_6 + 916020\bar{h}_7 + 671790\bar{h}_8 - 853440\bar{h}_9 + 211680\bar{h}_{10} - 641760\bar{h}_{11} + 30240\bar{h}_{12}$
$- 611520\bar{h}_{13} - 26460\bar{h}_{14} - 268800\bar{h}_{15} - 268800\bar{h}_{16} - 147840\bar{h}_{17} - 147840\bar{h}_{18}$
$+ 608160\bar{h}_{19} - 540960\bar{h}_{20} + 474069\bar{s}_1 - 114912\bar{s}_2 - 145131\bar{s}_3 - 14539\bar{s}_4 - 411810\bar{s}_5$
$- 192514\bar{s}_6 - 120561\bar{s}_7 + 6755\bar{s}_8 + 159782\bar{s}_9 - 1638\bar{s}_{10} + 122234\bar{s}_{11} + 13104\bar{s}_{12}$
$+ 171248\bar{s}_{13} - 11466\bar{s}_{14} + 29288\bar{s}_{15} + 29288\bar{s}_{16} - 119392\bar{s}_{17} + 110432\bar{s}_{18} + 35840\bar{s}_{19}$
$+ 35840\bar{s}_{20}) / 646380$

$q_{25} = \bar{u}_6 +$

$(-2118690\bar{f}_1 + 574560\bar{f}_2 + 1353240\bar{f}_3 + 725130\bar{f}_4 + 880740\bar{f}_5 + 88200\bar{f}_6$
$+ 1141560\bar{f}_7 + 443520\bar{f}_8 - 584640\bar{f}_9 - 268380\bar{f}_{10} - 853020\bar{f}_{11} - 7560\bar{f}_{12}$
$- 860580\bar{f}_{13} + 60480\bar{f}_{14} + 609840\bar{f}_{15} - 539280\bar{f}_{16} - 282240\bar{f}_{17} - 282240\bar{f}_{18}$
$- 40320\bar{f}_{19} - 40320\bar{f}_{20} + 3363570\bar{g}_1 - 1149120\bar{g}_2 - 2338560\bar{g}_3 - 954240\bar{g}_4$
$- 1832040\bar{g}_5 - 1235850\bar{g}_6 - 1431360\bar{g}_7 - 168000\bar{g}_8 + 1283520\bar{g}_9 - 60480\bar{g}_{10}$
$+ 1223040\bar{g}_{11} + 483840\bar{g}_{12} + 1706880\bar{g}_{13} - 423360\bar{g}_{14} + 295680\bar{g}_{15} + 295680\bar{g}_{16}$
$- 1216320\bar{g}_{17} + 1081920\bar{g}_{18} + 537600\bar{g}_{19} + 537600\bar{g}_{20} - 1208970\bar{h}_1 + 574560\bar{h}_2$
$+ 660240\bar{h}_3 + 16800\bar{h}_4 + 966420\bar{h}_5 + 152880\bar{h}_6 + 815220\bar{h}_7 + 249270\bar{h}_8 - 860160\bar{h}_9$
$+ 272160\bar{h}_{10} - 588000\bar{h}_{11} - 22680\bar{h}_{12} - 610680\bar{h}_{13} - 34020\bar{h}_{14} - 53760\bar{h}_{15}$
$- 53760\bar{h}_{16} - 144480\bar{h}_{17} - 144480\bar{h}_{18} + 581280\bar{h}_{19} - 567840\bar{h}_{20} - 577695\bar{s}_1$
$+ 114912\bar{s}_2 + 252861\bar{s}_3 + 122269\bar{s}_4 + 304080\bar{s}_5 + 84784\bar{s}_6 + 228291\bar{s}_7 + 100975\bar{s}_8$
$- 159782\bar{s}_9 + 1638\bar{s}_{10} - 122234\bar{s}_{11} - 13104\bar{s}_{12} - 171248\bar{s}_{13} + 11466\bar{s}_{14} - 29288\bar{s}_{15}$
$- 29288\bar{s}_{16} + 119392\bar{s}_{17} - 110432\bar{s}_{18} - 35840\bar{s}_{19} - 35840\bar{s}_{20}) / 646380$

$q_{26} = \bar{u}_7 +$

$(-1735650\bar{f}_1 + 574560\bar{f}_2 + 916020\bar{f}_3 + 671790\bar{f}_4 + 715680\bar{f}_5 + 84000\bar{f}_6$
$+ 953820\bar{f}_7 + 477120\bar{f}_8 - 611520\bar{f}_9 - 26460\bar{f}_{10} - 853440\bar{f}_{11} + 211680\bar{f}_{12}$
$- 641760\bar{f}_{13} + 30240\bar{f}_{14} + 608160\bar{f}_{15} - 540960\bar{f}_{16} - 268800\bar{f}_{17} - 268800\bar{f}_{18}$
$- 147840\bar{f}_{19} - 147840\bar{f}_{20} - 2645370\bar{g}_1 + 574560\bar{g}_2 + 1197000\bar{g}_3 + 510720\bar{g}_4$
$+ 1340640\bar{g}_5 + 1049370\bar{g}_6 + 981540\bar{g}_7 + 510720\bar{g}_8 - 853860\bar{g}_9 - 638400\bar{g}_{11}$
$- 215460\bar{g}_{12} - 853860\bar{g}_{13} - 255360\bar{g}_{15} - 255360\bar{g}_{16} + 606480\bar{g}_{17} - 542640\bar{g}_{18}$
$- 255360\bar{g}_{19} - 255360\bar{g}_{20} + 4225410\bar{h}_1 - 1149120\bar{h}_2 - 1431360\bar{h}_3 - 168000\bar{h}_4$
$- 2338560\bar{h}_5 - 954240\bar{h}_6 - 2262960\bar{h}_7 - 1666770\bar{h}_8 + 1706880\bar{h}_9 - 423360\bar{h}_{10}$
$+ 1283520\bar{h}_{11} - 60480\bar{h}_{12} + 1223040\bar{h}_{13} + 483840\bar{h}_{14} + 537600\bar{h}_{15} + 537600\bar{h}_{16}$
$+ 295680\bar{h}_{17} + 295680\bar{h}_{18} - 1216320\bar{h}_{19} + 1081920\bar{h}_{20} + 474069\bar{s}_1 - 114912\bar{s}_2$
$- 120561\bar{s}_3 + 6755\bar{s}_4 - 145131\bar{s}_5 - 14539\bar{s}_6 - 411810\bar{s}_7 - 192514\bar{s}_8 + 171248\bar{s}_9$
$- 11466\bar{s}_{10} + 159782\bar{s}_{11} - 1638\bar{s}_{12} + 122234\bar{s}_{13} + 13104\bar{s}_{14} + 35840\bar{s}_{15} + 35840\bar{s}_{16}$
$+ 29288\bar{s}_{17} + 29288\bar{s}_{18} - 119392\bar{s}_{19} + 110432\bar{s}_{20}) / 646380$

$q_{27} = \bar{u}_8 +$

$(-1208970\bar{f}_1 + 574560\bar{f}_2 + 815220\bar{f}_3 + 249270\bar{f}_4 + 660240\bar{f}_5 + 16800\bar{f}_6$
$+ 966420\bar{f}_7 + 152880\bar{f}_8 - 610680\bar{f}_9 - 34020\bar{f}_{10} - 860160\bar{f}_{11} + 272160\bar{f}_{12}$
$- 588000\bar{f}_{13} - 22680\bar{f}_{14} + 581280\bar{f}_{15} - 567840\bar{f}_{16} - 53760\bar{f}_{17} - 53760\bar{f}_{18}$
$- 144480\bar{f}_{19} - 144480\bar{f}_{20} - 2118690\bar{g}_1 + 574560\bar{g}_2 + 1141560\bar{g}_3 + 443520\bar{g}_4$
$+ 1353240\bar{g}_5 + 725130\bar{g}_6 + 880740\bar{g}_7 + 88200\bar{g}_8 - 860580\bar{g}_9 + 60480\bar{g}_{10}$
$- 584640\bar{g}_{11} - 268380\bar{g}_{12} - 853020\bar{g}_{13} - 7560\bar{g}_{14} - 40320\bar{g}_{15} - 40320\bar{g}_{16}$
$+ 609840\bar{g}_{17} - 539280\bar{g}_{18} - 282240\bar{g}_{19} - 282240\bar{g}_{20} + 3363570\bar{h}_1 - 1149120\bar{h}_2$
$- 1431360\bar{h}_3 - 168000\bar{h}_4 - 2338560\bar{h}_5 - 954240\bar{h}_6 - 1832040\bar{h}_7 - 1235850\bar{h}_8$
$+ 1706880\bar{h}_9 - 423360\bar{h}_{10} + 1283520\bar{h}_{11} - 60480\bar{h}_{12} + 1223040\bar{h}_{13} + 483840\bar{h}_{14}$
$+ 537600\bar{h}_{15} + 537600\bar{h}_{16} + 295680\bar{h}_{17} + 295680\bar{h}_{18} - 1216320\bar{h}_{19} + 1081920\bar{h}_{20}$
$- 577695\bar{s}_1 + 114912\bar{s}_2 + 228291\bar{s}_3 + 100975\bar{s}_4 + 252861\bar{s}_5 + 122269\bar{s}_6 + 304080\bar{s}_7$
$+ 84784\bar{s}_8 - 171248\bar{s}_9 + 11466\bar{s}_{10} - 159782\bar{s}_{11} + 1638\bar{s}_{12} - 122234\bar{s}_{13} - 13104\bar{s}_{14}$
$- 35840\bar{s}_{15} - 35840\bar{s}_{16} - 29288\bar{s}_{17} - 29288\bar{s}_{18} + 119392\bar{s}_{19} - 110432\bar{s}_{20}) / 646380$

$q_{28} = \bar{u}_9 +$

$(-1998990\bar{f}_1 + 143640\bar{f}_2 + 909720\bar{f}_3 + 833910\bar{f}_4 + 550620\bar{f}_5 + 510720\bar{f}_6$
$+ 550620\bar{f}_7 + 510720\bar{f}_8 - 207480\bar{f}_9 - 422940\bar{f}_{11} - 422940\bar{f}_{13} + 606480\bar{f}_{15}$
$- 542640\bar{f}_{16} - 255360\bar{f}_{17} - 255360\bar{f}_{18} - 255360\bar{f}_{19} - 255360\bar{f}_{20} + 4488750\bar{g}_1$
$- 1149120\bar{g}_2 - 2366280\bar{g}_3 - 987840\bar{g}_4 - 2256660\bar{g}_5 - 1828890\bar{g}_6 - 1266300\bar{g}_7$
$- 594720\bar{g}_8 + 1064700\bar{g}_9 - 30240\bar{g}_{10} + 1249920\bar{g}_{11} + 457380\bar{g}_{12} + 1707300\bar{g}_{13}$
$- 211680\bar{g}_{14} + 403200\bar{g}_{15} + 403200\bar{g}_{16} - 1214640\bar{g}_{17} + 1083600\bar{g}_{18} + 524160\bar{g}_{19}$
$+ 524160\bar{g}_{20} + 3579030\bar{h}_1 - 1149120\bar{h}_2 - 1431360\bar{h}_3 - 168000\bar{h}_4 - 1692180\bar{h}_5$
$- 954240\bar{h}_6 - 1832040\bar{h}_7 - 1451310\bar{h}_8 + 1275960\bar{h}_9 - 423360\bar{h}_{10} + 1283520\bar{h}_{11}$
$- 60480\bar{h}_{12} + 1223040\bar{h}_{13} + 268380\bar{h}_{14} + 537600\bar{h}_{15} + 537600\bar{h}_{16} + 295680\bar{h}_{17}$
$+ 295680\bar{h}_{18} - 1216320\bar{h}_{19} + 1081920\bar{h}_{20} + 1023891\bar{s}_1 - 229824\bar{s}_2 - 333417\bar{s}_3$
$- 78449\bar{s}_4 - 284466\bar{s}_5 - 341978\bar{s}_6 - 288246\bar{s}_7 - 345254\bar{s}_8 + 6160\bar{s}_9 + 2016\bar{s}_{10}$
$+ 295456\bar{s}_{11} - 1764\bar{s}_{12} + 293692\bar{s}_{13} - 252\bar{s}_{14} + 118888\bar{s}_{15} + 118888\bar{s}_{16} - 89264\bar{s}_{17}$
$+ 140560\bar{s}_{18} - 90272\bar{s}_{19} + 139552\bar{s}_{20}) / 646380$

$q_{29} = \bar{u}_{11} +$

$(3579030\bar{f}_1 - 1149120\bar{f}_2 - 1832040\bar{f}_3 - 1451310\bar{f}_4 - 1431360\bar{f}_5 - 168000\bar{f}_6$
$- 1692180\bar{f}_7 - 954240\bar{f}_8 + 1223040\bar{f}_9 + 268380\bar{f}_{10} + 1275960\bar{f}_{11} - 423360\bar{f}_{12}$
$+ 1283520\bar{f}_{13} - 60480\bar{f}_{14} - 1216320\bar{f}_{15} + 1081920\bar{f}_{16} + 537600\bar{f}_{17} + 537600\bar{f}_{18}$
$+ 295680\bar{f}_{19} + 295680\bar{f}_{20} - 1998990\bar{g}_1 + 143640\bar{g}_2 + 550620\bar{g}_3 + 510720\bar{g}_4$
$+ 909720\bar{g}_5 + 833910\bar{g}_6 + 550620\bar{g}_7 + 510720\bar{g}_8 - 422940\bar{g}_9 - 207480\bar{g}_{11}$
$- 422940\bar{g}_{13} - 255360\bar{g}_{15} - 255360\bar{g}_{16} + 606480\bar{g}_{17} - 542640\bar{g}_{18} - 255360\bar{g}_{19}$
$- 255360\bar{g}_{20} + 4488750\bar{h}_1 - 1149120\bar{h}_2 - 1266300\bar{h}_3 - 594720\bar{h}_4 - 2366280\bar{h}_5$
$- 987840\bar{h}_6 - 2256660\bar{h}_7 - 1828890\bar{h}_8 + 1707300\bar{h}_9 - 211680\bar{h}_{10} + 1064700\bar{h}_{11}$
$- 30240\bar{h}_{12} + 1249920\bar{h}_{13} + 457380\bar{h}_{14} + 524160\bar{h}_{15} + 524160\bar{h}_{16} + 403200\bar{h}_{17}$
$+ 403200\bar{h}_{18} - 1214640\bar{h}_{19} + 1083600\bar{h}_{20} + 1023891\bar{s}_1 - 229824\bar{s}_2 - 288246\bar{s}_3$
$- 345254\bar{s}_4 - 333417\bar{s}_5 - 78449\bar{s}_6 - 284466\bar{s}_7 - 341978\bar{s}_8 + 293692\bar{s}_9 - 252\bar{s}_{10}$
$+ 6160\bar{s}_{11} + 2016\bar{s}_{12} + 295456\bar{s}_{13} - 1764\bar{s}_{14} - 90272\bar{s}_{15} + 139552\bar{s}_{16} + 118888\bar{s}_{17}$
$+ 118888\bar{s}_{18} - 89264\bar{s}_{19} + 140560\bar{s}_{20}) / 646380$

$q_{30} = \bar{u}_{13} +$

$( 4488750\,\bar{f}_1 - 1149120\,\bar{f}_2 - 2256660\,\bar{f}_3 - 1828890\,\bar{f}_4 - 1266300\,\bar{f}_5 - 594720\,\bar{f}_6$
$- 2366280\,\bar{f}_7 - 987840\,\bar{f}_8 + 1249920\,\bar{f}_9 + 457380\,\bar{f}_{10} + 1707300\,\bar{f}_{11} - 211680\,\bar{f}_{12}$
$+ 1064700\,\bar{f}_{13} - 30240\,\bar{f}_{14} - 1214640\,\bar{f}_{15} + 1083600\,\bar{f}_{16} + 524160\,\bar{f}_{17} + 524160\,\bar{f}_{18}$
$+ 403200\,\bar{f}_{19} + 403200\,\bar{f}_{20} + 3579030\,\bar{g}_1 - 1149120\,\bar{g}_2 - 1692180\,\bar{g}_3 - 954240\,\bar{g}_4$
$- 1832040\,\bar{g}_5 - 1451310\,\bar{g}_6 - 1431360\,\bar{g}_7 - 168000\,\bar{g}_8 + 1283520\,\bar{g}_9 - 60480\,\bar{g}_{10}$
$+ 1223040\,\bar{g}_{11} + 268380\,\bar{g}_{12} + 1275960\,\bar{g}_{13} - 423360\,\bar{g}_{14} + 295680\,\bar{g}_{15} + 295680\,\bar{g}_{16}$
$- 1216320\,\bar{g}_{17} + 1081920\,\bar{g}_{18} + 537600\,\bar{g}_{19} + 537600\,\bar{g}_{20} - 1998990\,\bar{h}_1 + 143640\,\bar{h}_2$
$+ 550620\,\bar{h}_3 + 510720\,\bar{h}_4 + 550620\,\bar{h}_5 + 510720\,\bar{h}_6 + 909720\,\bar{h}_7 + 833910\,\bar{h}_8$
$- 422940\,\bar{h}_9 - 422940\,\bar{h}_{11} - 207480\,\bar{h}_{13} - 255360\,\bar{h}_{15} - 255360\,\bar{h}_{16} - 255360\,\bar{h}_{17}$
$- 255360\,\bar{h}_{18} + 606480\,\bar{h}_{19} - 542640\,\bar{h}_{20} + 1023891\,\bar{s}_1 - 229824\,\bar{s}_2 - 284466\,\bar{s}_3$
$- 341978\,\bar{s}_4 - 288246\,\bar{s}_5 - 345254\,\bar{s}_6 - 333417\,\bar{s}_7 - 78449\,\bar{s}_8 + 295456\,\bar{s}_9$
$- 1764\,\bar{s}_{10} + 293692\,\bar{s}_{11} - 252\,\bar{s}_{12} + 6160\,\bar{s}_{13} + 2016\,\bar{s}_{14} - 89264\,\bar{s}_{15} + 140560\,\bar{s}_{16}$
$- 90272\,\bar{s}_{17} + 139552\,\bar{s}_{18} + 118888\,\bar{s}_{19} + 118888\,\bar{s}_{20} ) / 646380$

$q_{31} = \bar{u}_1 +$

$( - 307230\,\bar{f}_1 + 95760\,\bar{f}_2 + 152670\,\bar{f}_3 + 129920\,\bar{f}_4 + 119280\,\bar{f}_5 + 14000\,\bar{f}_6$
$+ 194880\,\bar{f}_7 + 79520\,\bar{f}_8 - 101920\,\bar{f}_9 - 40320\,\bar{f}_{10} - 142240\,\bar{f}_{11} + 35280\,\bar{f}_{12}$
$- 106960\,\bar{f}_{13} + 5040\,\bar{f}_{14} + 101360\,\bar{f}_{15} - 90160\,\bar{f}_{16} - 44800\,\bar{f}_{17} - 44800\,\bar{f}_{18}$
$- 24640\,\bar{f}_{19} - 24640\,\bar{f}_{20} - 307230\,\bar{g}_1 + 95760\,\bar{g}_2 + 194880\,\bar{g}_3 + 79520\,\bar{g}_4 + 152670\,\bar{g}_5$
$+ 129920\,\bar{g}_6 + 119280\,\bar{g}_7 + 14000\,\bar{g}_8 - 106960\,\bar{g}_9 + 5040\,\bar{g}_{10} - 101920\,\bar{g}_{11} - 40320\,\bar{g}_{12}$
$- 142240\,\bar{g}_{13} + 35280\,\bar{g}_{14} - 24640\,\bar{g}_{15} - 24640\,\bar{g}_{16} + 101360\,\bar{g}_{17} - 90160\,\bar{g}_{18}$
$- 44800\,\bar{g}_{19} - 44800\,\bar{g}_{20} - 307230\,\bar{h}_1 + 95760\,\bar{h}_2 + 119280\,\bar{h}_3 + 14000\,\bar{h}_4 + 194880\,\bar{h}_5$
$+ 79520\,\bar{h}_6 + 152670\,\bar{h}_7 + 129920\,\bar{h}_8 - 142240\,\bar{h}_9 + 35280\,\bar{h}_{10} - 106960\,\bar{h}_{11} + 5040\,\bar{h}_{12}$
$- 101920\,\bar{h}_{13} - 40320\,\bar{h}_{14} - 44800\,\bar{h}_{15} - 44800\,\bar{h}_{16} - 24640\,\bar{h}_{17} - 24640\,\bar{h}_{18}$
$+ 101360\,\bar{h}_{19} - 90160\,\bar{h}_{20} - 31293\,\bar{s}_1 + 5985\,\bar{s}_3 + 5985\,\bar{s}_4 + 5985\,\bar{s}_5 + 5985\,\bar{s}_6$
$+ 5985\,\bar{s}_7 + 5985\,\bar{s}_8 ) / 53865$

$q_{32} = \bar{u}_3 +$

$( 3902220 \bar{f}_1 - 1149120 \bar{f}_2 - 2047500 \bar{f}_3 - 1559040 \bar{f}_4 - 1431360 \bar{f}_5 - 168000 \bar{f}_6$
$- 2338560 \bar{f}_7 - 954240 \bar{f}_8 + 1223040 \bar{f}_9 + 483840 \bar{f}_{10} + 1706880 \bar{f}_{11} - 423360 \bar{f}_{12}$
$+ 1283520 \bar{f}_{13} - 60480 \bar{f}_{14} - 1216320 \bar{f}_{15} + 1081920 \bar{f}_{16} + 537600 \bar{f}_{17} + 537600 \bar{f}_{18}$
$+ 295680 \bar{f}_{19} + 295680 \bar{f}_{20} - 1412460 \bar{g}_1 + 574560 \bar{g}_2 + 738360 \bar{g}_3 + 477120 \bar{g}_4$
$+ 700560 \bar{g}_5 + 564060 \bar{g}_6 + 715680 \bar{g}_7 + 84000 \bar{g}_8 - 641760 \bar{g}_9 + 30240 \bar{g}_{10}$
$- 611520 \bar{g}_{11} - 26460 \bar{g}_{12} - 637980 \bar{g}_{13} + 211680 \bar{g}_{14} - 147840 \bar{g}_{15} - 147840 \bar{g}_{16}$
$+ 608160 \bar{g}_{17} - 540960 \bar{g}_{18} - 268800 \bar{g}_{19} - 268800 \bar{g}_{20} - 2322180 \bar{h}_1 + 574560 \bar{h}_2$
$+ 766080 \bar{h}_3 + 510720 \bar{h}_4 + 1197000 \bar{h}_5 + 510720 \bar{h}_6 + 1125180 \bar{h}_7 + 941640 \bar{h}_8$
$- 853860 \bar{h}_9 - 638400 \bar{h}_{11} - 638400 \bar{h}_{13} - 215460 \bar{h}_{14} - 255360 \bar{h}_{15} - 255360 \bar{h}_{16}$
$- 255360 \bar{h}_{17} - 255360 \bar{h}_{18} + 606480 \bar{h}_{19} - 542640 \bar{h}_{20} + 413535 \bar{s}_1 - 114912 \bar{s}_2$
$- 343581 \bar{s}_3 - 156604 \bar{s}_4 - 138516 \bar{s}_5 - 11200 \bar{s}_6 - 163086 \bar{s}_7 - 32494 \bar{s}_8 + 122234 \bar{s}_9$
$+ 13104 \bar{s}_{10} + 171248 \bar{s}_{11} - 11466 \bar{s}_{12} + 159782 \bar{s}_{13} - 1638 \bar{s}_{14} - 119392 \bar{s}_{15} + 110432 \bar{s}_{16}$
$+ 35840 \bar{s}_{17} + 35840 \bar{s}_{18} + 29288 \bar{s}_{19} + 29288 \bar{s}_{20} ) / 323190$

$q_{33} = \bar{u}_5 +$

$( - 2322180 \bar{f}_1 + 574560 \bar{f}_2 + 1125180 \bar{f}_3 + 941640 \bar{f}_4 + 766080 \bar{f}_5 + 510720 \bar{f}_6$
$+ 1197000 \bar{f}_7 + 510720 \bar{f}_8 - 638400 \bar{f}_9 - 215460 \bar{f}_{10} - 853860 \bar{f}_{11} - 638400 \bar{f}_{13}$
$+ 606480 \bar{f}_{15} - 542640 \bar{f}_{16} - 255360 \bar{f}_{17} - 255360 \bar{f}_{18} - 255360 \bar{f}_{19} - 255360 \bar{f}_{20}$
$+ 3902220 \bar{g}_1 - 1149120 \bar{g}_2 - 2338560 \bar{g}_3 - 954240 \bar{g}_4 - 2047500 \bar{g}_5 - 1559040 \bar{g}_6$
$- 1431360 \bar{g}_7 - 168000 \bar{g}_8 + 1283520 \bar{g}_9 - 60480 \bar{g}_{10} + 1223040 \bar{g}_{11}\text{g} + 483840 \bar{g}_{12}$
$+ 1706880 \bar{g}_{13} - 423360 \bar{g}_{14} + 295680 \bar{g}_{15} + 295680 \bar{g}_{16} - 1216320 \bar{g}_{17} + 1081920 \bar{g}_{18}$
$+ 537600 \bar{g}_{19} + 537600 \bar{g}_{20} - 1412460 \bar{h}_1 + 574560 \bar{h}_2 + 715680 \bar{h}_3 + 84000 \bar{h}_4$
$+ 738360 \bar{h}_5 + 477120 \bar{h}_6 + 700560 \bar{h}_7 + 564060 \bar{h}_8 - 637980 \bar{h}_9 + 211680 \bar{h}_{10}$
$- 641760 \bar{h}_{11} + 30240 \bar{h}_{12} - 611520 \bar{h}_{13} - 26460 \bar{h}_{14} - 268800 \bar{h}_{15} - 268800 \bar{h}_{16}$
$- 147840 \bar{h}_{17} - 147840 \bar{h}_{18} + 608160 \bar{h}_{19} - 540960 \bar{h}_{20} + 413535 \bar{s}_1 - 114912 \bar{s}_2$
$- 163086 \bar{s}_3 - 32494 \bar{s}_4 - 343581 \bar{s}_5 - 156604 \bar{s}_6 - 138516 \bar{s}_7 - 11200 \bar{s}_8$
$+ 159782 \bar{s}_9 - 1638 \bar{s}_{10} + 122234 \bar{s}_{11} + 13104 \bar{s}_{12} + 171248 \bar{s}_{13} - 11466 \bar{s}_{14} + 29288 \bar{s}_{15}$
$+ 29288 \bar{s}_{16} - 119392 \bar{s}_{17} + 110432 \bar{s}_{18} + 35840 \bar{s}_{19} + 35840 \bar{s}_{20} ) / 323190$

$$q_{34} = \bar{u}_7 +$$
$$(-1412460\bar{f}_1 + 574560\bar{f}_2 + 700560\bar{f}_3 + 564060\bar{f}_4 + 715680\bar{f}_5 + 84000\bar{f}_6 + 738360\bar{f}_7$$
$$+ 477120\bar{f}_8 - 611520\bar{f}_9 - 26460\bar{f}_{10} - 637980\bar{f}_{11} + 211680\bar{f}_{12} - 641760\bar{f}_{13}$$
$$+ 30240\bar{f}_{14} + 608160\bar{f}_{15} - 540960\bar{f}_{16} - 268800\bar{f}_{17} - 268800\bar{f}_{18} - 147840\bar{f}_{19}$$
$$- 147840\bar{f}_{20} - 2322180\bar{g}_1 + 574560\bar{g}_2 + 1197000\bar{g}_3 + 510720\bar{g}_4 + 1125180\bar{g}_5$$
$$+ 941640\bar{g}_6 + 766080\bar{g}_7 + 510720\bar{g}_8 - 638400\bar{g}_9 - 638400\bar{g}_{11} - 215460\bar{g}_{12}$$
$$- 853860\bar{g}_{13} - 255360\bar{g}_{15} - 255360\bar{g}_{16} + 606480\bar{g}_{17} - 542640\bar{g}_{18} - 255360\bar{g}_{19}$$
$$- 255360\bar{g}_{20} + 3902220\bar{h}_1 - 1149120\bar{h}_2 - 1431360\bar{h}_3 - 168000\bar{h}_4 - 2338560\bar{h}_5$$
$$- 954240\bar{h}_6 - 2047500\bar{h}_7 - 1559040\bar{h}_8 + 1706880\bar{h}_9 - 423360\bar{h}_{10} + 1283520\bar{h}_{11}$$
$$- 60480\bar{h}_{12} + 1223040\bar{h}_{13} + 483840\bar{h}_{14} + 537600\bar{h}_{15} + 537600\bar{h}_{16} + 295680\bar{h}_{17}$$
$$+ 295680\bar{h}_{18} - 1216320\bar{h}_{19} + 1081920\bar{h}_{20} + 413535\bar{s}_1 - 114912\bar{s}_2 - 138516\bar{s}_3$$
$$- 11200\bar{s}_4 - 163086\bar{s}_5 - 32494\bar{s}_6 - 343581\bar{s}_7 - 156604\bar{s}_8 + 171248\bar{s}_9$$
$$- 11466\bar{s}_{10} + 159782\bar{s}_{11} - 1638\bar{s}_{12} + 122234\bar{s}_{13} + 13104\bar{s}_{14} + 35840\bar{s}_{15} + 35840\bar{s}_{16}$$
$$+ 29288\bar{s}_{17} + 29288\bar{s}_{18} - 119392\bar{s}_{19} + 110432\bar{s}_{20})/323190$$

$$q_{35} = \bar{u}_1 +$$
$$(-530670\bar{f}_1 + 191520\bar{f}_2 + 257460\bar{f}_3 + 223930\bar{f}_4 + 238560\bar{f}_5 + 28000\bar{f}_6 + 389760\bar{f}_7$$
$$+ 159040\bar{f}_8 - 203840\bar{f}_9 - 80640\bar{f}_{10} - 284480\bar{f}_{11} + 70560\bar{f}_{12} - 213920\bar{f}_{13}$$
$$+ 10080\bar{f}_{14} + 202720\bar{f}_{15} - 180320\bar{f}_{16} - 89600\bar{f}_{17} - 89600\bar{f}_{18} - 49280\bar{f}_{19}$$
$$- 49280\bar{f}_{20} - 530670\bar{g}_1 + 191520\bar{g}_2 + 389760\bar{g}_3 + 159040\bar{g}_4 + 257460\bar{g}_5 + 223930\bar{g}_6$$
$$+ 238560\bar{g}_7 + 28000\bar{g}_8 - 213920\bar{g}_9 + 10080\bar{g}_{10} - 203840\bar{g}_{11} - 80640\bar{g}_{12}$$
$$- 284480\bar{g}_{13} + 70560\bar{g}_{14} - 49280\bar{g}_{15} - 49280\bar{g}_{16} + 202720\bar{g}_{17} - 180320\bar{g}_{18}$$
$$- 89600\bar{g}_{19} - 89600\bar{g}_{20} - 530670\bar{h}_1 + 191520\bar{h}_2 + 238560\bar{h}_3 + 28000\bar{h}_4 + 389760\bar{h}_5$$
$$+ 159040\bar{h}_6 + 257460\bar{h}_7 + 223930\bar{h}_8 - 284480\bar{h}_9 + 70560\bar{h}_{10} - 213920\bar{h}_{11} + 10080\bar{h}_{12}$$
$$- 203840\bar{h}_{13} - 80640\bar{h}_{14} - 89600\bar{h}_{15} - 89600\bar{h}_{16} - 49280\bar{h}_{17} - 49280\bar{h}_{18}$$
$$+ 202720\bar{h}_{19} - 180320\bar{h}_{20} - 29241\bar{s}_1 + 5985\bar{s}_3 + 5985\bar{s}_4 + 5985\bar{s}_5 + 5985\bar{s}_6 + 5985\bar{s}_7$$
$$+ 5985\bar{s}_8)/71820$$

(C.2)

The fourth order ADER-CG formulation in nodal space requires three iterations of the fifteen evolutionary equations at each of the fifteen nodes. As before, we catalogue only the first two evolutionary equations in the jornal version of this paper, relegating the rest to the web version.

$$\bar{u}_{21} = q_{21} +$$
$$(23940\bar{f}_{21} - 59850\bar{f}_{22} + 35910\bar{f}_{23} - 11970\bar{f}_{31} + 11970\bar{f}_{32} + 23940\bar{g}_{21} - 59850\bar{g}_{24}$$
$$+ 35910\bar{g}_{25} - 11970\bar{g}_{31} + 11970\bar{g}_{33} + 23940\bar{h}_{21} - 59850\bar{h}_{26} + 35910\bar{h}_{27}$$
$$- 11970\bar{h}_{31} + 11970\bar{h}_{34} + 189468\bar{s}_{21} - 17955\bar{s}_{22} - 17955\bar{s}_{23} - 17955\bar{s}_{24}$$
$$- 17955\bar{s}_{25} - 17955\bar{s}_{26} - 17955\bar{s}_{27} - 39843\bar{s}_{31} + 11286\bar{s}_{35}) / 215460$$

$$\bar{u}_{22} = q_{22} +$$
$$(502740\bar{f}_{21} - 395010\bar{f}_{22} - 107730\bar{f}_{23} - 35910\bar{f}_{31} + 35910\bar{f}_{32} - 143640\bar{g}_{21}$$
$$+ 215460\bar{g}_{22} + 35910\bar{g}_{24} + 107730\bar{g}_{25} - 215460\bar{g}_{30} - 35910\bar{g}_{31} + 35910\bar{g}_{33}$$
$$- 143640\bar{h}_{21} + 215460\bar{h}_{22} + 35910\bar{h}_{26} + 107730\bar{h}_{27} - 215460\bar{h}_{29} - 35910\bar{h}_{31}$$
$$+ 35910\bar{h}_{34} - 164160\bar{s}_{21} + 517104\bar{s}_{22} + 107730\bar{s}_{23} - 53865\bar{s}_{24} - 53865\bar{s}_{25}$$
$$- 53865\bar{s}_{26} - 53865\bar{s}_{27} + 31293\bar{s}_{31} - 150822\bar{s}_{32} + 33858\bar{s}_{35}) / 646380$$

$$\bar{u}_{23} = q_{23} +$$
$$(-359100\bar{f}_{21} + 35910\bar{f}_{22} + 323190\bar{f}_{23} - 35910\bar{f}_{31} + 35910\bar{f}_{32} + 287280\bar{g}_{21}$$
$$- 215460\bar{g}_{22} - 395010\bar{g}_{24} + 107730\bar{g}_{25} + 215460\bar{g}_{30} - 35910\bar{g}_{31} + 35910\bar{g}_{33}$$
$$+ 287280\bar{h}_{21} - 215460\bar{h}_{22} - 395010\bar{h}_{26} + 107730\bar{h}_{27} + 215460\bar{h}_{29} - 35910\bar{h}_{31}$$
$$+ 35910\bar{h}_{34} + 439128\bar{s}_{21} - 193914\bar{s}_{22} + 215460\bar{s}_{23} - 53865\bar{s}_{24} - 53865\bar{s}_{25}$$
$$- 53865\bar{s}_{26} - 53865\bar{s}_{27} - 270351\bar{s}_{31} + 150822\bar{s}_{32} + 33858\bar{s}_{35}) / 646380$$

$$\bar{u}_{24} = q_{24} +$$
$$(-143640\bar{f}_{21} + 35910\bar{f}_{22} + 107730\bar{f}_{23} + 215460\bar{f}_{24} - 215460\bar{f}_{30} - 35910\bar{f}_{31}$$
$$+ 35910\bar{f}_{32} + 502740\bar{g}_{21} - 395010\bar{g}_{24} - 107730\bar{g}_{25} - 35910\bar{g}_{31} + 35910\bar{g}_{33}$$
$$- 143640\bar{h}_{21} + 215460\bar{h}_{24} + 35910\bar{h}_{26} + 107730\bar{h}_{27} - 215460\bar{h}_{28} - 35910\bar{h}_{31}$$
$$+ 35910\bar{h}_{34} - 164160\bar{s}_{21} - 53865\bar{s}_{22} - 53865\bar{s}_{23} + 517104\bar{s}_{24} + 107730\bar{s}_{25}$$
$$- 53865\bar{s}_{26} - 53865\bar{s}_{27} + 31293\bar{s}_{31} - 150822\bar{s}_{33} + 33858\bar{s}_{35}) / 646380$$

$$\bar{u}_{25} = q_{25} +$$
$$(\ 287280\bar{f}_{21} - 395010\bar{f}_{22} + 107730\bar{f}_{23} - 215460\bar{f}_{24} + 215460\bar{f}_{30} - 35910\bar{f}_{31}$$
$$+ 35910\bar{f}_{32} - 359100\bar{g}_{21} + 35910\bar{g}_{24} + 323190\bar{g}_{25} - 35910\bar{g}_{31} + 35910\bar{g}_{33}$$
$$+ 287280\bar{h}_{21} - 215460\bar{h}_{24} - 395010\bar{h}_{26} + 107730\bar{h}_{27} + 215460\bar{h}_{28} - 35910\bar{h}_{31}$$
$$+ 35910\bar{h}_{34} + 439128\bar{s}_{21} - 53865\bar{s}_{22} - 53865\bar{s}_{23} - 193914\bar{s}_{24} + 215460\bar{s}_{25}$$
$$- 53865\bar{s}_{26} - 53865\bar{s}_{27} - 270351\bar{s}_{31} + 150822\bar{s}_{33} + 33858\bar{s}_{35}\ ) / 646380$$

$$\bar{u}_{26} = q_{26} +$$
$$(\ -143640\bar{f}_{21} + 35910\bar{f}_{22} + 107730\bar{f}_{23} + 215460\bar{f}_{26} - 215460\bar{f}_{29} - 35910\bar{f}_{31}$$
$$+ 35910\bar{f}_{32} - 143640\bar{g}_{21} + 35910\bar{g}_{24} + 107730\bar{g}_{25} + 215460\bar{g}_{26} - 215460\bar{g}_{28}$$
$$- 35910\bar{g}_{31} + 35910\bar{g}_{33} + 502740\bar{h}_{21} - 395010\bar{h}_{26} - 107730\bar{h}_{27} - 35910\bar{h}_{31}$$
$$+ 35910\bar{h}_{34} - 164160\bar{s}_{21} - 53865\bar{s}_{22} - 53865\bar{s}_{23} - 53865\bar{s}_{24} - 53865\bar{s}_{25}$$
$$+ 517104\bar{s}_{26} + 107730\bar{s}_{27} + 31293\bar{s}_{31} - 150822\bar{s}_{34} + 33858\bar{s}_{35}\ ) / 646380$$

$$\bar{u}_{27} = q_{27} +$$
$$(\ 287280\bar{f}_{21} - 395010\bar{f}_{22} + 107730\bar{f}_{23} - 215460\bar{f}_{26} + 215460\bar{f}_{29} - 35910\bar{f}_{31}$$
$$+ 35910\bar{f}_{32} + 287280\bar{g}_{21} - 395010\bar{g}_{24} + 107730\bar{g}_{25} - 215460\bar{g}_{26} + 215460\bar{g}_{28}$$
$$- 35910\bar{g}_{31} + 35910\bar{g}_{33} - 359100\bar{h}_{21} + 35910\bar{h}_{26} + 323190\bar{h}_{27} - 35910\bar{h}_{31} + 35910\bar{h}_{34}$$
$$+ 439128\bar{s}_{21} - 53865\bar{s}_{22} - 53865\bar{s}_{23} - 53865\bar{s}_{24} - 53865\bar{s}_{25} - 193914\bar{s}_{26}$$
$$+ 215460\bar{s}_{27} - 270351\bar{s}_{31} + 150822\bar{s}_{34} + 33858\bar{s}_{35}\ ) / 646380$$

$$\bar{u}_{28} = q_{28} +$$
$$(\ -359100\bar{f}_{21} + 251370\bar{f}_{22} + 107730\bar{f}_{23} + 215460\bar{f}_{24} + 215460\bar{f}_{26} - 215460\bar{f}_{29}$$
$$- 215460\bar{f}_{30} - 35910\bar{f}_{31} + 35910\bar{f}_{32} + 287280\bar{g}_{21} - 179550\bar{g}_{24} - 107730\bar{g}_{25}$$
$$+ 215460\bar{g}_{26} - 215460\bar{g}_{28} - 35910\bar{g}_{31} + 35910\bar{g}_{33} + 287280\bar{h}_{21} + 215460\bar{h}_{24}$$
$$- 179550\bar{h}_{26} - 107730\bar{h}_{27} - 215460\bar{h}_{28} - 35910\bar{h}_{31} + 35910\bar{h}_{34} - 465804\bar{s}_{21}$$
$$- 53865\bar{s}_{22} - 53865\bar{s}_{23} + 86184\bar{s}_{24} + 107730\bar{s}_{25} + 86184\bar{s}_{26} + 107730\bar{s}_{27} + 430920\bar{s}_{28}$$
$$+ 182115\bar{s}_{31} - 150822\bar{s}_{33} - 150822\bar{s}_{34} + 33858\bar{s}_{35}\ ) / 646380$$

$$\bar{u}_{29} = q_{29} +$$
$$(287280\bar{f}_{21} - 179550\bar{f}_{22} - 107730\bar{f}_{23} + 215460\bar{f}_{26} - 215460\bar{f}_{29} - 35910\bar{f}_{31}$$
$$+ 35910\bar{f}_{32} - 359100\bar{g}_{21} + 215460\bar{g}_{22} + 251370\bar{g}_{24} + 107730\bar{g}_{25} + 215460\bar{g}_{26}$$
$$- 215460\bar{g}_{28} - 215460\bar{g}_{30} - 35910\bar{g}_{31} + 35910\bar{g}_{33} + 287280\bar{h}_{21} + 215460\bar{h}_{22}$$
$$- 179550\bar{h}_{26} - 107730\bar{h}_{27} - 215460\bar{h}_{29} - 35910\bar{h}_{31} + 35910\bar{h}_{34} - 465804\bar{s}_{21}$$
$$+ 86184\bar{s}_{22} + 107730\bar{s}_{23} - 53865\bar{s}_{24} - 53865\bar{s}_{25} + 86184\bar{s}_{26} + 107730\bar{s}_{27} + 430920\bar{s}_{29}$$
$$+ 182115\bar{s}_{31} - 150822\bar{s}_{32} - 150822\bar{s}_{34} + 33858\bar{s}_{35}) / 646380$$

$$\bar{u}_{30} = q_{30} +$$
$$(287280\bar{f}_{21} - 179550\bar{f}_{22} - 107730\bar{f}_{23} + 215460\bar{f}_{24} - 215460\bar{f}_{30} - 35910\bar{f}_{31}$$
$$+ 35910\bar{f}_{32} + 287280\bar{g}_{21} + 215460\bar{g}_{22} - 179550\bar{g}_{24} - 107730\bar{g}_{25} - 215460\bar{g}_{30}$$
$$- 35910\bar{g}_{31} + 35910\bar{g}_{33} - 359100\bar{h}_{21} + 215460\bar{h}_{22} + 215460\bar{h}_{24} + 251370\bar{h}_{26}$$
$$+ 107730\bar{h}_{27} - 215460\bar{h}_{28} - 215460\bar{h}_{29} - 35910\bar{h}_{31} + 35910\bar{h}_{34} - 465804\bar{s}_{21}$$
$$+ 86184\bar{s}_{22} + 107730\bar{s}_{23} + 86184\bar{s}_{24} + 107730\bar{s}_{25} - 53865\bar{s}_{26} - 53865\bar{s}_{27} + 430920\bar{s}_{30}$$
$$+ 182115\bar{s}_{31} - 150822\bar{s}_{32} - 150822\bar{s}_{33} + 33858\bar{s}_{35}) / 646380$$

$$\bar{u}_{31} = q_{31} +$$
$$(-23940\bar{f}_{21} - 11970\bar{f}_{22} + 35910\bar{f}_{23} + 11970\bar{f}_{31} - 11970\bar{f}_{32} - 23940\bar{g}_{21} - 11970\bar{g}_{24}$$
$$+ 35910\bar{g}_{25} + 11970\bar{g}_{31} - 11970\bar{g}_{33} - 23940\bar{h}_{21} - 11970\bar{h}_{26} + 35910\bar{h}_{27} + 11970\bar{h}_{31}$$
$$- 11970\bar{h}_{34} + 63954\bar{s}_{21} - 5985\bar{s}_{22} - 5985\bar{s}_{23} - 5985\bar{s}_{24} - 5985\bar{s}_{25} - 5985\bar{s}_{26}$$
$$- 5985\bar{s}_{27} + 1881\bar{s}_{31} + 1368\bar{s}_{35}) / 53865$$

$$\bar{u}_{32} = q_{32} +$$
$$(718200\bar{f}_{21} - 502740\bar{f}_{22} - 215460\bar{f}_{23} + 71820\bar{f}_{31} - 71820\bar{f}_{32} - 574560\bar{g}_{21}$$
$$+ 430920\bar{g}_{22} + 359100\bar{g}_{24} + 215460\bar{g}_{25} - 430920\bar{g}_{30} + 71820\bar{g}_{31} - 71820\bar{g}_{33}$$
$$- 574560\bar{h}_{21} + 430920\bar{h}_{22} + 359100\bar{h}_{26} + 215460\bar{h}_{27} - 430920\bar{h}_{29} + 71820\bar{h}_{31}$$
$$- 71820\bar{h}_{34} - 68742\bar{s}_{21} + 308826\bar{s}_{22} + 71820\bar{s}_{23} - 35910\bar{s}_{24} - 35910\bar{s}_{25} - 35910\bar{s}_{26}$$
$$- 35910\bar{s}_{27} + 22059\bar{s}_{31} - 10773\bar{s}_{32} + 8208\bar{s}_{35}) / 323190$$

$\bar{u}_{33} = q_{33} +$
$( - 574560\bar{f}_{21} + 359100\bar{f}_{22} + 215460\bar{f}_{23} + 430920\bar{f}_{24} - 430920\bar{f}_{30} + 71820\bar{f}_{31}$
$- 71820\bar{f}_{32} + 718200\bar{g}_{21} - 502740\bar{g}_{24} - 215460\bar{g}_{25} + 71820\bar{g}_{31} - 71820\bar{g}_{33}$
$- 574560\bar{h}_{21} + 430920\bar{h}_{24} + 359100\bar{h}_{26} + 215460\bar{h}_{27} - 430920\bar{h}_{28} + 71820\bar{h}_{31}$
$- 71820\bar{h}_{34} - 68742\bar{s}_{21} - 35910\bar{s}_{22} - 35910\bar{s}_{23} + 308826\bar{s}_{24} + 71820\bar{s}_{25} - 35910\bar{s}_{26}$
$- 35910\bar{s}_{27} + 22059\bar{s}_{31} - 10773\bar{s}_{33} + 8208\bar{s}_{35} ) / 323190$

$\bar{u}_{34} = q_{34} +$
$( - 574560\bar{f}_{21} + 359100\bar{f}_{22} + 215460\bar{f}_{23} + 430920\bar{f}_{26} - 430920\bar{f}_{29} + 71820\bar{f}_{31}$
$- 71820\bar{f}_{32} - 574560\bar{g}_{21} + 359100\bar{g}_{24} + 215460\bar{g}_{25} + 430920\bar{g}_{26} - 430920\bar{g}_{28}$
$+ 71820\bar{g}_{31} - 71820\bar{g}_{33} + 718200\bar{h}_{21} - 502740\bar{h}_{26} - 215460\bar{h}_{27} + 71820\bar{h}_{31}$
$- 71820\bar{h}_{34} - 68742\bar{s}_{21} - 35910\bar{s}_{22} - 35910\bar{s}_{23} - 35910\bar{s}_{24} - 35910\bar{s}_{25} + 308826\bar{s}_{26}$
$+ 71820\bar{s}_{27} + 22059\bar{s}_{31} - 10773\bar{s}_{34} + 8208\bar{s}_{35} ) / 323190$

$\bar{u}_{35} = q_{35} +$
$( - 215460\bar{f}_{21} + 107730\bar{f}_{22} + 107730\bar{f}_{23} + 107730\bar{f}_{31} - 107730\bar{f}_{32} - 215460\bar{g}_{21}$
$+ 107730\bar{g}_{24} + 107730\bar{g}_{25} + 107730\bar{g}_{31} - 107730\bar{g}_{33} - 215460\bar{h}_{21} + 107730\bar{h}_{26}$
$+ 107730\bar{h}_{27} + 107730\bar{h}_{31} - 107730\bar{h}_{34} + 69768\bar{s}_{21} - 5985\bar{s}_{22} - 5985\bar{s}_{23} - 5985\bar{s}_{24}$
$- 5985\bar{s}_{25} - 5985\bar{s}_{26} - 5985\bar{s}_{27} + 20007\bar{s}_{31} + 11286\bar{s}_{35} ) / 71820$

(C.3)

To complete our description we need two further things. First, we need a computationally efficient strategy for obtaining the first twenty nodal values from the first twenty modal values in eqn. (12). This is done as follows

$$q_1 = 0.25\, \hat{w}_2 - 0.125\, (\hat{w}_8 + \hat{w}_{10}) + (-21\, \hat{w}_{11} + 40\, \hat{w}_{16} + 40\, \hat{w}_{18} + 60\, \hat{w}_{20})/960\ ;$$
$$q_2 = 0.25\, \hat{w}_3 - 0.125\, (\hat{w}_8 + \hat{w}_9) + (-21\, \hat{w}_{12} + 40\, \hat{w}_{14} + 40\, \hat{w}_{19} + 60\, \hat{w}_{20})/960\ ;$$
$$q_3 = 0.25\, \hat{w}_4 - 0.125\, (\hat{w}_9 + \hat{w}_{10}) + (-21\, \hat{w}_{13} + 40\, \hat{w}_{15} + 40\, \hat{w}_{17} + 60\, \hat{w}_{20})/960\ ;$$
$$\bar{u}_1 = \hat{w}_1 - (\hat{w}_5 + \hat{w}_6 + \hat{w}_7)/12\ ;\quad \bar{u}_3 = \bar{u}_1 + 0.5\, \hat{w}_2 + 0.25\, \hat{w}_5 + 0.05\, \hat{w}_{11} - (\hat{w}_{16} + \hat{w}_{18})/24\ ;$$
$$\bar{u}_4 = 2\, \bar{u}_1 - \bar{u}_3 + 0.5\, \hat{w}_5\ ;\quad \bar{u}_5 = \bar{u}_1 + 0.5\, \hat{w}_3 + 0.25\, \hat{w}_6 + 0.05\, \hat{w}_{12} - (\hat{w}_{14} + \hat{w}_{19})/24\ ;$$
$$\bar{u}_6 = 2\, \bar{u}_1 - \bar{u}_5 + 0.5\, \hat{w}_6\ ;\quad \bar{u}_7 = \bar{u}_1 + 0.5\, \hat{w}_4 + 0.25\, \hat{w}_7 + 0.05\, \hat{w}_{13} - (\hat{w}_{15} + \hat{w}_{17})/24\ ;$$
$$\bar{u}_8 = 2\, \bar{u}_1 - \bar{u}_7 + 0.5\, \hat{w}_7\ ;\quad \bar{u}_9 = \bar{u}_5 + \bar{u}_7 - \bar{u}_1 + 0.25\, \hat{w}_9 + 0.125\, (\hat{w}_{17} + \hat{w}_{19})\ ;$$
$$\bar{u}_{11} = \bar{u}_3 + \bar{u}_7 - \bar{u}_1 + 0.25\, \hat{w}_{10} + 0.125\, (\hat{w}_{15} + \hat{w}_{18})\ ;$$
$$\bar{u}_{13} = \bar{u}_3 + \bar{u}_5 - \bar{u}_1 + 0.25\, \hat{w}_8 + 0.125\, (\hat{w}_{14} + \hat{w}_{16})\ ;$$
$$\bar{u}_{10} = 2\, \bar{u}_7 - \bar{u}_{11} + 0.5\, \hat{w}_5 + 0.25\, \hat{w}_{15}\ ;\quad \bar{u}_{12} = 2\, \bar{u}_3 - \bar{u}_{13} + 0.5\, \hat{w}_6 + 0.25\, \hat{w}_{16}\ ;$$
$$\bar{u}_{14} = 2\, \bar{u}_5 - \bar{u}_9 + 0.5\, \hat{w}_7 + 0.25\, \hat{w}_{19}\ ;\quad \bar{u}_2 = \bar{u}_{11} + \bar{u}_9 - \bar{u}_7 + 0.25\, \hat{w}_8 + 0.125\, (\hat{w}_{14} + \hat{w}_{16} + \hat{w}_{20})\ ;$$
$$\bar{u}_{15} = 2\, \bar{u}_8 - \bar{u}_{14} - q_1 + 0.0625\, \hat{w}_5 + 0.5\, \hat{w}_6 - 0.03125\, (\hat{w}_{14} + \hat{w}_{15}) - 0.25\, \hat{w}_{17}\ ;$$
$$\bar{u}_{17} = 2\, \bar{u}_4 - \bar{u}_{10} - q_2 + 0.0625\, \hat{w}_6 + 0.5\, \hat{w}_7 - 0.03125\, (\hat{w}_{16} + \hat{w}_{17}) - 0.25\, \hat{w}_{18}\ ;$$
$$\bar{u}_{19} = 2\, \bar{u}_6 - \bar{u}_{12} - q_3 + 0.5\, \hat{w}_5 + 0.0625\, \hat{w}_7 - 0.25\, \hat{w}_{14} - 0.03125\, (\hat{w}_{18} + \hat{w}_{19})\ ;$$
$$\bar{u}_{16} = \bar{u}_{15} + 2\, q_1\ ;\quad \bar{u}_{18} = \bar{u}_{17} + 2\, q_2\ ;\quad \bar{u}_{20} = \bar{u}_{19} + 2\, q_3$$

(C.4)

Notice that eqn. (C.3) provides the last fifteen nodal values. Second, we need a computationally efficient strategy for obtaining the space-time representation in eqn. (16) from the nodal values that we have built. The reconstruction has already yielded the first twenty modes in eqn. (16). The twenty-first through thirty-fifth modes in eqn. (16) are given by

$$\hat{u}_{21} = (-5\,\bar{u}_1 - \bar{u}_3 - \bar{u}_4 - \bar{u}_5 - \bar{u}_6 - \bar{u}_7 - \bar{u}_8 + 12\,\bar{u}_{21} + \bar{u}_{22} +$$
$$\bar{u}_{23} + \bar{u}_{24} + \bar{u}_{25} + \bar{u}_{26} + \bar{u}_{27} - 9\,\bar{u}_{31} + 2\,\bar{u}_{35})/2\ ;$$
$$\hat{u}_{22} = 4.5\,(2\,\bar{u}_1 - 5\,\bar{u}_{21} + 4\,\bar{u}_{31} - \bar{u}_{35})\ ;\ \hat{u}_{23} = -4.5\,(\bar{u}_1 - 3\,\bar{u}_{21} + 3\,\bar{u}_{31} - \bar{u}_{35})\ ;$$
$$\hat{u}_{24} = 3\,(\bar{u}_1 - 2\,\bar{u}_3 + \bar{u}_4 - 2\,\bar{u}_{21} + 3\,\bar{u}_{22} - \bar{u}_{23} + \bar{u}_{31} - \bar{u}_{32})\ ;$$
$$\hat{u}_{25} = 3\,(\bar{u}_1 - 2\,\bar{u}_5 + \bar{u}_6 - 2\,\bar{u}_{21} + 3\,\bar{u}_{24} - \bar{u}_{25} + \bar{u}_{31} - \bar{u}_{33})\ ;$$
$$\hat{u}_{26} = 3\,(\bar{u}_1 - 2\,\bar{u}_7 + \bar{u}_8 - 2\,\bar{u}_{21} + 3\,\bar{u}_{26} - \bar{u}_{27} + \bar{u}_{31} - \bar{u}_{34})\ ;$$
$$\hat{u}_{27} = -9\,(\bar{u}_1 - \bar{u}_3 - 2\,\bar{u}_{21} + 2\,\bar{u}_{22} + \bar{u}_{31} - \bar{u}_{32})\ ;$$
$$\hat{u}_{28} = -9\,(\bar{u}_1 - \bar{u}_5 - 2\,\bar{u}_{21} + 2\,\bar{u}_{24} + \bar{u}_{31} - \bar{u}_{33})\ ;$$
$$\hat{u}_{29} = -9\,(\bar{u}_1 - \bar{u}_7 - 2\,\bar{u}_{21} + 2\,\bar{u}_{26} + \bar{u}_{31} - \bar{u}_{34})\ ;\ \hat{u}_{30} = 6\,(2\,\bar{u}_1 - \bar{u}_3 - \bar{u}_4 - 2\,\bar{u}_{21} + \bar{u}_{22} + \bar{u}_{23})\ ;$$
$$\hat{u}_{31} = 6\,(2\,\bar{u}_1 - \bar{u}_5 - \bar{u}_6 - 2\,\bar{u}_{21} + \bar{u}_{24} + \bar{u}_{25})\ ;\ \hat{u}_{32} = 6\,(2\,\bar{u}_1 - \bar{u}_7 - \bar{u}_8 - 2\,\bar{u}_{21} + \bar{u}_{26} + \bar{u}_{27})\ ;$$
$$\hat{u}_{33} = -12\,(\bar{u}_1 - \bar{u}_3 - \bar{u}_5 + \bar{u}_{13} - \bar{u}_{21} + \bar{u}_{22} + \bar{u}_{24} - \bar{u}_{30})\ ;$$
$$\hat{u}_{34} = -12\,(\bar{u}_1 - \bar{u}_5 - \bar{u}_7 + \bar{u}_9 - \bar{u}_{21} + \bar{u}_{24} + \bar{u}_{26} - \bar{u}_{28})\ ;$$
$$\hat{u}_{35} = -12\,(\bar{u}_1 - \bar{u}_3 - \bar{u}_7 + \bar{u}_{11} - \bar{u}_{21} + \bar{u}_{22} + \bar{u}_{24} - \bar{u}_{29})$$

(C.5)

The space-time integral of the source term, which is needed in eqn. (6) for each zone "$i,j,k$", is then given by

$$\bar{S}_{i,j,k} = (15\,\bar{s}_1 - 2\,\bar{s}_3 - 2\,\bar{s}_4 - 2\,\bar{s}_5 - 2\,\bar{s}_6 - 2\,\bar{s}_7 - 2\,\bar{s}_8 - 27\,\bar{s}_{21} + 6\,\bar{s}_{22} + 6\,\bar{s}_{23} + 6\,\bar{s}_{24}$$
$$+ 6\,\bar{s}_{25} + 6\,\bar{s}_{26} + 6\,\bar{s}_{27} + 9\,\bar{s}_{31} + 3\,\bar{s}_{35})/(24\,\Delta t)$$

(C.6)

This completes our description of the ADER-CG scheme in nodal space at fourth order.

## Appendix D

In Section V we detailed a strategy for obtaining numerical fluxes (and electric fields for MHD) when the space-time representation of the conserved variables, $\hat{u}$, and the fluxes $\hat{f}$, $\hat{g}$ and $\hat{h}$ are stored in each zone. It is more efficient to build the space-time representation of the flux within each face and use that to obtain the integrals in eqns. (55) and (57). We describe that procedure at second, third and fourth orders below.

At second order, we define two sets of one-sided nodes within each face. Consequently, for each face, such nodes are defined from the two zones on either side of it. We focus on the lower side of the top *x*-face of zone "*i,j,k*" where we define the set of facial nodes as

$$\{(1/2,1/2,0,1/2);\ (1/2,-1/2,0,1/2);\ (1/2,0,1/2,1/2);\ (1/2,0,-1/2,1/2);\ (1/2,0,0,1)\} \tag{D.1}$$

Notice that the nodes in eqn. (D.1) are defined relative to the reference element in zone "*i,j,k*". A similar set of nodes are defined at the upper side of the bottom *x*-face of the same zone. Recall the dashed lines in the lower panel of Fig. 1. The nodes in eqn. (D.1) are defined in the dashed line that yields $\langle u \rangle_{x+;i,j,k}$. Within the face under consideration, we define a space-time representation of the flux that is given by

$$\begin{aligned} f(\eta,\zeta,\tau) &= \widehat{f}_1\, P_0(\eta)P_0(\zeta)Q_0(\tau) + \widehat{f}_2\, P_1(\eta)P_0(\zeta)Q_0(\tau) \\ &+ \widehat{f}_3\, P_0(\eta)P_1(\zeta)Q_0(\tau) + \widehat{f}_4\, P_0(\eta)P_0(\zeta)Q_1(\tau) \end{aligned} \tag{D.2}$$

Compare eqn. (D.2) with eqn. (14) to realize that the *x*-variation has been eliminated in the above equation. The modes in eqn. (D.2) can be obtained by first evaluating the conserved variables at the nodes in eqn. (D.1) as

$$\begin{aligned} &q_1 = \hat{w}_1 + 0.5\,(\hat{w}_2 + \hat{u}_5)\ ;\ \tilde{u}_1 = q_1 + 0.5\,\hat{w}_3\ ;\ \tilde{u}_2 = q_1 - 0.5\,\hat{w}_3\ ;\\ &\tilde{u}_3 = q_1 + 0.5\,\hat{w}_4\ ;\ \tilde{u}_4 = q_1 - 0.5\,\hat{w}_4\ ;\ \tilde{u}_5 = q_1 + 0.5\,\hat{u}_5 \end{aligned} \tag{D.3}$$

The conserved variables at the facial nodes in eqn. (D.3) are denoted with a tilde. The terms with a caret in eqn. (D.3) pertain to the space-time modes at the zone center. Once the conserved variables are evaluated at the facial nodes, we can obtain the fluxes at those nodes, denoted here with a tilde over the flux terms. The modes in eqn. (D.2) are now obtained by

$$\widehat{f}_2 = \tilde{f}_1 - \tilde{f}_2 \; ; \; \widehat{f}_3 = \tilde{f}_3 - \tilde{f}_4 \; ; \; \widehat{f}_4 = 2\tilde{f}_5 - 0.5\left(\tilde{f}_1 + \tilde{f}_2 + \tilde{f}_3 + \tilde{f}_4\right) \; ; \; \widehat{f}_1 = \tilde{f}_5 - \widehat{f}_4 \quad \text{(D.4)}$$

The space-time integrals at the lower side of the top $x$-face of zone "$i,j,k$" in eqns. (55) and (57) are now given by

$$\langle f \rangle_{x+} = \widehat{f}_1 + 0.5\,\widehat{f}_4 \; ; \; \langle f \rangle_{x+;y+} = \langle f \rangle_{x+} + 0.5\,\widehat{f}_2 \; ; \; \langle f \rangle_{x+;y-} = \langle f \rangle_{x+} - 0.5\,\widehat{f}_2 \; ;$$
$$\langle f \rangle_{x+;z+} = \langle f \rangle_{x+} + 0.5\,\widehat{f}_3 \; ; \; \langle f \rangle_{x+;z-} = \langle f \rangle_{x+} - 0.5\,\widehat{f}_3 \quad \text{(D.5)}$$

The same approach described here can be used to obtain the space-time integrals of the $x$-flux at the lower $x$-boundary of each zone. This completes our description for obtaining the flux and electric field integrals at second order.

At third order, we again define two sets of one-sided nodes within each face. We focus on the lower side of the top $x$-face of zone "$i,j,k$" where we define the set of facial nodes as

$$\{(1/2,0,0,0);\; (1/2,1/2,1/2,0);\; (1/2,-1/2,1/2,0);\; (1/2,1/2,-1/2,0);\; (1/2,-1/2,-1/2,0);$$
$$(1/2,1/2,0,1/2);\; (1/2,-1/2,0,1/2);\; (1/2,0,1/2,1/2);\; (1/2,0,-1/2,1/2);\; (1/2,0,0,1)\}$$
$$\text{(D.6)}$$

Recall the dashed lines in the lower panel of Fig. 1. The nodes in eqn. (D.6) are defined in the dashed line that yields $\langle u \rangle_{x+;i,j,k}$. Within the face under consideration, we define a space-time representation of the flux that is given by

$$\begin{aligned} f(\eta,\zeta,\tau) =\;& \widehat{f}_1\, P_0(\eta)P_0(\zeta)Q_0(\tau) + \widehat{f}_2\, P_1(\eta)P_0(\zeta)Q_0(\tau) + \widehat{f}_3\, P_0(\eta)P_1(\zeta)Q_0(\tau) \\ &+ \widehat{f}_4\, P_2(\eta)P_0(\zeta)Q_0(\tau) + \widehat{f}_5\, P_0(\eta)P_2(\zeta)Q_0(\tau) + \widehat{f}_6\, P_1(\eta)P_1(\zeta)Q_0(\tau) \\ &+ \widehat{f}_7\, P_0(\eta)P_0(\zeta)Q_1(\tau) + \widehat{f}_8\, P_0(\eta)P_0(\zeta)Q_2(\tau) + \widehat{f}_9\, P_1(\eta)P_0(\zeta)Q_1(\tau) \\ &+ \widehat{f}_{10}\, P_0(\eta)P_1(\zeta)Q_1(\tau) \end{aligned}$$
$$\text{(D.7)}$$

Compare eqn. (D.7) with eqn. (15) to realize that the *x*-variation has been eliminated in the above equation. The modes in eqn. (D.7) can be obtained by first evaluating the conserved variables at the nodes in eqn. (D.6) as

$$q_1 = 0.5\ \hat{u}_{11} + 0.25\ (\hat{u}_{12} + \hat{u}_{13})\ ;$$
$$\tilde{u}_1 = \hat{w}_1 + 0.5\ \hat{w}_2 + (2\ \hat{w}_5 - \hat{w}_6 - \hat{w}_7)/12\ ;$$
$$\tilde{u}_2 = \tilde{u}_1 + 0.5\ (\hat{w}_3 + \hat{w}_4) + 0.25\ (\hat{w}_6 + \hat{w}_7 + \hat{w}_8 + \hat{w}_9 + \hat{w}_{10})\ ;$$
$$\tilde{u}_3 = \tilde{u}_2 - \hat{w}_3 - 0.5\ (\hat{w}_8 + \hat{w}_9)\ ;\ \tilde{u}_4 = \tilde{u}_2 - \hat{w}_4 - 0.5\ (\hat{w}_9 + \hat{w}_{10})\ ;$$
$$\tilde{u}_5 = \tilde{u}_4 - \hat{w}_3 - 0.5\ (\hat{w}_8 - \hat{w}_9)\ ;\ \tilde{u}_6 = 0.5\ (\tilde{u}_2 + \tilde{u}_4) + q_1 + 0.25\ (-\hat{w}_7 + \hat{u}_{14})\ ;$$
$$\tilde{u}_7 = 0.5\ (\tilde{u}_3 + \tilde{u}_5) + q_1 - 0.25\ (\hat{w}_7 + \hat{u}_{14})\ ;\ \tilde{u}_8 = 0.5\ (\tilde{u}_2 + \tilde{u}_3) + q_1 + 0.25\ (-\hat{w}_6 + \hat{u}_{15})\ ;$$
$$\tilde{u}_9 = 0.5\ (\tilde{u}_4 + \tilde{u}_5) + q_1 - 0.25\ (\hat{w}_6 + \hat{u}_{15})\ ;\ \tilde{u}_{10} = \tilde{u}_1 + \hat{u}_{11} + \hat{u}_{12} + 0.5\ \hat{u}_{13}$$
$$(D.8)$$

As before, the conserved variables at the facial nodes in eqn. (D.8) are denoted with a tilde. The terms with a caret in eqn. (D.8) pertain to the space-time modes at the zone center. As in Section IV, we see that the evaluation of each nodal value can be used to simplify the evaluation of the ones that follow it. Once the conserved variables are evaluated at the facial nodes, we can use them to obtain the fluxes at those nodes, denoted as before with a tilde over the flux terms. The modes in eqn. (D.7) are now obtained by

$$q_2 = 0.5\ (\tilde{f}_2 + \tilde{f}_4 - 4\tilde{f}_1 + \tilde{f}_3 + \tilde{f}_5)\ ;\ q_3 = \tilde{f}_6 + \tilde{f}_7 - \tilde{f}_8 - \tilde{f}_9\ ;$$
$$\hat{f}_1 = 2\ (\tilde{f}_1 + 0.125\ (\tilde{f}_2 + \tilde{f}_3 + \tilde{f}_4 + \tilde{f}_5))/3\ ;\ \hat{f}_2 = 0.5\ (\tilde{f}_2 + \tilde{f}_4 - \tilde{f}_3 - \tilde{f}_5)\ ;$$
$$\hat{f}_3 = 0.5\ (\tilde{f}_2 + \tilde{f}_3 - \tilde{f}_4 - \tilde{f}_5)\ ;\ \hat{f}_4 = q_2 + q_3\ ;\ \hat{f}_5 = q_2 - q_3\ ;\qquad (D.9)$$
$$\hat{f}_6 = \tilde{f}_2 - \tilde{f}_3 - \tilde{f}_4 + \tilde{f}_5\ ;\ \hat{f}_8 = 2\ (\tilde{f}_1 + \tilde{f}_{10}) - (\tilde{f}_6 + \tilde{f}_7 + \tilde{f}_8 + \tilde{f}_9) + q_2\ ;$$
$$\hat{f}_7 = \tilde{f}_{10} - \tilde{f}_1 - \hat{f}_8\ ;\ \hat{f}_9 = 2\ (\tilde{f}_6 - \tilde{f}_7 - \hat{f}_2)\ ;\ \hat{f}_{10} = 2\ (\tilde{f}_8 - \tilde{f}_9 - \hat{f}_3)$$

The space-time integrals at the lower side of the top *x*-face of zone "*i,j,k*" in eqns. (55) and (57) are now given by

$$\langle f \rangle_{x+} = \widehat{f}_1 + 0.5\, \widehat{f}_7 + \widehat{f}_8/3 \;;$$
$$q_4 = 0.5\, \widehat{f}_2 + 0.25\, \widehat{f}_9 \;;\; q_5 = 0.5\, \widehat{f}_3 + 0.25\, \widehat{f}_{10} \;;$$
$$\langle f \rangle_{x+;y+} = \langle f \rangle_{x+} + q_4 + \widehat{f}_4/6 \;;\; \langle f \rangle_{x+;y-} = \langle f \rangle_{x+;y+} - 2\, q_4 \;;$$
$$\langle f \rangle_{x+;z+} = \langle f \rangle_{x+} + q_5 + \widehat{f}_5/6 \;;\; \langle f \rangle_{x+;z-} = \langle f \rangle_{x+;z+} - 2\, q_5$$

(D.10)

The same approach described here can be used to obtain the space-time integrals of the $x$-flux at the lower $x$-boundary of each zone. This completes our description for obtaining the flux and electric field integrals at third order.

At fourth order, we again define two sets of one-sided nodes within each face. We focus on the lower side of the top $x$-face of zone "$i,j,k$" where we define the set of facial nodes as

$\{(1/2,0,0,0);\, (1/2,1/2,0,0);\, (1/2,1/4,0,0);\, (1/2,-1/4,0,0);\, (1/2,-1/2,0,0);$
$(1/2,0,1/2,0);\, (1/2,0,1/4,0);\, (1/2,0,-1/4,0);\, (1/2,0,-1/2,0);\, (1/2,1/2,1/2,0);$
$(1/2,-1/2,1/2,0);\, (1/2,1/2,-1/2,0);\, (1/2,-1/2,-1/2,0);\, (1/2,0,0,1/3);\, (1/2,1/2,1/2,1/3);$
$(1/2,-1/2,1/2,1/3);\, (1/2,1/2,-1/2,1/3);\, (1/2,-1/2,-1/2,1/3);\, (1/2,0,0,2/3);\, (1/2,1/2,0,2/3);$
$(1/2,-1/2,0,2/3);\, (1/2,0,1/2,2/3);\, (1/2,0,-1/2,2/3);\, (1/2,0,0,1)\}$

(D.11)

Recall the dashed lines in the lower panel of Fig. 1. The nodes in eqn. (D.11) are defined in the dashed line that yields $\langle u \rangle_{x+;i,j,k}$. Within the face under consideration, we define a space-time representation of the flux that is given by

$$\begin{aligned}
f(\eta,\zeta,\tau) = &\ \widehat{f}_1\, P_0(\eta)P_0(\zeta)Q_0(\tau) + \widehat{f}_2\, P_1(\eta)P_0(\zeta)Q_0(\tau) + \widehat{f}_3\, P_0(\eta)P_1(\zeta)Q_0(\tau) \\
&+ \widehat{f}_4\, P_2(\eta)P_0(\zeta)Q_0(\tau) + \widehat{f}_5\, P_0(\eta)P_2(\zeta)Q_0(\tau) + \widehat{f}_6\, P_1(\eta)P_1(\zeta)Q_0(\tau) \\
&+ \widehat{f}_7\, P_3(\eta)P_0(\zeta)Q_0(\tau) + \widehat{f}_8\, P_0(\eta)P_3(\zeta)Q_0(\tau) \\
&+ \widehat{f}_9\, P_2(\eta)P_1(\zeta)Q_0(\tau) + \widehat{f}_{10}\, P_1(\eta)P_2(\zeta)Q_0(\tau) \\
&+ \widehat{f}_{11}\, P_0(\eta)P_0(\zeta)Q_1(\tau) + \widehat{f}_{12}\, P_0(\eta)P_0(\zeta)Q_2(\tau) + \widehat{f}_{13}\, P_0(\eta)P_0(\zeta)Q_3(\tau) \\
&+ \widehat{f}_{14}\, P_1(\eta)P_0(\zeta)Q_1(\tau) + \widehat{f}_{15}\, P_0(\eta)P_1(\zeta)Q_1(\tau) \\
&+ \widehat{f}_{16}\, P_1(\eta)P_0(\zeta)Q_2(\tau) + \widehat{f}_{17}\, P_0(\eta)P_1(\zeta)Q_2(\tau) \\
&+ \widehat{f}_{18}\, P_2(\eta)P_0(\zeta)Q_1(\tau) + \widehat{f}_{19}\, P_0(\eta)P_2(\zeta)Q_1(\tau) + \widehat{f}_{20}\, P_1(\eta)P_1(\zeta)Q_1(\tau)
\end{aligned}$$

(D.12)

Compare eqn. (D.12) with eqn. (16) to realize that the *x*-variation has been eliminated in the above equation. The modes in eqn. (D.12) can be obtained by first evaluating the conserved variables at the nodes in eqn. (D.11) as

$$\tilde{u}_1 = \hat{w}_1 + 0.5\,\hat{w}_2 + 0.05\,\hat{w}_{11} + \left(4\,\hat{w}_5 - 2\,\hat{w}_6 - 2\,\hat{w}_7 - \hat{w}_{16} - \hat{w}_{18}\right)/24 \;;$$

$$\tilde{u}_2 = \tilde{u}_1 + 0.5\,\hat{w}_3 + 0.25\,(\hat{w}_6 + \hat{w}_8) + 0.05\,\hat{w}_{12} + 0.125\,\hat{w}_{16} + \left(0.25\,\hat{w}_{14} - 0.125\,\hat{w}_{19}\right)/3 \;;$$

$$\tilde{u}_3 = 0.5\,(\tilde{u}_1 + \tilde{u}_2) - 0.0625\,\hat{w}_6 - 0.046875\,\hat{w}_{12} - 0.03125\,\hat{w}_{16} \;;$$

$$\tilde{u}_4 = 3\,\tilde{u}_1 + \tilde{u}_2 - 3\,\tilde{u}_3 - 0.09375\,\hat{w}_{12} \;;\quad \tilde{u}_5 = 6\,\tilde{u}_1 + 3\,\tilde{u}_2 - 8\,\tilde{u}_3 - 0.375\,\hat{w}_{12} \;;$$

$$\tilde{u}_6 = \tilde{u}_1 + 0.5\,\hat{w}_4 + 0.25\,(\hat{w}_7 + \hat{w}_{10}) + 0.05\,\hat{w}_{13} + 0.125\,\hat{w}_{18} + \left(0.25\,\hat{w}_{15} - 0.125\,\hat{w}_{17}\right)/3 \;;$$

$$\tilde{u}_7 = 0.5\,(\tilde{u}_1 + \tilde{u}_6) - 0.0625\,\hat{w}_7 - 0.046875\,\hat{w}_{13} - 0.03125\,\hat{w}_{18} \;;$$

$$\tilde{u}_8 = 3\,\tilde{u}_1 + \tilde{u}_6 - 3\,\tilde{u}_7 - 0.09375\,\hat{w}_{13} \;;\quad \tilde{u}_9 = 6\,\tilde{u}_1 + 3\,\tilde{u}_6 - 8\,\tilde{u}_7 - 0.375\,\hat{w}_{13} \;;$$

$$\tilde{u}_{10} = \tilde{u}_2 + \tilde{u}_6 - \tilde{u}_1 + 0.25\,\hat{w}_9 + 0.125\,(\hat{w}_{17} + \hat{w}_{19} + \hat{w}_{20}) \;;$$

$$\tilde{u}_{11} = \tilde{u}_{10} + \tilde{u}_5 - \tilde{u}_2 - 0.5\,\hat{w}_9 - 0.25\,(\hat{w}_{19} + \hat{w}_{20}) \;;$$

$$\tilde{u}_{12} = \tilde{u}_{10} + \tilde{u}_9 - \tilde{u}_6 - 0.5\,\hat{w}_9 - 0.25\,(\hat{w}_{17} + \hat{w}_{20}) \;;\quad \tilde{u}_{13} = \tilde{u}_{11} + \tilde{u}_{12} - \tilde{u}_{10} + \hat{w}_9 + 0.5\,\hat{w}_{20} \;;$$

$$\tilde{u}_{14} = \tilde{u}_1 - \left(\hat{u}_{31} + \hat{u}_{32}\right)/36 + \left(9\,\hat{u}_{21} + 3\,\hat{u}_{22} + \hat{u}_{23}\right)/27 + \hat{u}_{24}/6 + \left(\hat{u}_{27} + \hat{u}_{30}\right)/18 \;;$$

$$\tilde{u}_{15} = \tilde{u}_{14} + \tilde{u}_{10} - \tilde{u}_1 + \left(6\,\hat{u}_{25} + 6\,\hat{u}_{26} + 2\,\hat{u}_{28} + 2\,\hat{u}_{29} + 3\,\hat{u}_{31} + 3\,\hat{u}_{32} + 3\,\hat{u}_{33} + 3\,\hat{u}_{34} + 3\,\hat{u}_{35}\right)/36 \;;$$

$$\tilde{u}_{16} = \tilde{u}_{15} + \tilde{u}_{11} - \tilde{u}_{10} - \left(6\,\hat{u}_{25} + 2\,\hat{u}_{28} + 3\,\hat{u}_{33} + 3\,\hat{u}_{34}\right)/18 \;;$$

$$\tilde{u}_{17} = \tilde{u}_{15} + \tilde{u}_{12} - \tilde{u}_{10} - \left(6\,\hat{u}_{26} + 2\,\hat{u}_{29} + 3\,\hat{u}_{34} + 3\,\hat{u}_{35}\right)/18 \;;$$

$$\tilde{u}_{18} = \tilde{u}_{16} + \tilde{u}_{17} - \tilde{u}_{15} + \hat{u}_9 + 0.5\,\hat{u}_{20} + \hat{u}_{34}/3 \;;\quad \tilde{u}_{19} = 2\,\tilde{u}_{14} - \tilde{u}_1 + \left(2\,\hat{u}_{22} + 2\,\hat{u}_{23} + \hat{u}_{27}\right)/9 \;;$$

$$\tilde{u}_{20} = \tilde{u}_{19} + \tilde{u}_2 - \tilde{u}_1 + \left(6\,\hat{u}_{25} + 4\,\hat{u}_{28} + 3\,\hat{u}_{31} + 3\,\hat{u}_{33}\right)/18 \;;$$

$$\tilde{u}_{21} = 2\,\tilde{u}_{19} - \tilde{u}_{20} + 0.5\,\hat{u}_6 + 0.25\,\hat{u}_{16} + \hat{u}_{31}/3 \;;$$

$$\tilde{u}_{22} = \tilde{u}_{19} + \tilde{u}_6 - \tilde{u}_1 + \left(6\,\hat{u}_{26} + 4\,\hat{u}_{29} + 3\,\hat{u}_{32} + 3\,\hat{u}_{35}\right)/18 \;;$$

$$\tilde{u}_{23} = 2\,\tilde{u}_{19} - \tilde{u}_{22} + 0.5\,\hat{u}_7 + 0.25\,\hat{u}_{18} + \hat{u}_{32}/3 \;;\quad \tilde{u}_{24} = \tilde{u}_1 - 3\,\tilde{u}_{14} + 3\,\tilde{u}_{19} + 2\,\hat{u}_{23}/9$$

(D.13)

As before, the conserved variables at the facial nodes in eqn. (D.13) are denoted with a tilde. The terms with a caret in eqn. (D.13) pertain to the space-time modes at the zone center. Just as we saw in the previous paragraph, the evaluation of each nodal value can be used to simplify the evaluation of the ones that follow it. Once the conserved variables are evaluated at the facial nodes, we can use them to obtain the fluxes at those nodes, denoted as before with a tilde over the flux terms. The modes in eqn. (D.12) are now obtained by

$$q_1 = 3\,(\tilde{f}_{20} - \tilde{f}_{21}) - 1.5\,(\tilde{f}_{15} - \tilde{f}_{16} + \tilde{f}_{17} - \tilde{f}_{18})\ ;\ \ q_2 = 1.5\,(\tilde{f}_{20} - \tilde{f}_{21} - \tilde{f}_2 + \tilde{f}_5)\ ;$$

$$q_3 = 3\,(\tilde{f}_{22} - \tilde{f}_{23}) - 1.5\,(\tilde{f}_{15} + \tilde{f}_{16} - \tilde{f}_{17} - \tilde{f}_{18})\ ;\ \ q_4 = 1.5\,(\tilde{f}_{22} - \tilde{f}_{23} - \tilde{f}_6 + \tilde{f}_9)\ ;$$

$$\widehat{f}_1 = \left(2\tilde{f}_1 + \tilde{f}_2 + \tilde{f}_5 + \tilde{f}_6 + \tilde{f}_9\right)/6\ ;\ \ \widehat{f}_4 = 2\,(\tilde{f}_2 - 2\tilde{f}_1 + \tilde{f}_5)\ ;\ \ \widehat{f}_5 = 2\,(\tilde{f}_6 - 2\tilde{f}_1 + \tilde{f}_9)\ ;$$

$$\widehat{f}_6 = \tilde{f}_{10} - \tilde{f}_{11} - \tilde{f}_{12} + \tilde{f}_{13}\ ;\ \ \widehat{f}_7 = 16\left(\tilde{f}_2 - 2\tilde{f}_3 + 2\tilde{f}_4 - \tilde{f}_5\right)/3\ ;$$

$$\widehat{f}_8 = 16\left(\tilde{f}_6 - 2\tilde{f}_7 + 2\tilde{f}_8 - \tilde{f}_9\right)/3\ ;\ \ \widehat{f}_9 = 2\,(\tilde{f}_{10} - 2\tilde{f}_6 + \tilde{f}_{11} - \tilde{f}_{12} + 2\tilde{f}_9 - \tilde{f}_{13})\ ;$$

$$\widehat{f}_{10} = 2\,(\tilde{f}_{10} - 2\tilde{f}_2 + \tilde{f}_{12} - \tilde{f}_{11} + 2\tilde{f}_5 - \tilde{f}_{13})\ ;\ \ \widehat{f}_2 = \tilde{f}_2 - \tilde{f}_5 + \widehat{f}_{10}/12 - 0.1\,\widehat{f}_7\ ;$$

$$\widehat{f}_3 = \tilde{f}_6 - \tilde{f}_9 + \widehat{f}_9/12 - 0.1\,\widehat{f}_8\ ;\ \ \widehat{f}_{13} = 4.5\,(\tilde{f}_{24} - \tilde{f}_1) - 13.5\,(\tilde{f}_{19} - \tilde{f}_{14})\ ;$$

$$\widehat{f}_{12} = 2.25\,(\tilde{f}_{24} - \tilde{f}_{19} - \tilde{f}_{14} + \tilde{f}_1 - 2\,\widehat{f}_{13}/3)\ ;$$

$$\widehat{f}_{11} = 1.5\,\tilde{f}_{24} - 2.5\,\tilde{f}_{14} + 0.25\,(\tilde{f}_{15} + \tilde{f}_{16} + \tilde{f}_{17} + \tilde{f}_{18}) - 0.25\,\widehat{f}_4 - 0.25\,\widehat{f}_5 - 4\,\widehat{f}_{12}/3 - 13\,\widehat{f}_{13}/9\ ;$$

$$\widehat{f}_{14} = 3\,q_2 - 2\,q_1 - 1.5\,\widehat{f}_{10}\ ;\ \ \widehat{f}_{16} = 3\,(q_1 - q_2) + 2.25\,\widehat{f}_{10}\ ;\ \ \widehat{f}_{15} = 3\,q_4 - 2\,q_3 - 1.5\,\widehat{f}_9\ ;$$

$$\widehat{f}_{17} = 3\,(q_3 - q_4) + 2.25\,\widehat{f}_9\ ;\ \ \widehat{f}_{18} = 3\,(\tilde{f}_{20} - 2\tilde{f}_{19} + \tilde{f}_{21}) - 1.5\,\widehat{f}_4\ ;$$

$$\widehat{f}_{19} = 3\,(\tilde{f}_{22} - 2\tilde{f}_{19} + \tilde{f}_{23}) - 1.5\,\widehat{f}_5\ ;\ \ \widehat{f}_{20} = 3\,(\tilde{f}_{15} - \tilde{f}_{16} - \tilde{f}_{17} + \tilde{f}_{18}) - 3\,\widehat{f}_6$$

(D.14)

The space-time integrals at the lower side of the top *x*-face of zone "*i,j,k*" in eqns. (55) and (57) are now given by

$$\langle f \rangle_{x+} = \widehat{f}_1 + 0.5\,\widehat{f}_{11} + \widehat{f}_{12}/3 + 0.25\,\widehat{f}_{13}\ ;$$

$$q_5 = \langle f \rangle_{x+} + (2\,\widehat{f}_4 + \widehat{f}_{18})/12\ ;\ \ q_6 = 0.5\,\widehat{f}_2 + 0.05\,\widehat{f}_7 + 0.25\,\widehat{f}_{14} + \widehat{f}_{16}/6\ ;$$

$$q_7 = \langle f \rangle_{x+} + (2\,\widehat{f}_5 + \widehat{f}_{19})/12\ ;\ \ q_8 = 0.5\,\widehat{f}_3 + 0.05\,\widehat{f}_8 + 0.25\,\widehat{f}_{15} + \widehat{f}_{17}/6\ ;$$

$$\langle f \rangle_{x+;y+} = q_5 + q_6\ ;\ \ \langle f \rangle_{x+;y-} = q_5 - q_6\ ;\ \ \langle f \rangle_{x+;z+} = q_7 + q_8\ ;\ \ \langle f \rangle_{x+;z-} = q_7 - q_8$$

(D.15)

The same approach described here can be used to obtain the space-time integrals of the *x*-flux at the lower *x*-boundary of each zone. This completes our description for obtaining the flux and electric field integrals at fourth order.

# References

[1]  D. S. Balsara, Linearized formulation of the Riemann problem for adiabatic and isothermal magnetohydrodynamics, Astrophysical Journal Supplement 116 (1998) 119

[2]  D. S. Balsara, Total variation diminishing algorithm for adiabatic and isothermal magnetohydrodynamics, Astrophysical Journal Supplement 116 (1998) 133

[3]  D. S. Balsara and D. S. Spicer, Maintaining pressure positivity in magnetohydrodynamic simulations, Journal of Computational Physics 148 (1999) 133-148

[4]  D. S. Balsara and D. S. Spicer, A staggered mesh algorithm using high order Godunov fluxes to ensure solenoidal magnetic fields in magnetohydrodynamic simulations, Journal of Computational Physics 149 (1999) 270-292

[5]  D. S. Balsara and C.-W. Shu, Monotonicity Preserving Weighted Non-oscillatory schemes wirh increasingly High Order of Accuracy, Journal of Computational Physics 160 (2000) 405-452

[6]  D. S. Balsara, Divergence-free adaptive mesh refinement for magnetohydrodynamics, Journal of Computational Physics 174 (2001) 614-648

[7]  D. S. Balsara, Total variation diminishing scheme for relativistic magneto-hydrodynamics, Astrophysical Journal Supplement 132 (2001) 83

[8]  D. S. Balsara, Second-order-accurate schemes for magnetohydrodynamics with divergence-free reconstruction, Astrophysical Journal Supplement 151 (2004) 149-184

[9] D. S. Balsara, C. Altmann, C.D. Munz, M. Dumbser, A sub-cell based indicator for troubled zones in RKDG schemes and a novel class oh hybrid RKDG+HWENO schemes, Journal of Computational Physics 226 (2007) 586-620

[10] D. S. Balsara, Divergence-free reconstruction of magnetic fields and WENO schemes for magnetohydrodynamics, Journal of Computational Physics 228 (2009) 5040

[11] D. S. Balsara, T. Rumpf, M. Dumbser & C.-D. Munz, Efficient, high-accuracy ADER-WENO schemes for hydrodynamics and divergence-free magnetohydrodynamics, Journal Computational Physics 228 (2009) 2480

[12] D.S. Balsara, Multidimensional Extension of the HLLE Riemann Solver; Application to Euler and Magnetohydrodynamical Flows, J. Comp. Phys., 229 (2010) 1970-1993

[13] T. J .Barth and P. O. Frederickson, Higher order solution of the Euler equations on unstructured grids using quadratic reconstruction, AIAA Paper no. 90-0013, 28[th] Aerospace Sciences Meeting, January (1990)

[14] M. Ben-Artzi and J. Falcovitz, A second-order Godunov-type scheme for compressible fluid dynamics, Journal of Computational Physics 55 (1984) 1-32

[15] M. Berger and P. Colella, Local adaptive mesh refinement for shock hydrodynamics, Journal of Computational Physics 82 (1989) 64-84

[16] J. U. Brackbill and D. C. Barnes, The effect of nonzero $\nabla \cdot \mathbf{B}$ on the numerical solution of the magnetohydrodynamic equations, Journal of Computational Physics 35 (1980) 426-430

[17]  S. H. Brecht, J. G. Lyon, J. A. Fedder, K. Hain, A simulation study of east-west IMF effects on the magnetosphere, Geophysical Reserach Lett. 8 (1981) 397

[18]  M. Brio and C.-C. Wu, An Upwind Differencing Scheme for the Equations of MHD, Journal of Computational Physics, 75 (1988) 400

[19]  B. Cockburn and C.-W. Shu, The Runge-Kutta discontinuous Galerkin method for Conservation Laws V, Journal of Computational Physics 141 (1998) 199-224

[20]  P. Colella and P. Woodward, The piecewise parabolic method (PPM) for gas-dynamical simulations, Journal of Computational Physics, 54 (1984) 174-201

[21]  P. Colella and M.D. Sekora, A limiter for PPM that preserves accuracy at smooth extrema, Journal of Computational Physics, 227 (2008) 7069

[22]  R. K. Crockett, P. Colella, R. T. Fisher, R. I. Klein & C. F. McKee, An unsplit cell-centered Godunov method for ideal MHD, Journal of Computational Physics 203 (2005) 422

[23]  W. Dai and P.R. Woodward, An approximate Riemann solver for ideal magnetohydrodynamics, Journal of Computational Physics 111 (1994) 354-372

[24]  W. Dai and P.R. Woodward, On the divergence-free condition and conservation laws in numerical simulations for supersonic magnetohydrodynamic flows, Astrophysical Journal 494 (1998) 317-335

[25]  A. Dedner, F. Kemm, D. Kröner, C.-D. Munz, T. Schnitzer, M. Wesenberg, Hyperbolic divergence cleaning for MHD equations, Journsl of Computational Physics 175 (2002) 645-673

[26]   C. R. DeVore, Flux-corrected transport techniques for multidimensional compressible magnetohydrodynamics, Journal of Computational Physics 92 (1991) 142-160

[27]   M. Dubiner, Spectral methods on triangles and other domains, Journal of Scientific Computing, 6 (1991) 345-390

[28]   M. Dumbser, M. Käser, Arbitary high order non-oscillatory finite volume schmes on unstuructured meshes for linear hyperbolic systems, Journal of Computational Physcis 221 (2007) 693-723

[29]   M. Dumbser, M. Käser, V.A. Titarev & E.F. Toro, Quadrature-free non-oscillatory finite volume schemes on unstructured meshes for nonlinear hyperbolic systems, Journal of Computational Physics 226 (2007) 204-243

[30]   M. Dumbser, D.S. Balsara, E.F. Toro, C.D. Munz, A unified framework for the construction of one-step finite volume and discontinuous Galerkin schemes on unstructured meshes, Journal of Computational Physics 227 (2008) 8209-8253

[31]   M. Dumbser, C. Enaux, E.F. Toro, Finite volume schemes of very high order of accuracy for stiff hyperbolic balance laws, Journal of Computational Physics 227 (2008) 3971-4001

[32]   M. . Dumbser and D.S. Balsara, High-Order Unstructured One-Step PNPM Schemes for the Viscous and Resistive MHD Equations, accepted, Computer Modeling for Engineers and Scientists

[33]   C.R. Evans and J.F. Hawley, Simulation of magnetohydrodynamic flows: a constrained transport method, Astrophys.J., vol. 332, pg. 659, (1989)

[34]   B. Einfeldt, C.-D. Munz, P.L. Roe and B. Sjogreen, On Godunov type methods near low densities, Journal of Computational Physics 92 (1991) 273-295

[35] S. A. E. G. Falle, S. S. Komissarov and P. Joarder, A multidimensional upwind scheme for magnetohydrodynamics, Monthly Notices of the Royal Astronomical Society 297 (1998), 265-277

[36] F. Fuchs, S. Mishra & N.H. Risebro, Splitting based finite volume schemes for the ideal MHD equations, J. Comput. Phys., 228(3) (2009) 641-660

[37] T. Gardiner and J.M. Stone, An unsplit Godunov method for ideal MHD via constrained transport, Journal of Computational Physics, 205 (2005) 509

[38] S. Gottlieb and C. Shu , Total-variation-diminishing Runge–Kutta schemes. Math. Comput. 67 (1998), pp. 73–85

[39] Harten, A. Lax, P.D. and van Leer, B, On upstream differencing and Godunov-type schemes for hyperbolic conservation laws, SIAM Rev. 25 (1983) 289-315

[40] A. Harten and J. Hyman, Self-adjusting grid methods for one-dimensional hyperbolic conservation laws, J. Comput. Phys., 50 (1983) 297-322

[41] A. Harten, B. Engquist, S.Osher and S. Chakravarthy, Uniformly high order essentially non-oscillatory schemes III, Journal of Computational Physics, 71 (1987) 231-303

[42] M.K. Horn, Fourth and fifth order, scaled Runge-Kutta algorithms for treating dense output, SIAM J. Numer. Anal., 20(3) (1983) 588

[43] A. Jeffrey and T. Taniuti, Non-linear Wave Propagation, Academic Press (1964)

[44] G.-S. Jiang and C.-W. Shu, Efficient implementation of weighted ENO schemes, Journal of Computational Physics 126 (1996) 202-228

[45]   X.-D. Liu, S. Osher and T. Chan, Weighted essentially non-oscillatory schemes, Journal of Computational Physics 115 (1994) 200-212

[46]   P. Londrillo and L. DelZanna, On the divergence-free condition in Godunov-type schemes for ideal magnetohydrodynamics: the upwind constrained transport method, Journal of Computational Physics 195 (2004) 17-48

[47]   K. G. Powell, An approximate Riemann solver for MHD (That actually works in more than one dimension), ICASE Report 94-24

[48]   J. Qiu & C.-W. Shu, Hermite WENO schemes and their application as limiters for Runge-Kutta discontinuous Galerkin schemes : The one dimensional case, Journal of Computational Physics, 193 (2004) 115

[49]   P.L. Roe, Approximate Riemann Solvers, Parameter Vectors, and Difference Schemes, Journal of Computational Physics, 43 (1981) 357-372

[50]   P. L. Roe and D. S. Balsara, Notes on the eigensystem of magnetohydrodynamics, SIAM Journal of applied Mathematics 56 (1996), 57

[51]   D. Ryu and T.W. Jones, Numerical MHD in astrophysics: algorithm and tests for one-dimensional flow, Astrophysical Journal 442 (1995) 228

[52]   D. Ryu, F. Miniati, T. W. Jones, and A. Frank, A divergence-free upwind code for multidimensional magnetohydrodynamic flows, Astrophysical Journal 509 (1998) 244-255

[53]   T. Schwartzkopff, M. Dumbser & C.-D. Munz, Fast high order ADER schemes for linear hyperbolic equations, Journal of Computational Physics 197 (2004) 532

[54] C.-W. Shu and S. J. Osher, Efficient implementation of essentially non-oscillatory shock capturing schemes, Journal of Computational Physics 77 (1988) 439-471

[55] C.-W. Shu and S. J. Osher, Efficient implementation of essentially non-oscillatory shock capturing schemes II, Journal of Computational Physics 83 (1989) 32-78

[56] R.J. Spiteri and S.J. Ruuth, A new class of optimal high-order strong-stability-preserving time-stepping schemes. *SIAM J. Numer. Anal.* 40 (2002), pp. 469–491

[57] R.J. Spiteri & S.J. Ruuth, Non-linear evolution using optimal fourth-order strong-stability-preserving Runge-Kutta methods, Mathematics and Computers in Simulation 62 (2003) 125-135

[58] A. Suresh and H.T. Huynh, Accurate monotonicity preserving scheme with Runge-Kutta time-stepping, Journal of Computational Physics 136 (1997) 83-99.

[59] A. Taube, M. Dumbser, D. S. Balsara and C. D. Munz, Arbitrary high order discontinuous Galerkin schemes fort the magnetohydrodynamic equations, Journal of Scientific Computing 30 (2007) 441-464

[60] V.A. Titarev and E.F. Toro, ADER: arbitrary high order Godunov approach, Journal of Scientific Computing 17 (1-4) (2002) 609-618

[61] V.A. Titarev and E.F. Toro, ADER schemes for three-dimensional nonlinear hyperbolic systems, Journal of Computational Physics 204 (2005) 715-736

[62] E.F. Toro and V.A. Titarev, Derivative Riemann solvers for systems of conservation laws and ADER methods, Journal of Computational Physics 212 (1) (2006) 150-    165

[63]   S.D. Ustyugov, M.V. Popov, A.G. Kritsuk & M.L. Norman, Piecewise parabolic method on a local stencil for supersonic turbulence simulations, J. Comput. Phys. 228 (2009) 7614

[64]   B. van Leer, Towards the Ultimate Conservative Difference Scheme V. A Second Order Sequel to Godunov's Method, J. Computational Phys., 32 (1979) 101-136

[65]   P. Woodward and P. Colella, The numerical simulation of two-dimensional fluid flow with strong shocks, Journal of Computational Physics 54 (1984), 115-173

[66]   Z. Xu, Y. Liu and C.-W. Shu, Hierarchical reconstruction for discontinuous Galerkin methods on unstructured grids with a WENO-type linear reconstruction and partial neighboring cells, Journal of Computational Physics 228 (2009a) 2194

[67]   Z. Xu, Y. Liu and C.-W. Shu, Hierarchical reconstruction for spectral volume method on unstructured grids, Journal of Computational Physics 228 (2009b) 5787

[68]   K.S. Yee, Numerical Solution of Initial Boundary Value Problems Involving Maxwell Equation in an Isotropic Media, IEEE Trans. Antenna Propagation 14 (1966) 302

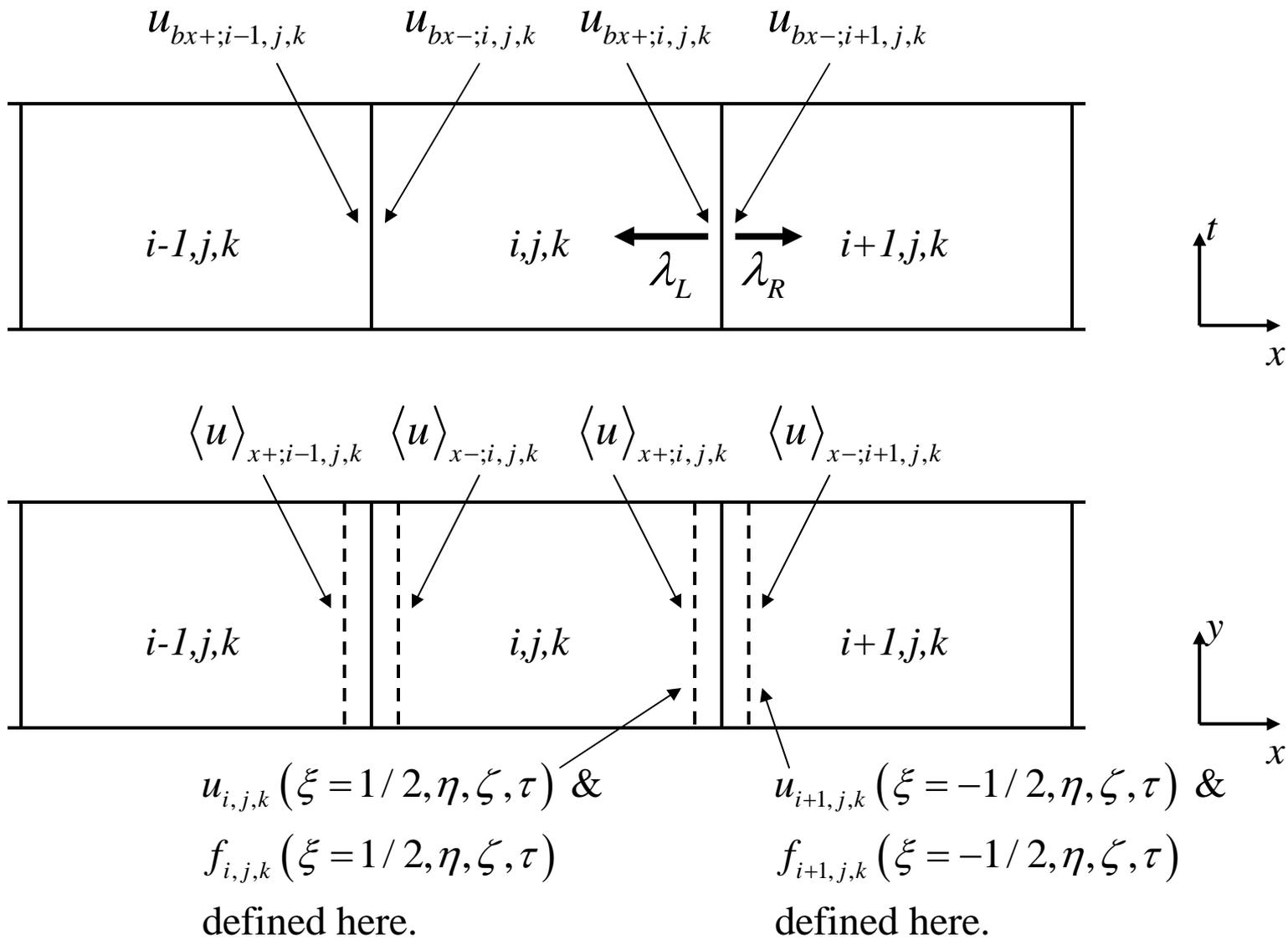

*Fig.1) The upper panel shows the space-time centered barycentric variables at the x-faces of a one dimensional mesh. The lower panel shows the facially averaged integrals at the x-faces of a two dimensional mesh. The dashed lines show where the integrals are done. Because we only need to focus on the x-directional variation, we only show one row of zones in this figure.*

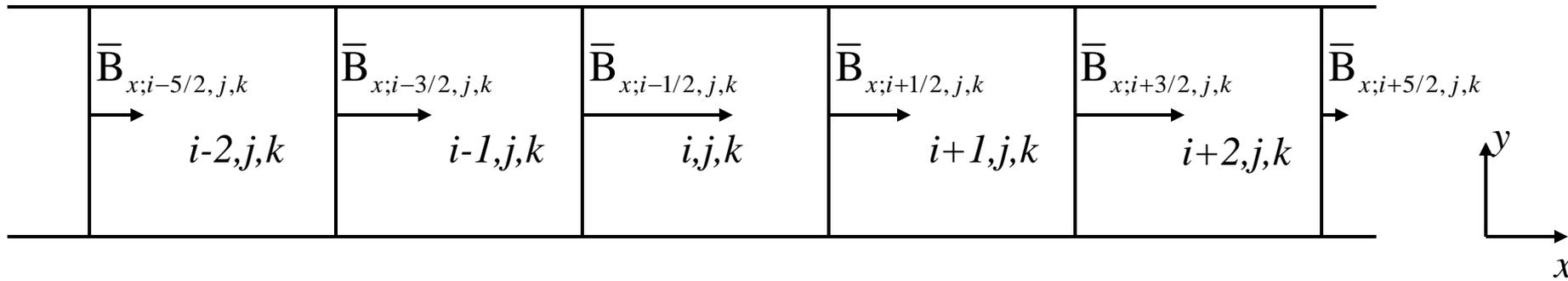

*Fig. 2) shows the facially averaged x-components of the magnetic field on a two dimensional mesh. Because we only need to focus on the x-directional variation, we only show one row of zones in this figure.*

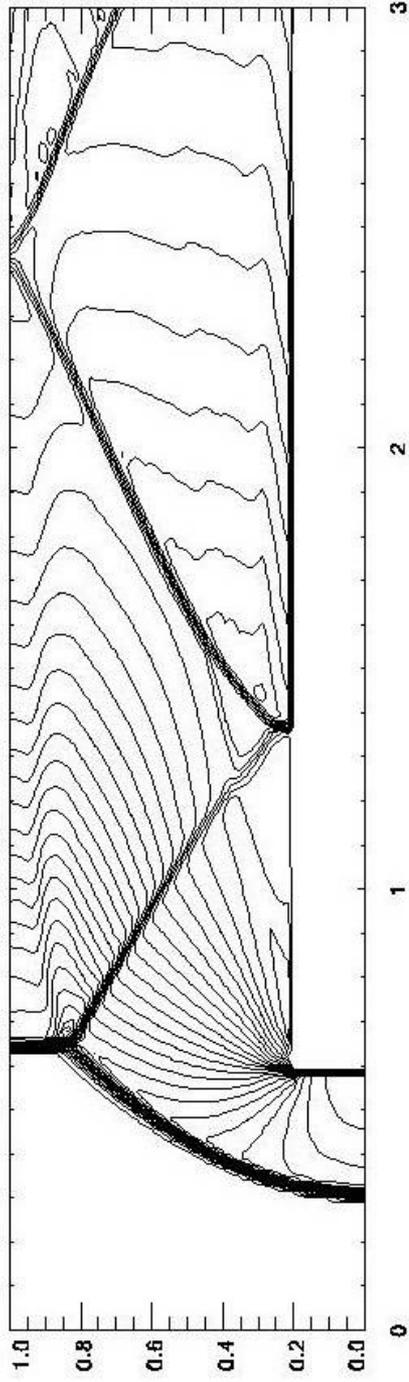 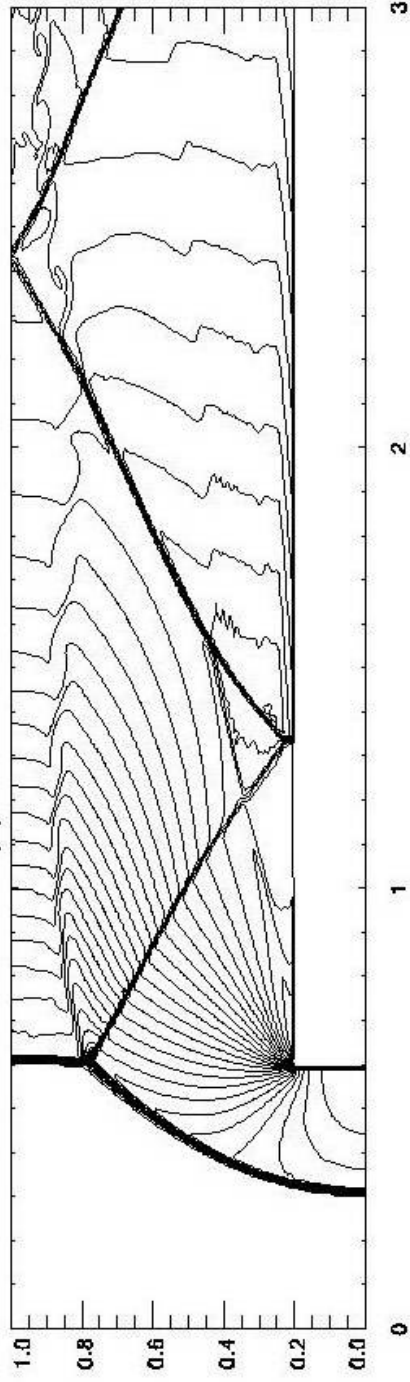 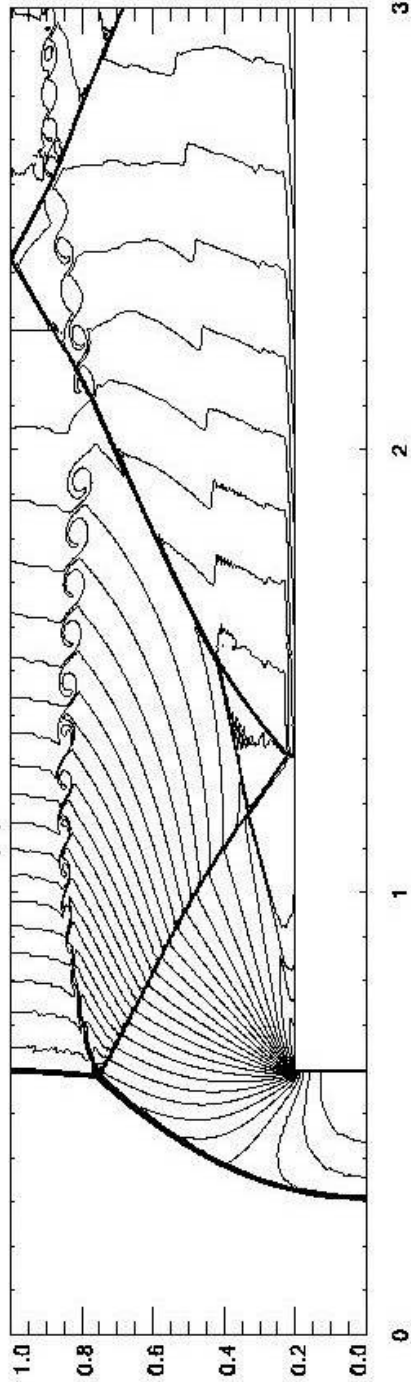

*Fig. 3) This resolution study shows the density variable from the forward facing step problem at resolutions of 240X80, 480X160 and 960X320 zones at a time of 4 units. Thirty equally spaced contours are shown in the density variable ranging from 0.105 to 6.699. The fourth order scheme with a linearized Riemann solver was used. We see the beginnings of the vortex sheet roll-up at a resolution of 480X160 zones and the 960X320 zone simulation captures the roll-up very clearly.*

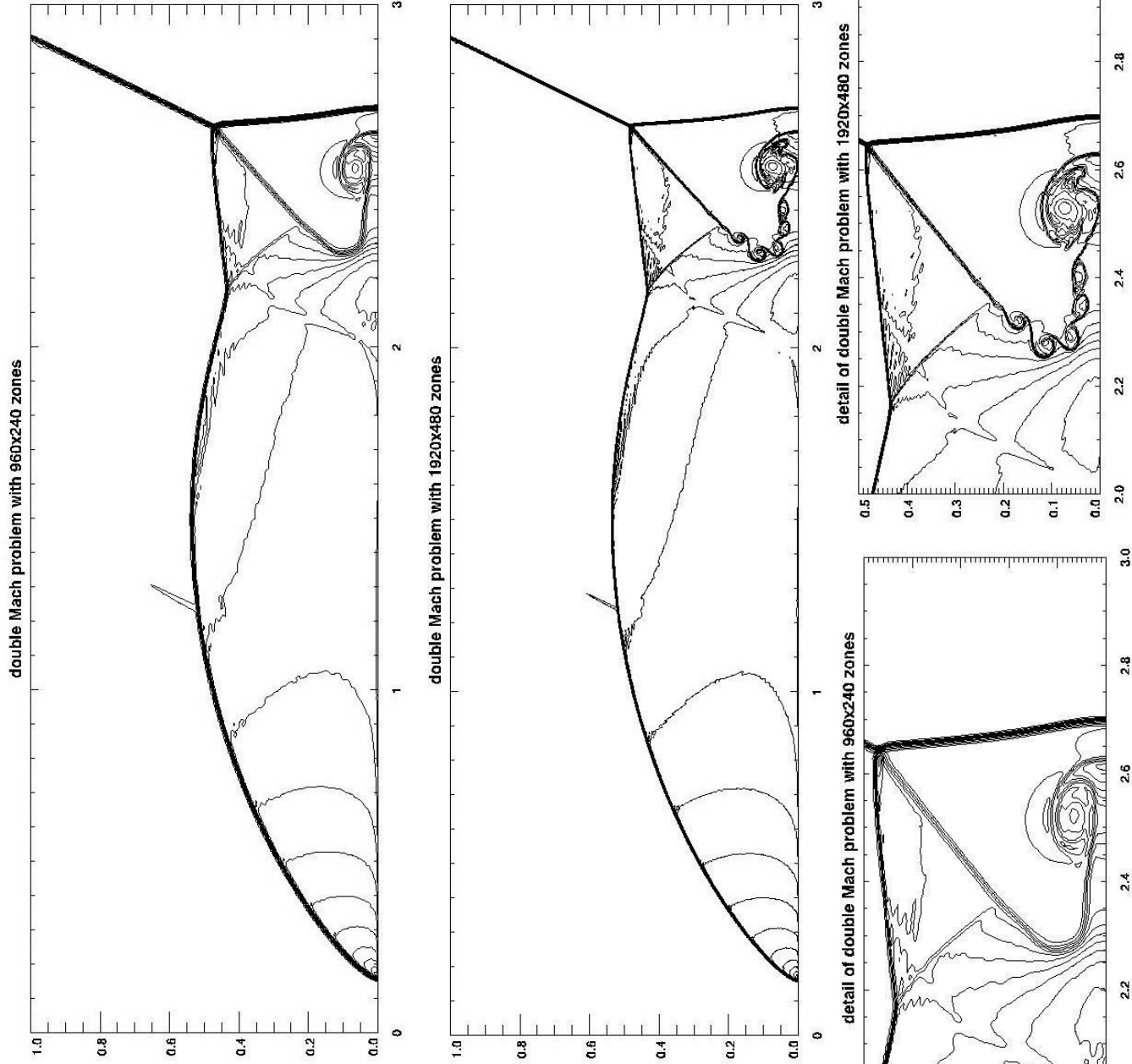

*Fig. 4) Shows a resolution study of the double Mach reflection of a strong shock. The 1st and 2nd panels show the density from 960X240 and 1920X480 zone simulations. The lowest two panels show details at the Mach stem for each of those two simulations with the lowest left panel corresponding to the lower resolution simulation. The 4rd order ADER-WENO scheme with a linearized Riemann solver was used. 30 contours were fit between a range of 1.4 and 20.975. We clearly see the roll up of the Mach stem due to Kelvin-Helmholtz instability in the higher resolution simulation.*

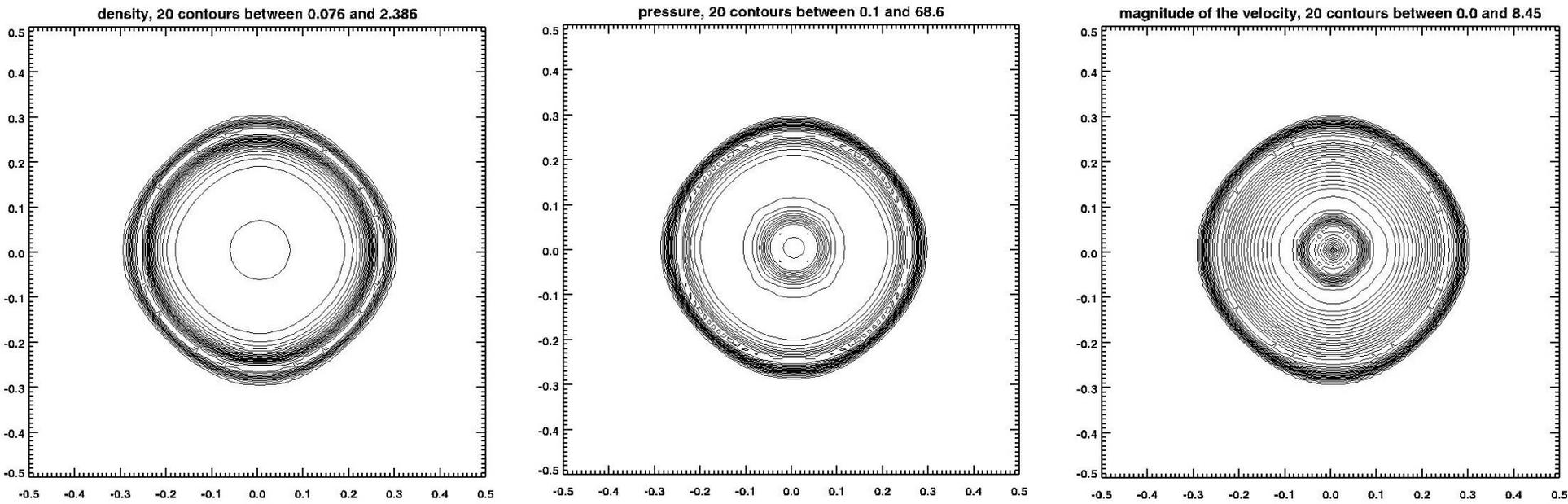

*Fig. 5) Shows the density, pressure and magnitude of the velocity for the hydrodynamical blast wave problem run with an ADER-WENO scheme and linearized Riemann solver at fourth order. 20 contours were fit between the lowest and highest values shown on each figure.*

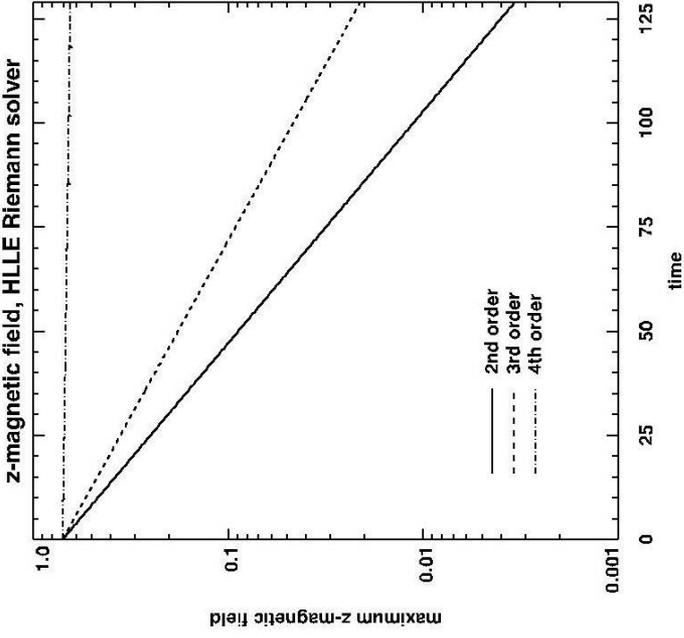 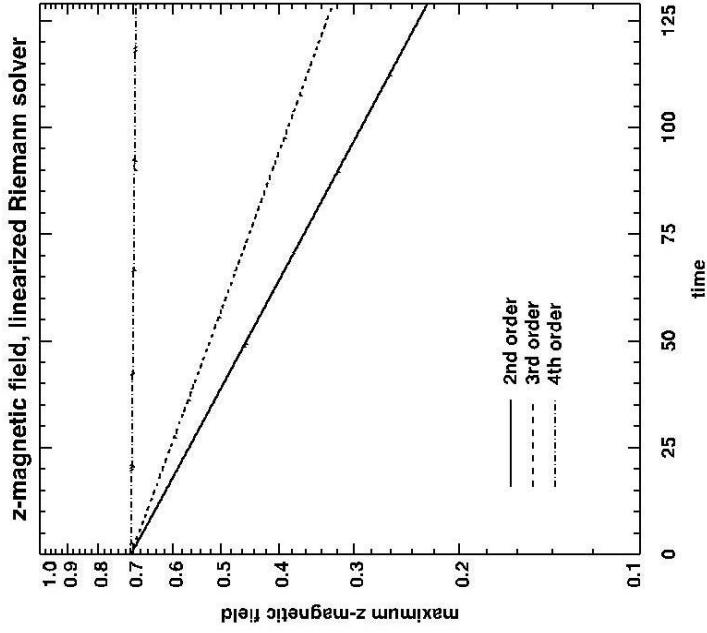

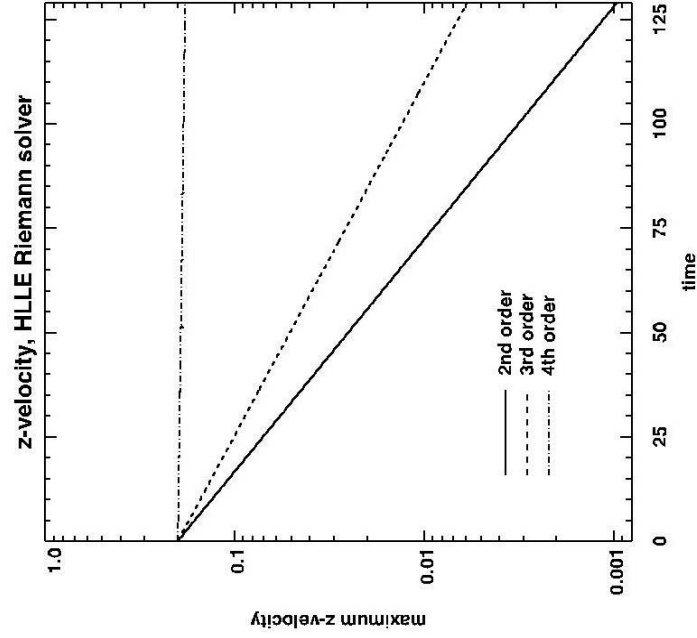 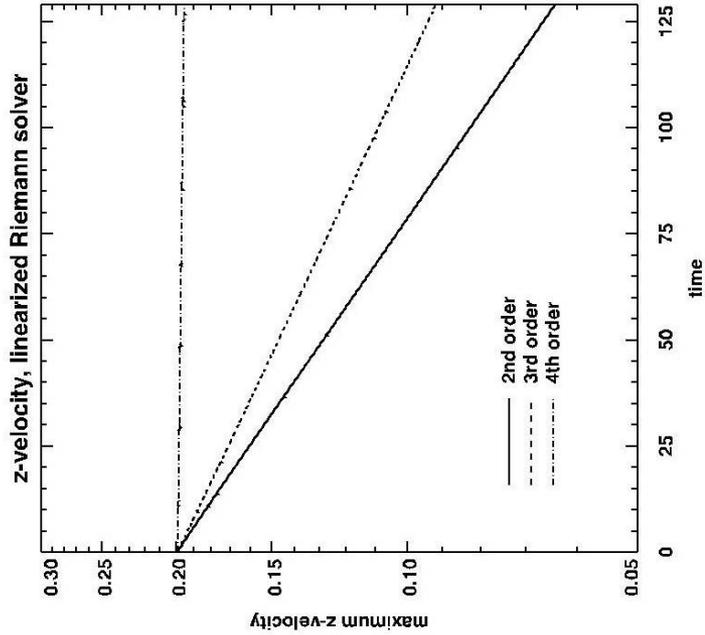

*Fig 6) The log-linear plots show the decay of torsional Alfven waves that are made to propagate obliquely on a two dimensional square. The above two panels show the decay of the maximum z-velocity and the maximum z-component of the magnetic field when second, third and fourth order schemes are used with an HLLE Riemann solver. The lower two panels show the same information when a linearized Riemann solver is used. Notice that the decay is substantially reduced with increasing order. Notice too that the linearized Riemann solver provides a substantial improvement to the solution, especially at lower orders.*

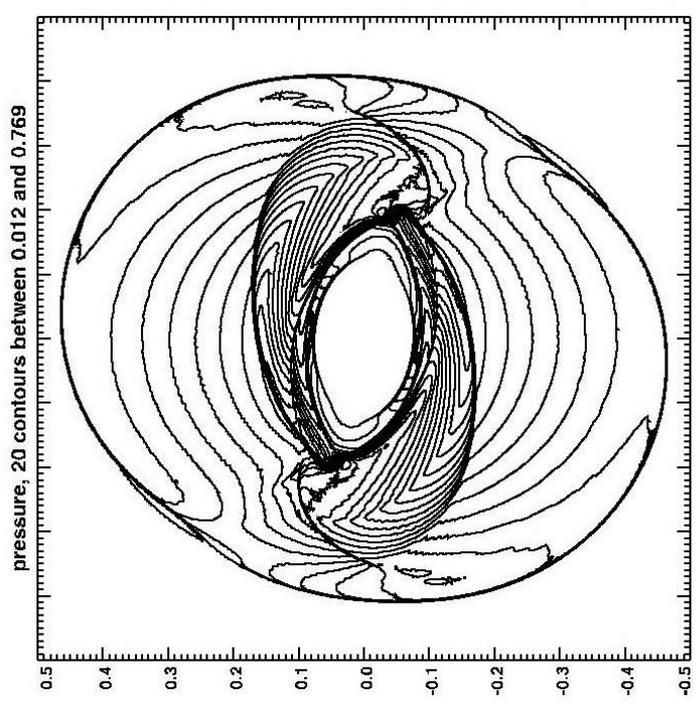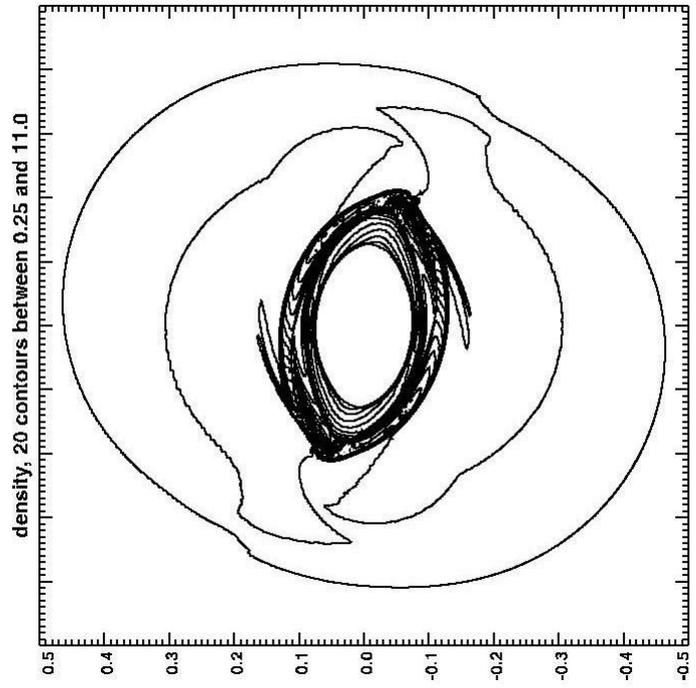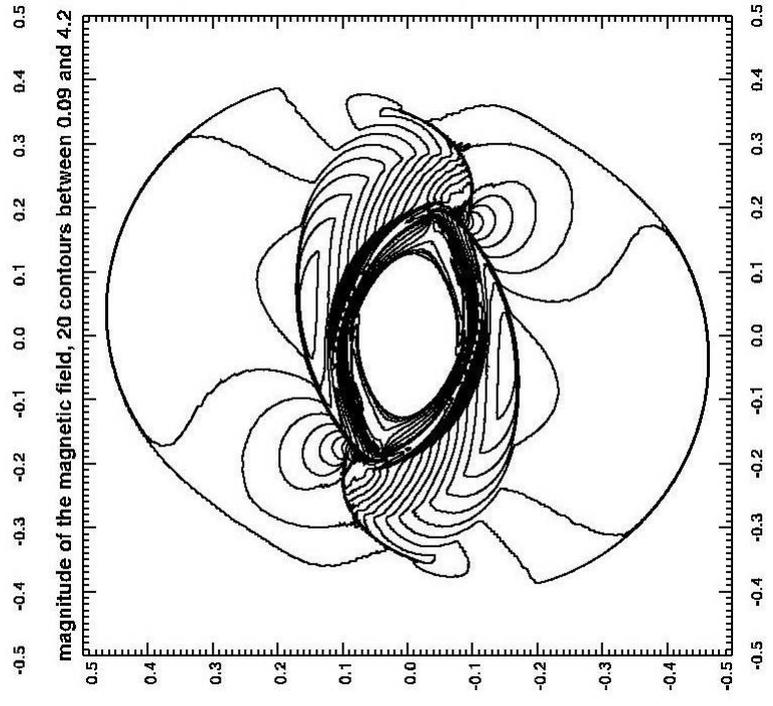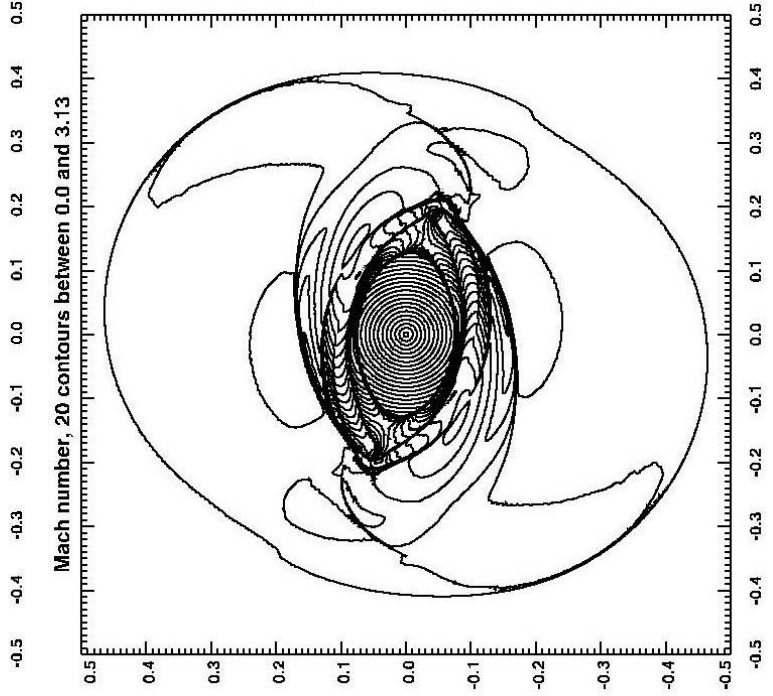

Fig 7) Rotor problem where the top two panels show the density and pressure and the bottom two panels show the Mach number and the magnitude of the magnetic field. The fourth order nodal ADER-WENO scheme with a linearized Riemann solver was used. 20 contours are shown for each figure with the min and max values catalogued above the panels.

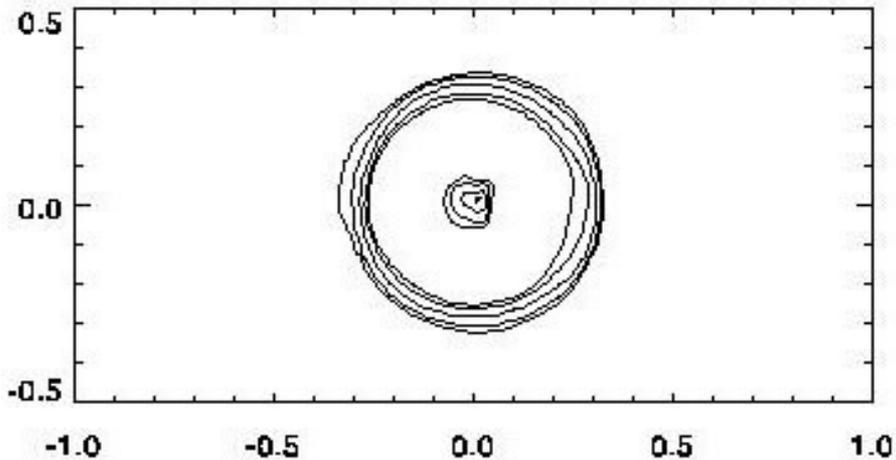
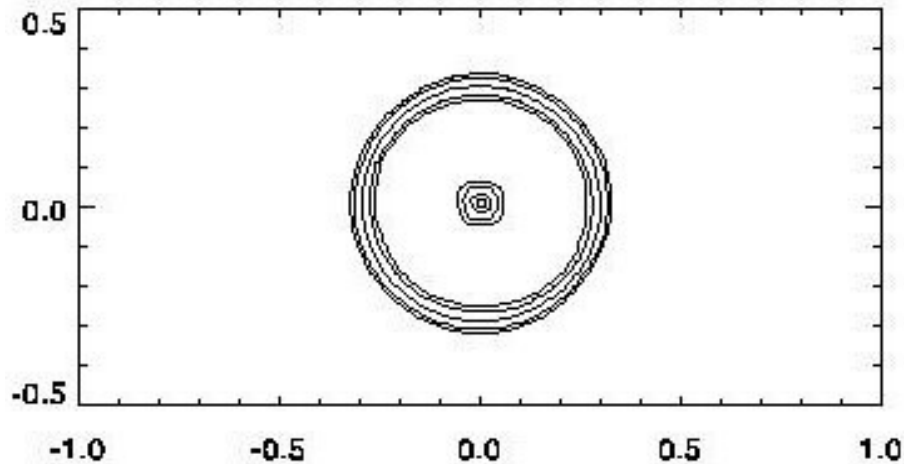
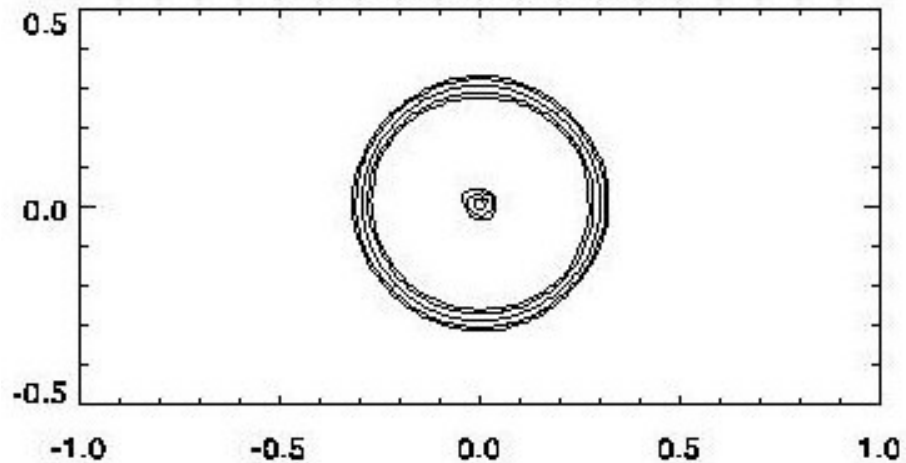

Fig. 8 shows the magnitude of the magnetic field for the field loop advection problem. Five contours were fitted between the minimum and the maximum values of the magnetic field. The panels show the ADER-WENO schemes with HLL Riemann solvers at second, third and fourth orders. The loop is advected along the diagonal of the rectangular domain shown here. A 128×64 zone mesh was used. The plot shows the field loop after it has executed one complete orbit around the computational domain.

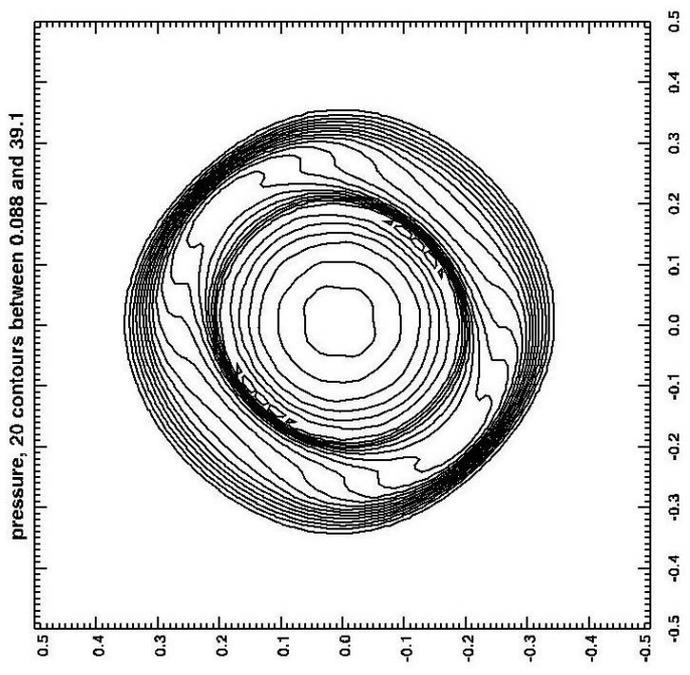
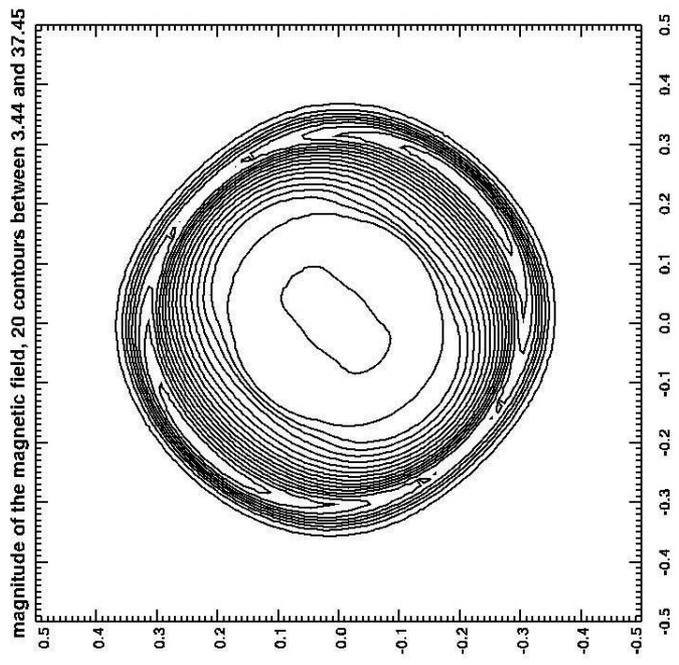
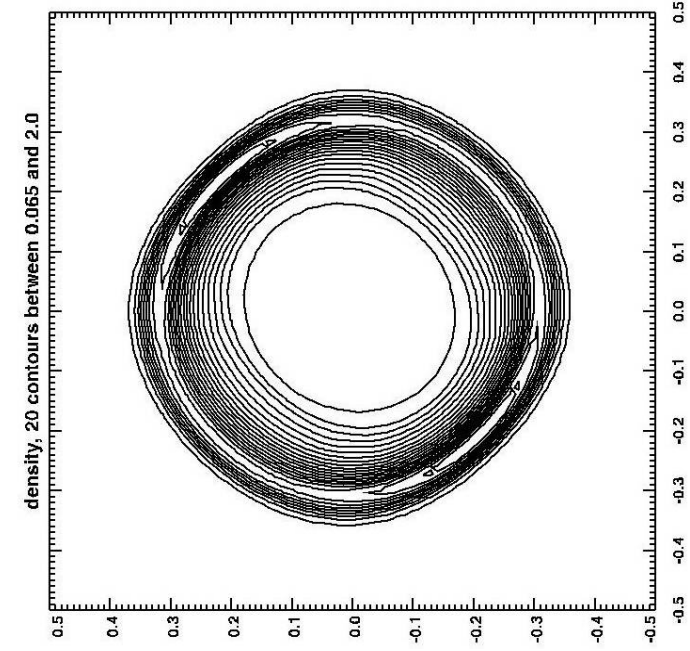
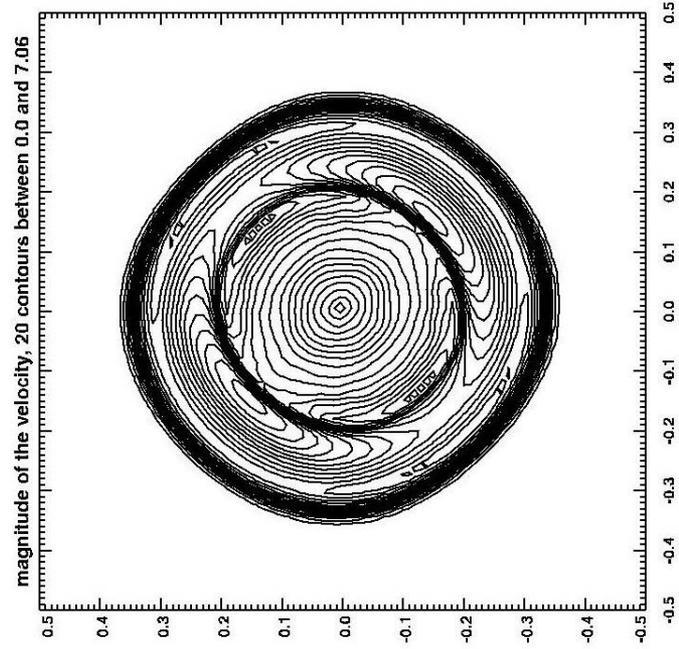

*Fig. 9) Shows the density, pressure, magnitude of the velocity and magnitude of the magnetic field for the MHD blast wave problem run with an ADER-WENO scheme and linearized Riemann solver at fourth order. 20 contours were fit between the lowest and highest values shown on each figure.*